\documentclass[%
reprint,%
 amssymb, amsmath,%
 aip,jcp,%
superscriptaddress,
]{revtex4-1}

\usepackage{amsmath}
\usepackage{docs}%
\usepackage{bm}%
\usepackage[colorlinks=true,linkcolor=blue,citecolor=blue,urlcolor=black]{hyperref}%
\usepackage{graphicx}
\usepackage[utf8]{inputenc}
\usepackage{epstopdf}
\usepackage{mdframed}
\usepackage{xcolor}
\usepackage{bold-extra}
\mdfdefinestyle{MyFrame}{%
linecolor=black,
outerlinewidth=20pt,
roundcorner=30pt,
innertopmargin=\baselineskip,
innerbottommargin=\baselineskip,
innerrightmargin=30pt,
innerleftmargin=30pt,
backgroundcolor=gray!10!white}
\expandafter\ifx\csname package@font\endcsname\relax\else
\expandafter\expandafter
\expandafter\usepackage
\expandafter\expandafter
\expandafter{\csname package@font\endcsname}%
\fi
\begin{document}

\title[]{
Interaction between water and carbon nanostructures: How good are current density functional approximations?
} 

\author{Jan Gerit Brandenburg}
\email{j.g.brandenburg@gmx.de}
\affiliation{
Interdisciplinary Center for Scientific Computing,
University of Heidelberg, Im Neuenheimer Feld 205A, 69120 Heidelberg, Germany}
\author{Andrea Zen}
\affiliation{Department of Earth Sciences, University College London, Gower Street, London WC1E 6BT, United Kingdom}
\affiliation{Thomas Young Centre and London Centre for Nanotechnology, 17-19 Gordon Street, London WC1H 0AH, United Kingdom}
\author{Dario Alf\`{e}}
\affiliation{Department of Earth Sciences, University College London, Gower Street, London WC1E 6BT, United Kingdom}
\affiliation{Thomas Young Centre and London Centre for Nanotechnology, 17-19 Gordon Street, London WC1H 0AH, United Kingdom}
\affiliation{Dipartimento di Fisica Ettore Pancini, Università di Napoli Federico II, Monte S. Angelo, I-80126 Napoli, Italy}
\author{Angelos Michaelides}
\email{angelos.michaelides@ucl.ac.uk}
\affiliation{Department of Physics and Astronomy, University College London, Gower Street, London WC1E 6BT, United Kingdom}
\affiliation{Thomas Young Centre and London Centre for Nanotechnology, 17-19 Gordon Street, London WC1H 0AH, United Kingdom}
\date{\today}

\begin{abstract}
Due to their current and future technological applications, including realisation of water filters and desalination membranes, water adsorption on graphitic sp$^{2}$-bonded carbon is of overwhelming interest. However, these systems are notoriously challenging to model, even for electronic structure methods such as density functional theory (DFT), because of the crucial role played by London dispersion forces and non-covalent interactions in general.  Recent efforts have established reference quality interactions of several carbon nanostructures interacting with water. Here, we compile a new benchmark set (dubbed \textbf{WaC18}), which includes a single water molecule interacting with a broad range of carbon structures, and various bulk (3D) and two dimensional (2D) ice polymorphs.  The performance of 28 approaches, including semi-local exchange-correlation functionals, non-local (Fock) exchange contributions, and long-range van der Waals (vdW) treatments, are tested by computing the deviations from the reference interaction energies.  The calculated mean absolute deviations on the WaC18 set depends crucially on the DFT approach, ranging from 135 meV for LDA to 12 meV for PBE0-D4. We find that modern vdW corrections to DFT significantly improve over their precursors.  Within the 28 tested approaches, we identify the best performing within the functional classes of: generalized gradient approximated (GGA), meta-GGA, vdW-DF, and hybrid DF, which are BLYP-D4, TPSS-D4, rev-vdW-DF2, and PBE0-D4, respectively. 
\end{abstract}

\maketitle 

\bigskip


\section{Introduction}
\label{sec:intro}

The interaction of molecules or liquids with nanostructured surfaces is central to many real-life applications, including 
catalysis, gas storage, desalination, and more. 
Interfaces involving water and carbon show unique and fascinating behavior, which can be employed in important applications, for instance in water purification devices.\cite{Fumagalli2018_dielectric, Abraham2017, Joshi2014_MolSieving, Nair2012, AlgaraSiller:2015_2Dice, water_cnt_secchi, md_w@capillaries,Hummer:2001gd, Strogatz:2005kj, Holt:2006kr, water_cnt_secchi}
Topologically similar materials can have substantially different properties~\cite{Tocci2014_friction, Michaelides:2016_Nature_news, Esfandiar:2017_transport, Bocquet-NatRevChem-2017, bocquet-naturecomm-2018} emphasizing that the understanding of the nature of the interaction has to be sought at a quantum mechanical electronic structure level.

Density functional theory (DFT) is the simulation method of choice for many materials applications due to its favorable accuracy to computational cost ratio.\cite{Kieron-JCP,Becke-JCP,Truhlar-JCP,dft-materials-rev} 
Modern density functional approximations (DFAs) combine semi-local expansions of the exchange-correlation with long-range corrections for missing London dispersion interactions, i.e.\ the attractive part of the van der Waals (vdW) forces. The approximations are physically motivated, but additionally require adjustment of a small number of parameters, which are either based on exact constraints or on empirical data. 
Adjusting the parameters to optimally describe short, long, and middle-ranges of interactions is challenging,\cite{scand3,dfa_disp_balance} and a solution that is good for interactions between small molecules is not necessarily good for the interaction of a molecule with extended surfaces or within the bulk (e.g., molecules in solutions, molecular crystals).\cite{dft-materials-rev,water_at_hbn}
In either case, it is mandatory to carefully benchmark the DFT methods, especially as the 'zoo' of methodologies is growing and it is often unclear what is the expected reliability of a possible DFT setup, so how to pick the best DFT flavor for a specific application.
In the past decade, these DFA benchmarks mainly focused on molecular properties with recent studies testing more than 200 DFAs on thousands of references including thermochemistry, kinetics, and non-covalent interactions.\cite{gmtkn55,wb97mv,b97mv,Mardirossian2017,revm06l}
Recently, some focus of DFT benchmarking moved to the description of equilibrium geometries.\cite{pbeh3c,headgordan_geom,ccse21,rot25}
Similar large-scale benchmarks for condensed phase properties are much more rare, which is mainly due to the lack of theoretical reference data. 
While for bulk solids experimental lattice constants and cohesive energies have been used successfully for DFT benchmarking,\cite{dftbench-bulk,ss20,pob,dftbench_bs64}
similar data for more weakly bound molecular crystals have substantially higher uncertainty.\cite{c21, x23, ice10}
This is due to the experimental measurement uncertainty,\cite{Chickos2003_DHerror} indirect measurements that cannot directly be compared to simple equilibrium geometries and energies, or the challenge to do the measurement itself. 
The latter point holds for a single water molecule adsorbed at surfaces as water readily forms clusters.\cite{water-interface}

Concerning theoretical reference calculations, exciting progress has been made in the field of high-level wavefunction methods. On the one hand, embedding techniques~\cite{cryscor_embedding,manby_embedding, csp_beran}  and local approaches of coupled cluster theories~\cite{dlpnoccsdt,dlpno_compact,ao_lmp2, werner_lcc,lno_ccsdt, i-ccsdt} have made the gold standard of quantum chemistry applicable to molecular systems with a few hundred atoms and molecular crystals of small molecules.\cite{beran_chemrev, bzsolid_ccsdt, ccsdt_surfaces}
On the other hand, new algorithmic developments in the field of diffusion Monte Carlo (DMC, a quantum Monte Carlo technique) have made the computation of chemically accurate lattice energies of small molecular crystals feasible within reasonable computational effort.\cite{dmc_sizeconsistent,molcryst_dmc}
These developments together with an increased capacity of available computational resources  makes the interaction energy determination of extended systems feasible. This is an important step towards the better understanding of large non-covalently interacting systems as recently highlighted.\cite{advancesQC_editorial,largeNCI_perspective,dft-materials-rev}

Here, we capitalize on this by gathering the benchmark quality interaction energies of water with carbon nanomaterials that we have studied in the past few years. This involves the adsorption of water on graphene,\cite{water-graphene-dmc} water on benzene and coronene as two representative aromatic hydrocarbons (abbreviated as AH),\cite{water-graphene-dmc} and water on a carbon nanotube (CNT).\cite{yasmine_water_cnt}
In practical applications it is important to describe correctly also the interaction between water molecules, thus we additionally analyze different phases of two-dimensional (2D) ice\cite{2dice-dmc} as well as bulk (3D) ice polymorphs.\cite{molcryst_dmc}
Therefore, we obtain a dataset of eighteen configurations and associated reference interaction energies, dubbed \textbf{WaC18} set.
We test as many as 28 DFAs on the WaC18 set, including several  recently developed vdW corrections to DFAs.
Some DFT benchmarks already exist on these or related system types,\cite{water_graphene_dmc_old, water_graphene_paesani,2dice-dmc, yasmine_water_cnt, ice_dmc_dft, hirata-ice, hirata-fragmentation-ice, graphene-metal-dft, water@graphene_paulus, water-cnt_ccsdt, water@graphene_dft1, water@graphene_dft2,water@graphene_dft2b, water@graphene_dft3, water-graphene-pbe, water-graphene-hesselmann, water@graphene_dft/cc, water@graphene_sapt2,water_graphene_jordancomment}
but they typically include a limited set of DFAs, the recent vdW developments are not included, and the used reference data is not equally well converged. 
Here, we will address all of these issues.

The WaC18 benchmark test will help in understanding the essential ingredients needed to describe seamlessly both the strong hydrogen bonds and the weak interaction with surfaces. 
Furthermore, the identification of the most accurate DFAs can be used by researchers aiming to describe these widely spread system types, complementing and updating the perspective in Ref.~\onlinecite{dftforwater_perspective} that focused on DFT recommendations for water.

In Section~\ref{sec:setup}, we describe the benchmark systems considered, discuss the best estimates of the interaction energies including additional DMC calculations to have equally well converged reference data, and give the computational details of the DFT calculations. 
Following this, we report the results of a variety of DFAs and vdW corrections from several functional classes and analyze the critical aspects determining the DFA-vdW performance (\ref{sec:results}).
Conclusions and a future perspective are given in section~\ref{sec:conclusion}.

\begin{figure*}[bht]
\centering
\includegraphics[width=0.99\textwidth]{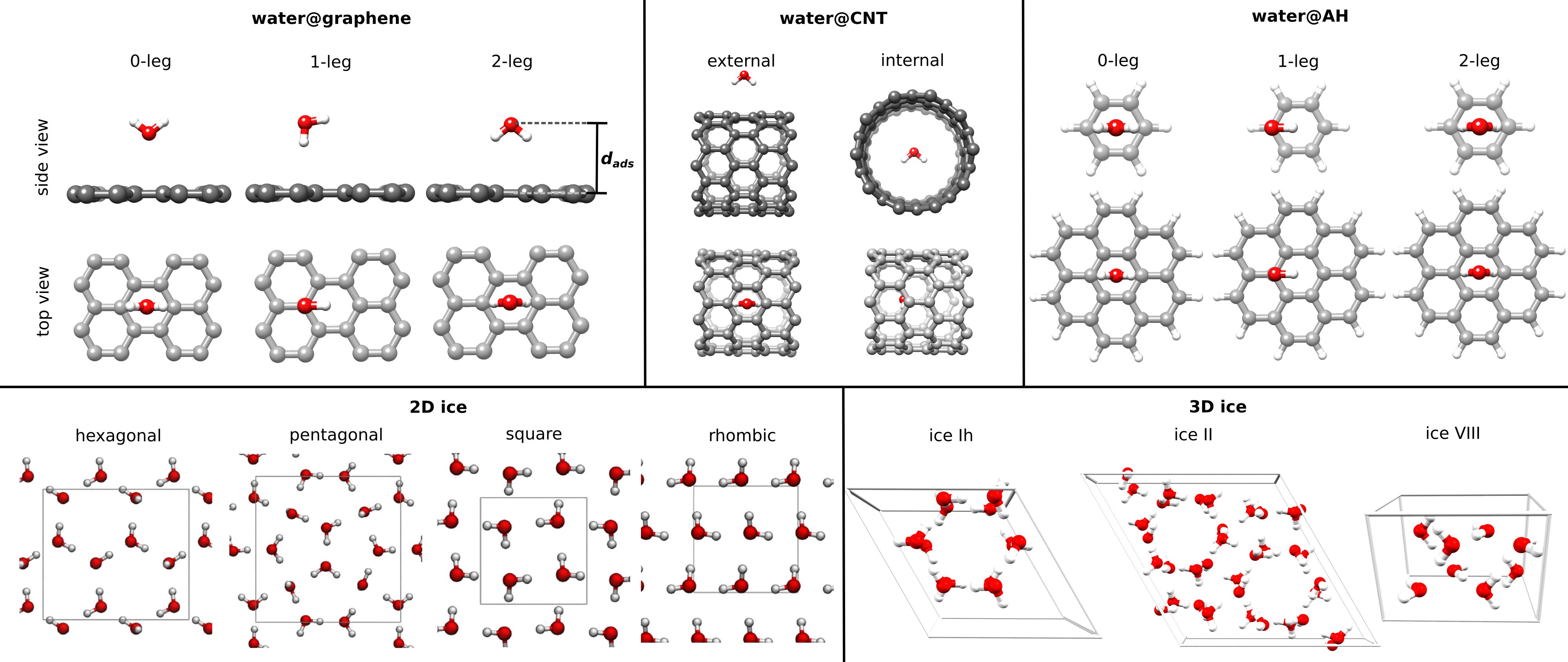}
\caption{\label{fig:system} 
\textbf{WaC18} benchmark set: Structures are shown as provided in the Supporting Information~\cite{suppinfo}. Top panel shows the water adsorption structures: 0-leg, 1-leg, and 2-leg motif on graphene, benzene, and coronene, as well as the adsorption inside and outside of a CNT. 
Lower panel shows the primitive unit cells of 2D and 3D ice structures. 
}
\end{figure*}


\section{Benchmark Setup}
\label{sec:setup}

\subsection{Systems under consideration}
\label{subsec:systems}

In our current benchmark, we will focus on water interacting with carbon nanostructures and ice. 
We separate the analysis into a single water molecule interacting with graphene, a carbon nanotube (CNT), aromatic hydrocarbons (AHs), and the interaction of solid water in three-dimensional and two-dimensional ice
(see Fig.~\ref{fig:system}).
Water adsorption on graphene and on the AHs is tested with three different water orientations, dubbed 0-leg, 1-leg, and 2-leg. Adsorption on the CNT is considered outside the CNT in a 2-leg configuration (external) and inside the CNT in a 2-leg configuration (internal). The two-dimensional ice polymorphs have been constructed in Ref.~\onlinecite{2dice-dmc} by confining a single water layer resulting in four stable polymorphs of hexagonal, pentagonal, square, and rhombic ice structures. The three bulk ice phases cover the subtle balance of the competing polymorphs Ih, II and the high-pressure form VIII.
Overall this benchmark dataset has been designed to investigate both the water surface interaction with different adsorption motifs, different surface sizes and curvature, as well as the transferability to many-water systems at different pressures and confinements.

For all systems under consideration in our study, we use the interaction energy at fixed equilibrium geometry, where we have well converged theoretical reference energies available. 
We provide the reference geometries and energies as supporting information files to make our compiled benchmark easily accessible to other researchers.\cite{suppinfo}

\subsection{Reference interaction energies}\label{sec:dmc-ref}

The interaction energies considered are defined as the difference between a bound and an unbound configuration. Adsorption on nanostructure are computed as
\begin{align}
    E_{\text{int}} = E_{\text{w@nano}}- (E_{\text{w}} + E_{\text{nano}}),
\end{align}
where $ E_{\text{w@nano}}$ is in the bound configuration and the geometries of the individual fragments $E_{\text{w}}$ and $E_{\text{nano}}$ for the unbound configuration are kept frozen.\cite{geoms}
The interaction energies for 2D and 3D ice are given per molecule as
\begin{align}
    E_{\text{int}} = E_{\text{solid}}/N_{\text{mol}}- E_{\text{w}}.
\end{align}

The reference interaction energies used in the following analysis are obtained 
from studies published in the past four years.
All the reference values are reported in Table~\ref{tab:refs}.
The reference values for the 2-dimensional ice crystals are taken from Ref.~\onlinecite{2dice-dmc}, while for the three bulk-ice crystals are given in Ref.~\onlinecite{molcryst_dmc}. 
The reference values for water on benzene, coronene, and graphene with different orientations are taken from Ref.~\onlinecite{water-graphene-dmc}.
The reference interaction energy between water and CNT where previously investigated in Ref.~\onlinecite{water-cnt-dmc}, but here we report new values that are as tightly converged as the other references.

The computational approaches used to obtain the reference values are CCSD(T) and fixed-node DMC.
In particular, the values for water on benzene and on coronene are from CCSD(T) (where DMC yields identical results within the stochastic error), and all the other values are from fixed-node DMC.
Indeed, most of these systems are very challenging or even out-of-reach for CCSD(T) due to their large size and the presence of periodic boundary conditions.
With DMC it is easier to assess the binding energy in large complexes,
because it is straightforward to simulate periodic systems, and
DMC has favorable scaling
with system size.
In terms of accuracy, 
there is generally very good agreement between CCSD(T) and DMC in the evaluation of non-covalent interactions, provided that care  is taken to ensure that an accurate setup is used for each method.\cite{dmc_sizeconsistent, Dubecky_chemrev_2016, AlHamdani_JCP-perspective_2019}
Agreement between methods is not expected beyond a given precision.
To this aim, in Table~\ref{tab:refs} we report the estimated uncertainty associated with any evaluation.
For the water on benzene and coronene systems, where both CCSD(T) and DMC are affordable, there is excellent agreement among them.\cite{water-graphene-dmc}

\begin{table}[htb]
\caption{
WaC18 reference interaction energies given per water molecule in meV. The reported error represents the uncertainty on the evaluation.
}\label{tab:refs}
\begin{ruledtabular}
\begin{tabular}{lr c lr c lr}
\multicolumn{2}{l}{{\bf w@graphene}~\cite{water-graphene-dmc}}
 && \multicolumn{2}{l}{{\bf w@benzene}~\cite{water-graphene-dmc}}
 && \multicolumn{2}{l}{{\bf w@coronene}~\cite{water-graphene-dmc}}\\
\cline{1-2}
\cline{4-5}
\cline{7-8}
0-leg  & $-90\pm6$ && 0-leg  &  $+43\pm1$     && 0-leg  & $-61\pm3$\\
1-leg  & $-92\pm6$ && 1-leg~\footnotemark[2]  & $-124\pm3$ && 1-leg  & $-118\pm5$\\
2-leg  & $-99\pm6$ && 2-leg~\footnotemark[2]  & $-136\pm2$ && 2-leg  & $-143\pm4$ \\[0.1cm]
\multicolumn{2}{l}{{\bf w@CNT}~\footnotemark[1]}\\
 \cline{1-2}
external & $-85\pm18$ \\
internal & $-287\pm16$\\[0.1cm]
\multicolumn{2}{l}{{\bf 3D ice}~\cite{molcryst_dmc}}
&&\multicolumn{5}{l}{{\bf 2D ice}~\cite{2dice-dmc}}\\
\cline{1-2}
\cline{4-5}
\cline{7-8}
Ih   & $-615\pm5$ && hex.   & $-423\pm3$ && pent.    & $-419\pm3$\\
II   & $-613\pm6$ && square & $-404\pm3$ && rhombic  & $-389\pm3$\\
VIII & $-594\pm6$ && \\
%
\end{tabular}
\footnotetext[1]{Evaluation based on Ref.~\onlinecite{water-cnt-dmc}, but recomputed in this work. See details in Section~\ref{sec:dmc-ref}.}
\footnotetext[2]{Some earlier studies reported the 1-leg structure as most stable, which might be due to small differences in numerical and geometrical setups.\cite{feller-wbz,slipchenko-wbz}}
\end{ruledtabular}
\end{table}%
Although reported DMC results are coming from different studies, the setup is consistent among them.
DMC simulations were carried out with the {\sc casino} code.\cite{casino}
Dirac-Fock pseudopotentials~\cite{trail05_NCHF, trail05_SRHF} with the localization approximation~\cite{mitas91} (LA) are used. 
The trial wavefunctions were of the Slater-Jastrow type with single Slater determinants and the single particle orbitals obtained from DFT-LDA plane-wave calculations performed with {\sc pwscf}~\cite{espresso} and re-expanded in terms of B-splines.\cite{alfe04}
The Jastrow factor included electron-electron, electron-nucleus and electron-electron-nucleus terms. 
The parameters of the Jastrow were carefully optimized by minimizing the variance, within a variational QMC scheme. 
The recently developed size-consistent DMC algorithm (ZSGMA)\cite{dmc_sizeconsistent} was used. 
Finite time-step errors are carefully minimized by performing simulations with different values of the time-step, untill the bias appears safely smaller than the stochastic error.
In periodic calculations, finite-size corrections are applied either using the model periodic Coulomb interaction\cite{MPC:Fraser1996, MPC:Will1997, MPC:Kent1999} or the \citet{KZK:prl2008} approach. 

We reevaluate here the binding energy of water on the CNT because one specific aspect of the setup used in Ref.~\onlinecite{water-cnt-dmc} is now known to be a possible issue: the optimization of the Jastrow factor for the configuration of water inside the CNT was not optimal.
Since in the standard LA approach the Jastrow factor plays a major role in the pseudopotential error, we have developed a new method (to be reported elsewhere\cite{zen_DMCwDLA}) that removes the LA  bias.
In repeating the evaluation, we also tuned some other aspect of the setup to be in line with the actual state-of-the-art.
In particular, we used the recently developed eCEPP pseudopotentials\cite{eCEPP} and the ZSGMA algorithm.\cite{dmc_sizeconsistent} 
The reported binding values are obtained with time step $\tau=0.01$~a.u., which gives errors less than the stochastic uncertainty  (standard deviation of 18~meV).

\subsection{DFT Computational Details}
\label{subsec:compdetails}
DFT test calculations based on the PBE functional~\cite{pbe} were done on selected systems to ensure convergence of all relevant numerical settings.
Several different electronic structure codes and orbital basis expansions have been employed: {\sc Orca\,4}~\cite{orca} with large aug-cc-pVQZ, aug-cc-pV5Z~\cite{dunning, augdunning}, and def2-QZVPPD~\cite{qzvp} basis sets, {\sc Crystal17}~\cite{crystal17, crystal17_wire} with a def2-QZVPPD(-f) basis;
and {\sc Vasp\,5.4}~\cite{vasp2,vasp3} with projector-augmented plane waves (hard PAWs~\cite{paw1,paw2}) with an energy cutoff of 1000\,eV.
In all codes, tight self-consistent field settings and large integration (and fine FFT) grids are used. The Brillouin zone sampling  has been increased to converge the interaction energy to 1\,meV. 
For the adsorption on graphene (5$\times$5 supercell with 53 atoms in the unit cell) and the CNT (non-metallic (10,0) configuration with 83 atoms in the unit cell), this reduces to a $\Gamma$-point calculation. For all PAW calculations the non-periodic directions use a vacuum spacing of 20\,\AA\ for the absorbed geometries and the same unit cell is used for the individual fragments, which compensates possible remaining image interactions for the binding energies. {\sc Crystal} and {\sc Orca} use open conditions in the non-periodic directions consistent with the reference calculations.

For the water@AH system, we established that the PBE interaction energy is converged within 2\,meV using the def2-QZVPPD basis and the codes {\sc Orca} and {\sc CRYSTAL} yield results within 1\,meV. We additionally compared the PBE/def2-QZVPPD interaction energy with the PBE/hard\,PAW/1000\,eV ones for water@graphene, water@AH, and ice Ih with a maximum deviation of 2\,meV.
To reach the DFT convergence for these systems, unusually tight thresholds are needed. In Table~\ref{tab:convergence} we list the convergence of the PBE ice Ih lattice energy for three complementary basis set expansions and software codes. For the PAW code {\sc Vasp}, hard PAWs (i.e.\ small potential core) are needed as well as a minimum PW cutoff of 700\,eV. {\sc Crystal} employs an all-electron basis set with atom-centered functions. Here, the interaction energy is neither converged employing a counterpoise-corrected def2-TZVPP, nor a def2-QZVPP calculation. For fully converged values, a counterpoise-corrected def2-QZVPP calculation is needed and an extrapolation to the basis set limit reduced the binding by only 1.2\,meV.
\begin{table}[bht]
\caption{
Lattice energy convergence in meV of ice Ih using various numerical settings based on PBE and a $\Gamma$-centered 3$\times$3$\times$3 $k$-grid.\cite{pawconvergence,pawimpact} The best estimates are highlighted in bold.
}\label{tab:convergence}
\begin{ruledtabular}
\begin{tabular}{l rrr}
\multicolumn{1}{l}{\bfseries{\scshape{Vasp}}} & \multicolumn{3}{c}{PAW}\\
PW cutoff [eV] &  soft & normal & hard\\
\cline{2-4}
500  & -798.4 & -665.3 & -646.2\\
700  & -798.1 & -664.6 & -636.8\\
1000 & -798.0 & -664.7 & {\bf -637.1}\\
\multicolumn{2}{l}{\bfseries{\scshape{Crystal}}} & w.o.c.\footnotemark[1] & CP-corrected\\
\cline{2-4}
def2-mSVP  && -1104.2 & -858.8\\
def2-TZVPP &&  -721.2 & -646.7\\
def2-QZVPP &&  -665.6 & -638.2 \\
CBS(TZ,QZ)\footnotemark[2] &&  -657.9 & {\bf -637.0} \\
\end{tabular}
\footnotetext[1]{Supramolecular approach without counterpoise (CP) correction.}
\footnotetext[2]{Basis set extrapolation using optimized exponents.\cite{extrapolation}}
\end{ruledtabular}
\end{table}%
Production level calculations for all reported DFT interactions on the full WaC18 benchmark are performed with {\sc Vasp\,5.4} using hard PAWs and PW energy cutoff of 1000\,eV.

DFAs from several rungs are tested: local density approximation (LDA), generalized gradient approximation (GGA), meta-GGA, and hybrid functionals (incorporation of nonlocal Fock exchange).
The semi-local DFAs are corrected for missing long-range correlation effects by means of a variety of different semi-classical and nonlocal density based corrections (D2~\cite{dftd2}, D3~\cite{dftd3,dftd3bj}, D4~\cite{dftd3.5,dftd4}, TS~\cite{ts}, MBD~\cite{ts-mbd}, VV10~\cite{vv10}, dDsC~\cite{ddsc}, vdW-DF~\cite{vdwdf}).
TS and MBD are used with the non-iterative Hirshfeld partitioning, D3 is used in the rational damping variant including Axilrod-Teller-Muto type three-body contributions, for D4 partitioned charges are generated by the electronegativity equilibration procedure (EEQ) and many-body contributions are covered by a standard coupled fluctuating dipole expression retaining an RPA-like expression. See Refs.~\onlinecite{vdw_perspective,disp_chemrev, chemrev_tkatchenko, vdwdf_review} for further overview on vdW interactions in the DFT framework.


\section{Results and Discussion}
\label{sec:results}

\begin{table*}[ptb]
\caption{
Interaction energies and RMS deviations (meV) from the reference data for various electronic structure methods for the WaC18 benchmark set. All values have been consistently computed in the present study. Systems are as listed in Table~\ref{tab:refs} and shown in Fig.~\ref{fig:system}.
The smallest RMS in each category is highlighted in bold.
Apart from reference data taken from Refs~\onlinecite{water-graphene-dmc,molcryst_dmc,2dice-dmc}, all other data has been computed as part of this study. References to the DFAs used are also given.
}\label{tab:water-carbon-energy}
\begin{ruledtabular}
\begin{tabular}{l rrr c rr c rrr c rrr c rrrr c rrr rr}
       & \multicolumn{3}{c}{w@graphene} && \multicolumn{2}{c}{w@CNT}
       &&\multicolumn{3}{c}{w@benzene}  &&\multicolumn{3}{c}{w@coronene}
       &&\multicolumn{4}{c}{2D-ice}     &&\multicolumn{3}{c}{3D-ice}\\ 
       & 0-leg & 1-leg & 2-leg && ext. & int.
       && 0-leg & 1-leg & 2-leg && 0-leg & 1-leg & 2-leg
       && hex. & pent. & squ. & rhom. && Ih & II & VIII
       && RMS\\
\cline{2-4}\cline{6-7}\cline{9-11}\cline{13-15}\cline{17-20}\cline{22-24}\cline{26-26}
Reference & \multicolumn{3}{c}{DMC} && \multicolumn{2}{c}{DMC} && \multicolumn{3}{c}{CCSD(T)} && \multicolumn{3}{c}{CCSD(T)} && \multicolumn{4}{c}{DMC} && \multicolumn{3}{c}{DMC}\\
$E_{\text{int}}$\footnotemark[1]  & -90 & -92 & -99 && -85 & -287 && 43 & -124 & -136 && -61 & -118 & -143 && -423 & -419 & -404 & -389 && -615 & -613 & -594 && --\\
$\Delta$\footnotemark[2]  &
6 & 6 & 6 && 18 & 16 && 1 & 3 & 2 && 3 & 5 & 4 && 3 & 3 & 3 & 3 && 5 & 6 & 6 && --\\[0.1cm]
\multicolumn{15}{l}{\bf Local density approximation}\\
LDA &
-96 & -114 & -125 && 
-121 & -235 && 
4 & -172 & -197 &&
-62 & -135 & -155 &&
-710 & -699 & -657 & -637 &&
-1016 & -978 & -876 &&
193\\[0.1cm]
\multicolumn{15}{l}{\bf PBE with various vdW corrections}\\
         PBE\cite{pbe} &   -9 &  -26 &  -19 &&  -26 &  -82 &&    86 &  -89 &  -81 &&    21 &  -44 &  -50 && -434 & -416 & -370 & -354 && -639 & -571 & -462 &&   79 \\ 
    PBE-VV10\cite{pbe,vv10} & -123 & -131 & -139 && -121 & -375 &&    22 & -140 & -149 &&   -78 & -138 & -154 && -498 & -490 & -455 & -437 && -755 & -733 & -672 &&   63 \\ 
    PBE-dDsC\cite{pbe,ddsc} & -109 & -132 & -143 && -123 & -390 &&    26 & -142 & -155 &&   -70 & -142 & -160 && -482 & -476 & -447 & -427 && -739 & -738 & -688 &&   61 \\ 
      PBE-TS\cite{pbe,ts} & -116 & -141 & -160 && -136 & -408 &&    28 & -143 & -162 &&   -71 & -145 & -175 && -467 & -462 & -429 & -410 && -712 & -698 & -621 &&   52 \\ 
     PBE-MBD\cite{pbe,ts-mbd} &  -93 & -120 & -126 && -112 & -310 &&    41 & -137 & -146 &&   -51 & -128 & -145 && -478 & -472 & -437 & -420 && -721 & -694 & -619 &&   41 \\ 
      PBE-D2\cite{pbe,dftd2} &  -89 & -128 & -135 && -119 & -303 &&    35 & -147 & -161 &&   -53 & -140 & -154 && -489 & -482 & -447 & -432 && -731 & -698 & -637 &&   47 \\ 
      PBE-D3\cite{pbe,dftd3} &  -85 & -117 & -124 && -112 & -291 &&    39 & -139 & -147 &&   -49 & -127 & -144 && -476 & -467 & -430 & -412 && -716 & -679 & -597 &&   36 \\ 
      PBE-D4\cite{pbe,dftd4} & -104 & -115 & -118 && -107 & -314 &&    30 & -132 & -138 &&   -65 & -122 & -137 && -474 & -464 & -426 & -409 && -711 & -677 & -593 &&  {\bf 35} \\[0.1cm]
\multicolumn{15}{l}{\bf GGAs with D4\cite{dftd4} vdW correction}\\
     RPBE-D4\cite{rpbe} & -102 & -110 & -108 && -100 & -322 &&    34 & -115 & -117 &&   -62 & -114 & -124 && -412 & -404 & -371 & -355 && -632 & -594 & -519 &&   26 \\ 
   revPBE-D4\cite{revpbe} &  -97 & -105 & -105 &&  -94 & -308 &&    43 & -109 & -113 &&   -56 & -110 & -121 && -402 & -392 & -357 & -340 && -620 & -585 & -515 &&   29 \\ 
     BLYP-D4\cite{b88,lyp} & -109 & -117 & -118 && -104 & -332 &&    31 & -114 & -119 &&   -74 & -116 & -128 && -439 & -432 & -403 & -386 && -659 & -645 & -574 &&   {\bf 22} \\[0.1cm]
\multicolumn{15}{l}{\bf meta-GGA (with D4\cite{dftd4} vdW correction)}\\ 
        M06L\cite{m06l} &  -55 &  -64 &  -67 &&  -58 & -383 &&    53 &  -93 & -111 &&   -29 &  -76 &  -95 && -338 & -339 & -343 & -321 && -516 & -545 & -577 &&   56 \\ 
        SCAN\cite{scan} &  -63 &  -74 &  -84 &&  -78 & -197 &&    45 & -123 & -144 &&   -29 &  -92 & -116 && -464 & -459 & -439 & -421 && -667 & -655 & -615 &&   35 \\ 
     SCAN-D4\cite{scan,scand3} & -106 & -116 & -129 && -113 & -304 &&    31 & -136 & -158 &&   -62 & -122 & -150 && -476 & -473 & -455 & -436 && -766 & -694 & -661 &&   52 \\
     TPSS-D4\cite{tpss} & -103 & -110 & -113 &&  -99 & -324 &&    40 & -119 & -125 &&   -62 & -114 & -131 && -446 & -433 & -387 & -370 && -664 & -638 & -549 &&  {\bf 22} \\[0.1cm]     
\multicolumn{15}{l}{\bf 1st and 2nd generation vdW-DFs}\\ 
     vdW-DF1\cite{vdwdf} & -136 & -134 & -133 && -107 & -455 &&    26 & -117 & -109 &&   -99 & -136 & -147 && -346 & -346 & -326 & -309 && -564 & -557 & -522 &&   63 \\ 
 optB86b-vdW\cite{optpbevdw} & -142 & -144 & -150 && -121 & -454 &&    23 & -132 & -137 &&   -98 & -150 & -167 && -448 & -444 & -413 & -395 && -708 & -704 & -661 &&   59 \\ 
     vdW-DF2\cite{vdwdf2} & -120 & -123 & -128 && -110 & -401 &&    15 & -124 & -125 &&   -92 & -128 & -140 && -404 & -406 & -397 & -378 && -624 & -624 & -598 &&   33 \\ 
 rev-vdW-DF2\cite{revvdwdf2} & -105 & -110 & -115 &&  -95 & -360 &&    38 & -120 & -122 &&   -67 & -122 & -137 && -446 & -439 & -406 & -388 && -685 & -666 & -610 &&   {\bf 29} \\[0.1cm]
\multicolumn{15}{l}{\bf Hybrid functionals (with D4\cite{dftd4} vdW correction)}\\ 
          HF &   33 &   38 &   46 &&   42 &    9 &&   166 &  -41 &  -29 &&    75 &   29 &   -2 && -227 & -226 & -216 & -201 && -298 & -271 & -292 &&  198 \\ 
       HF-D4 & -109 &  -95 & -106 &&  -88 & -315 &&    68 & -111 & -133 &&   -56 &  -93 & -137 && -332 & -347 & -356 & -339 && -468 & -491 & -568 &&   57 \\ 
  revPBE0-D4\cite{revpbe,pbe0} &  -99 & -101 & -105 &&  -91 & -301 &&    51 & -115 & -126 &&   -52 & -106 & -130 && -389 & -383 & -353 & -337 && -583 & -560 & -516 &&   32 \\ 
    B3LYP-D4\cite{b3lypa,b3lypb} & -111 & -115 & -119 && -103 & -328 &&    36 & -122 & -132 &&   -71 & -115 & -136 && -442 & -438 & -413 & -397 && -641 & -639 & -588 &&   18 \\
     PBE0-D4\cite{pbe0} & -103 & -108 & -114 && -100 & -305 &&    44 & -131 & -142 &&   -57 & -115 & -140 && -449 & -443 & -411 & -395 && -643 & -623 & -582 &&   {\bf 14} \\[0.1cm]
\multicolumn{15}{l}{\bf Simplified density functional approximations}\\ 
      sHF-3c\cite{hf3c,shf3c} &  -63 &  -90 & -108 && -102 & -242 &&    65 & -119 & -151 &&    -7 & -123 & -166 && -467 & -451 & -425 & -412 && -692 & -670 & -563 &&   34 \\
      HSE-3c\cite{hse3c} & -114 &  -97 & -123 && -130 & -260 &&    70 & -158 & -194 &&   -43 & -129 & -185 && -511 & -495 & -454 & -419 && -735 & -709 & -627 &&   54 \\ 
      B97-3c\cite{b973c} & -112 & -133 & -137 && -131 & -321 &&    40 & -145 & -155 &&   -74 & -138 & -166 && -459 & -446 & -412 & -388 && -689 & -656 & -590 &&  {\bf 32} \\ 
%
\end{tabular}
\footnotetext[1]{Interaction energy at fixed equilibrium geometry provided as Supporting Information.\cite{suppinfo}}
\footnotetext[2]{Uncertainty estimation of reference interaction energy.}
\end{ruledtabular}
\end{table*}%

Using the DMC and CCSD(T) high-quality interaction energies, we can now test the capability of standard and new DFAs and a few simplified electronic structure methods for water adsorption on nanostructured surfaces, and within ice polymorphs. While discussing the performances of each method, it is important to keep in mind the numerical DFT uncertainty of about 2\,meV as well as the reference uncertainty, ranging from 1\,meV (w@benzene, 0-leg) to 18\,meV (w@CNT, external).

Table~\ref{tab:water-carbon-energy} summarizes the individual interaction energies from several electronic structure approaches and gives a root-mean-square (RMS) error over the full WaC18 set.
Additional statistical information on the performance of the methods is given in Table~\ref{tab:water-carbon-stats}.

\subsection{Impact of vdW interactions}
\label{sec:vdw}
\begin{table}[htb]
\caption{
Mean deviation (MD) and mean absolute deviation (MAD) in meV of the interaction energies from the references, see Table~\ref{tab:refs} and Fig.~\ref{fig:system}.
}\label{tab:water-carbon-stats}
\begin{ruledtabular}
\begin{tabular}{l rr c rr c rr}
       & \multicolumn{2}{c}{w@nano\footnotemark[1]} && \multicolumn{2}{c}{2D/3D-ice}
       &&\multicolumn{2}{c}{all}\\ 
Method & MD\footnotemark[1] & MAD &\phantom{p}& MD & MAD &\phantom{p}& MD & MAD \\
\cline{2-3}\cline{5-6}\cline{8-9}
\multicolumn{9}{l}{\bf Local density approximation}\\
          LDA &    -20 &     29 &&   -302 &    302 &&   -130 &    135 \\[0.1cm]
\multicolumn{9}{l}{\bf PBE with various vdW corrections}\\
          PBE &     79 &     79 &&     30 & {\bf 40} &&     60 &     64 \\ 
     PBE-VV10 &    -30 &     30 &&    -83 &     83 &&    -51 &     51 \\ 
     PBE-dDsC &    -32 &     32 &&    -77 &     77 &&    -49 &     49 \\ 
       PBE-TS &    -40 &     40 &&    -49 &     49 &&    -43 &     43 \\ 
      PBE-MBD &    -12 &     14 &&    -55 &     55 &&    -29 &     30 \\ 
       PBE-D2 &    -18 &     20 &&    -65 &     65 &&    -37 &     38 \\ 
       PBE-D3 &    -10 &    {\bf 13} &&    -46 &     46 &&    -24 &     26 \\ 
       PBE-D4 &    -12 &    {\bf 13} &&    -42 &     43 &&    -24 & {\bf 25} \\[0.1cm]
\multicolumn{9}{l}{\bf GGAs with D4 vdW correction}\\
      RPBE-D4 &     -4 &     14 &&     24 &     29 &&      7 &     20 \\ 
    revPBE-D4 &      2 &{\bf 12} &&     35 &     37 &&     15 &     21 \\ 
      BLYP-D4 &    -10 &     18 &&    -12 &{\bf 18} &&    -10 &{\bf 18} \\[0.1cm]
\multicolumn{9}{l}{\bf meta-GGA (with D4 vdW correction)}\\
         M06L &     19 &     37 &&     68 &     68 &&     38 &     49 \\ 
         SCAN &     22 &     23 &&    -38 &     38 &&     -1 &     29 \\ 
      SCAN-D4 &    -16 &     16 &&    -72 &     72 &&    -38 &     38 \\
      TPSS-D4 &     -6 &{\bf 12} &&     -5 &{\bf 27} &&     -6 &{\bf 18} \\[0.1cm]
\multicolumn{9}{l}{\bf 1st and 2nd generation vdW-DFs}\\       
      vdW-DF1 &    -32 &     39 &&     70 &     70 &&      7 &     51 \\ 
  optB86b-vdW &    -44 &     44 &&    -45 &     45 &&    -44 &     44 \\ 
      vdW-DF2 &    -26 &     28 &&      4 &{\bf 11} &&    -14 &     21 \\ 
  rev-vdW-DF2 &    -11 &{\bf 16} &&    -26 &     26 &&    -17 &{\bf 20} \\[0.1cm]
\multicolumn{9}{l}{\bf Hybrids (with D4 vdW correction)}\\    
           HF &    142 &    142 &&    247 &    247 &&    182 &    182 \\ 
        HF-D4 &      1 &     12 &&     79 &     79 &&     32 &     39 \\ 
   revPBE0-D4 &      2 &     10 &&     48 &     48 &&     20 &     25 \\ 
     B3LYP-D4 &    -11 &     14 &&    -14 &{\bf 16} &&    -12 &     15 \\
      PBE0-D4 &     -7 &{\bf 9} &&    -13 &{\bf 16} &&     -9 &{\bf 12} \\[0.1cm]
\multicolumn{9}{l}{\bf Simplified density functional approximations}\\  
       sHF-3c &      8 &{\bf 20} &&    -32 &     40 &&     -8 &     28 \\
       HSE-3c &    -16 &     29 &&    -70 &     70 &&    -37 &     45 \\
       B97-3c &    -25 &     25 &&    -26 &{\bf 28} &&    -26 &{\bf 26} \\       
\end{tabular}
\footnotetext[1]{Combination of the three adsorption sets water@graphene, water@CNT, and water@AH.}
\footnotetext[2]{Negative MD means too strongly bounded system}
\end{ruledtabular}
\end{table}%

From looking at Tables~\ref{tab:water-carbon-energy} and~\ref{tab:water-carbon-stats}, it is clear that Hartree-Fock (HF) misses all (Coulomb) correlation effects and cannot describe any of these non-covalently bound systems appropriately.\cite{london1937,stone} 
All systems are systematically underbound by up to 342\,meV. While there is some weak binding for water on benzene, this diminishes for larger substrates and becomes repulsive for the adsorption on graphene and on the CNT. Thus, exchange repulsion, induction, and electrostatics are not sufficient to lead to a net binding of water on the carbon nanostructures considered. This is consistent with our symmetry adapted perturbation theory analysis in Ref.~\onlinecite{water-graphene-dmc} as well as many studies on molecular dimers.\cite{s22}
That the local density approximation (LDA) yields an inconsistent description of small vdW complexes has been known since the mid-90s.\cite{kristyan1994} This is confirmed here, where LDA systematically overbinds all systems, in particular the ice polymorphs. The LDA results for water adsorption, on the other hand, seem reasonably good. However, this is a fortuitous event due to the fixed geometries. While all other DFAs give equilibrium adsorption distances within 0.1 to 0.2\,\AA\ (see Ref.~\onlinecite{yasmine_water_cnt,water_at_hbn, water_graphene_dmc_old}), the LDA minimum is at substantially smaller distance and cannot be recommended for either geometry or stability estimates of vdW bound systems.

Including semi-local exchange-correlation effects as in the popular PBE GGA functional improves the behavior, although most systems are bound a bit too weakly. Clearly, the long-range correlation effects leading to vdW attraction are missing. In the past decade, several methods have been developed for including these missing interactions (see e.g.\ Refs.~\onlinecite{vdw_perspective,disp_chemrev,vdwdf_review}).
We test a broad range of these vdW corrections coupled with PBE (see Table~\ref{tab:water-carbon-energy}, \ref{tab:water-carbon-stats}). 
The error spread is still substantial and in particular the older effective pair-wise schemes (PBE-VV10, PBE-dDsC, PBE-TS, PBE-D2) do not perform well. 
Recent vdW developments pay off and we can see a clear improvement of PBE-MBD~\cite{ts-mbd} and PBE-D4~\cite{dftd4} over their predecessors.  
The many-body vdW contributions decrease the binding yielding better agreement with the references. Most of this effect is already covered at the Axilrod-Teller-Muto type three-body level\cite{atm1,atm2} as included in the D3 method.\cite{dftd3} 
At the PBE-vdW level, only D4 and VV10 are able to reproduce the relative stability of the water adsorption, i.e. 0-leg vs. 2-leg and 1-leg vs. 2-leg stability, to good accuracy (coming within the error of the reference energy). 
The best PBE based method (PBE-D4) yields an excellent description of the water adsorption with mean absolute deviation (MAD) from the reference of 13\,meV. The description of the ice polymorphs is less satisfactory, which can be traced back to the intrinsic overpolarization and thus overestimated induction interaction of the PBE functional.\cite{peach2011,model-overpolarization, water27, s66-bench, beran-threebody} 
For instance, uncorrected PBE already overbinds hexagonal ice Ih by 24\,meV, which clearly is not corrected by a (mostly) attractive vdW interaction.
Overall, PBE-vdW does not perform well for strong hydrogen bonded systems.\cite{dftforwater_perspective}

DFAs with a nonlocal kernel to describe vdW interactions have been pioneered by Dion \textit{et al.} (vdW-DF1).\cite{dion2004,dion2005}
This first-generation nonlocal functional gives unsatisfactory results on our benchmarks, the adsorption strength is overestimated and the ice lattice energy is underestimated. The revised version with optimized semi-local exchange-correlation optB86b-vdW improves upon this, but the overall MAD is at 44\,meV still rather high. Binding to graphene and the CNT seems to be extremely challenging for the nonlocal functionals with maximum deviation of 168\,meV for water inside the CNT, as already noted in Ref.~\onlinecite{water-cnt-dmc}.
Consistent with previous studies,\cite{vdwdf2,revvdwdf2} the second generation of nonlocal functionals significantly improves the results on all benchmark systems and the revised variant rev-vdW-DF2 more than halves the overall MAD to 20\,meV. Only water inside the CNT remains challenging being overestimated by 72\,meV, which is worse than all other vdW corrected semi-local DFAs (with the exception of PBE-TS).

\subsection{Performance by Jacob's ladder classification}
\label{sec:ladder}
\begin{figure}[htb]
\centering
\includegraphics[width=0.49\textwidth]{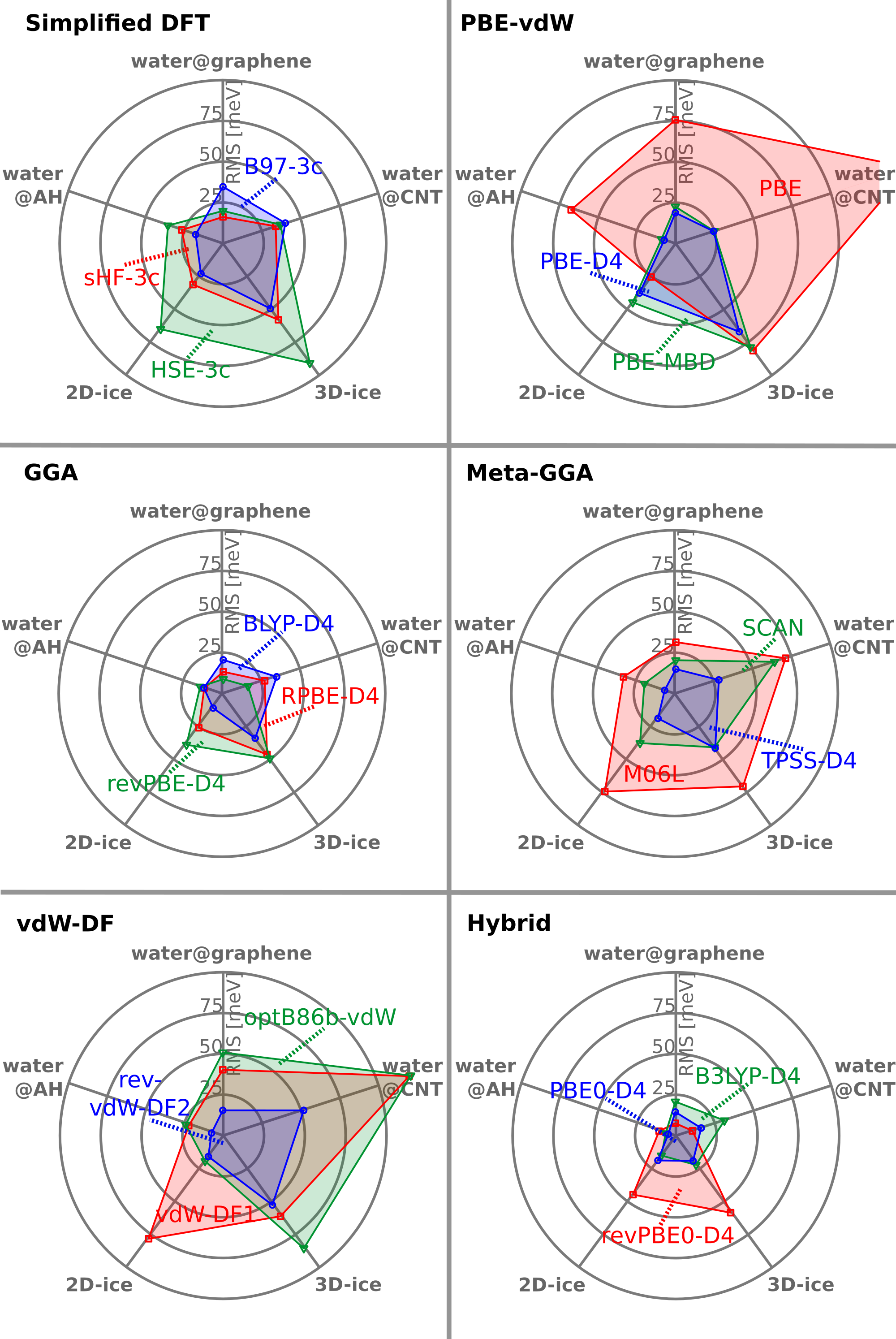}
\caption{\label{fig:polar_rms} 
Root-mean-square (RMS) errors in meV of the computed interaction energies for the WaC18 benchmark, separated into five subsets with various theoretical methods.
The systems are defined in Fig.~\ref{fig:system} and Table~\ref{tab:refs}.
}
\end{figure}
For a better visual comparison of the performance of the DFAs for the different WaC18 subsets, we show a graphical representation of the individual RMS errors separated into the different functional classes in Fig.~\ref{fig:polar_rms}.
LDA is not reported, as it yields very bad results.
In the GGA panel we show the results for the three most accurate GGA functionals, after inclusion of D4 for long-range vdW interactions.
PBE-D4 is not included, as some of the various modifications of the PBE exchange enhancement factors prove better, most notably RPBE and revPBE that are both known to give more reasonable hydrogen bond strengths.\cite{s66-bench,ice10}
While the water adsorption computed by RPBE-D4 and revPBE-D4 is very similar to PBE-D4, we see a clear improvement for the 2D/3D ice polymorphs. However, the MADs at 29 and 37\,meV are still unexpectedly high. Especially the denser ice structures (rhombic 2D-ice and high-pressure ice VIII) are systematically underbound.
Note that the use of normal PAWs or smaller basis set expansions results in a systematic shift towards more strongly bound systems (see Table~\ref{tab:convergence}), removing most of the bias for RPBE-D4 and revPBE-D4 and giving artificially better results.\cite{ice-errorcompensation}
The most successful GGA tested by us is BLYP-D4 giving a very consistent performance with MADs below 20\,meV for all considered benchmark sets.

Including higher derivatives (like the kinetic energy density $\tau$) in the exchange-correlation enhancement factors give rise to the meta-GGA class. Formally, their computational cost scales with system size as the GGAs, but a stable self-consistent field convergence can be numerically more involved and typically requires larger integration grids.\cite{minnesota_grid,scand3}
TPSS is based on the PBE GGA and has a rather weak $\tau$-dependency, but still improves over PBE for many physical properties.\cite{gmtkn55}
This also holds for our benchmark systems, TPSS-D4 has a rather balanced description of very accurate adsorption energies and decently good ice lattice energies, its overall MAD of 18\,meV is identical that of BLYP-D4.
SCAN and M06L are modern meta-GGAs that can cover part of the medium-range correlation and have been shown to yield good structures and energies for diversely bonded systems.\cite{scan_natchem, m06l} 
Still, both underbind all water adsorption systems, which can be partially compensated by correction schemes (see SCAN-D4).
However, since SCAN already includes some vdW forces, it is non-trivial to combine SCAN with correction schemes.\cite{scand3,dfa_disp_balance} This is particularly relevant for 2D/3D-ice, where plain SCAN already overbinds all systems.

The next DFA rung requires the inclusion of non-local (Fock) exchange resulting in hybrid functionals.
The most widely used hybrid DFAs are PBE0 and B3LYP and while they are typically only of medium quality for many chemical properties,\cite{gmtkn55} PBE0-D4 and B3LYP-D4 consistently improve over their GGA parents. In particular the improvement for the 2D/3D ice polymorphs is significant. Of all tested methods PBE0-D4 has the closest agreement with the reference interaction energies for all considered systems with an overall smallest MAD of 12\,meV. 
Importantly, all tested relative stability sequences (ice polymorphs and the adsorption motifs) of PBE0-D4 are correct. The relative adsorption energies of the different binding motifs on graphene and the CNT are actually within the stochastic uncertainty of the DMC references.

The simplified DFAs (sHF-3c~\cite{hf3c,shf3c}, HSE-3c~\cite{hse3c}, B97-3c~\cite{b973c}, see Ref.~\onlinecite{3c_review} for an overview) give mixed results. Overall their accuracy is similar to the average vdW corrected DFA. Especially HSE-3c has problems at describing the strong hydrogen bonds in 2D/3D-ice, most likely due to remaining basis set superposition errors that cannot be fully  compensated. B97-3c, on the other hand, employs a slightly larger basis set expansion and gives reasonably balanced results.
As those methods have been designed for increased computational speed (speedup of up to 2 orders of magnitude compared to converged basis set DFT\cite{3c_review}), they might still be useful for screening applications.

\subsection{Essential ingredients for well-balanced DFA}
\label{sec:ingredient}
We find the following points essential for a well-balanced description of both water adsorbed on nanostructures as well as within ice polymorphs:
\begin{itemize}
    \item Correlation effects beyond the local density approximation (avoid HF and LDA).
    \item Use of a modern vdW correction (D3/D4, MBD, or vdW-DF2)
    \item Converged numerical settings with hard cores for PAWs or expansions beyond triple-$\zeta$ quality for atom-centered basis sets. 
    \item GGA, meta-GGA, and hybrid DFAs are similarly good for adsorption. 
    \item Fock exchange improves strong H-bonds in ice polymorphs. 
\end{itemize}
As shown in Fig.~\ref{fig:polar_rms}, the best GGA, meta-GGA, vdW-DF, and simplified DFT methods (BLYP-D4, TPSS-D4, rev-vdW-DF2, and B97-3c, respectively), fulfil the first three points and perform rather similarly. Significantly more accurate are the best hybrid functionals mainly due to their improved description of 2D/3D ice.
In terms of computational cost, GGAs seem to have the best accuracy vs. effort ratio, while hybrids should be used when aiming for the highest accuracy.
The most successfull DFAs have errors consistently well below chemical accuracy (1\,kcal/mol = 43\,meV),
challenging experimental errors of sublimation enthalpies.\cite{whalley_ice,Chickos2003_DHerror}
\begin{figure}[bth]
\centering
\includegraphics[width=0.49\textwidth]{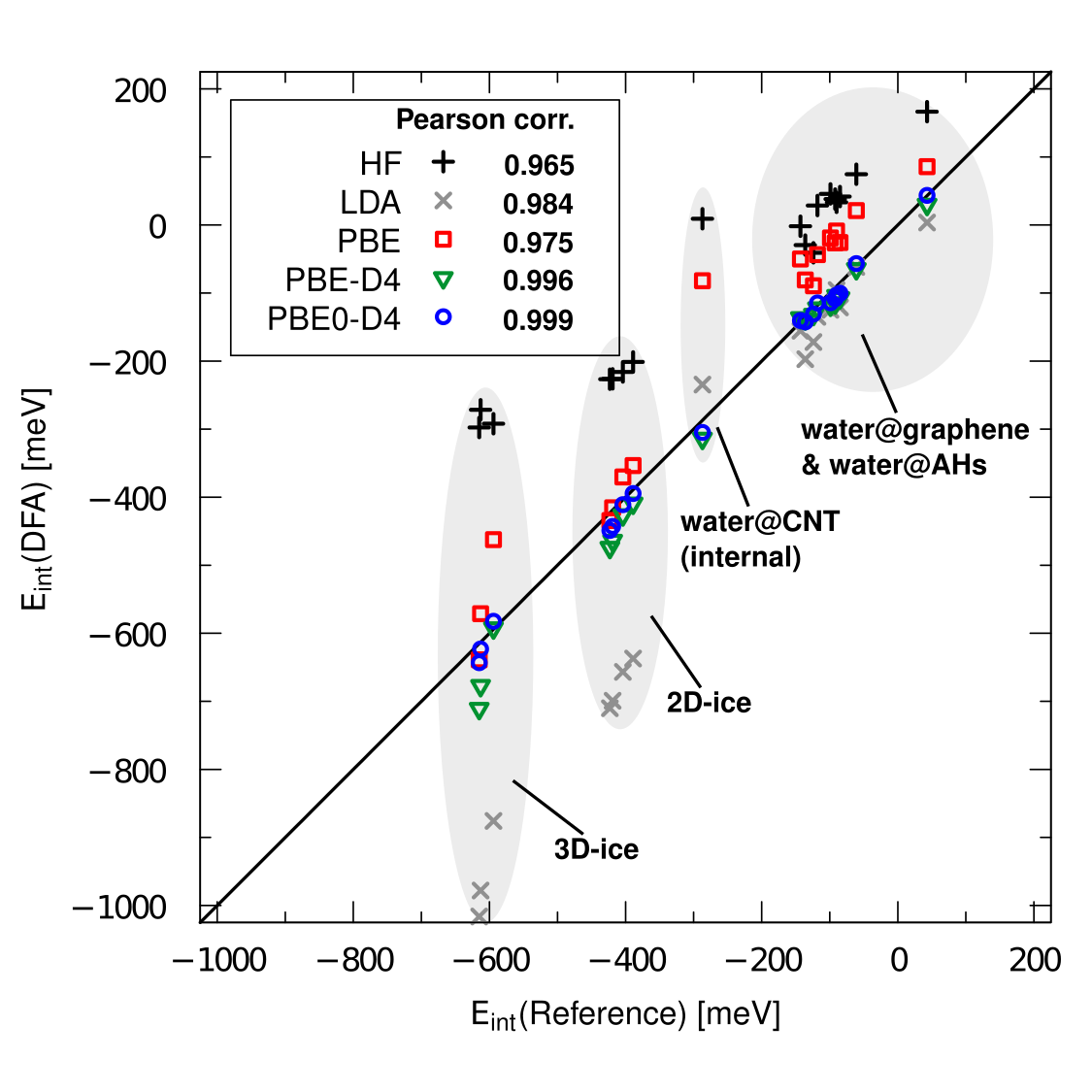}
\caption{\label{fig:energy_corr} 
Correlation between reference energies and interaction energies computed with a selection of density functional approximations and HF. Explicit data are given in Table~\ref{tab:water-carbon-energy}.
}
\end{figure}

\subsection{Comparison to water scoring scheme}
\begin{table*}[bht]
\caption{Percentage scores of selected DFAs using the scoring scheme from Ref.~\onlinecite{dftforwater_perspective}. Physical quantities scored are monomer symmetric
stretch frequency $f_{ss}^{\text{mono}}$, dimer binding energy $E_{b}^{\text{dim}}$, ring-hexamer binding energy per monomer $E_{b}^{\text{ring}}$, ice Ih lattice energy $E_{\text{Ih}}$, difference 
$\Delta E_{b}^{\text{pr-ring}}$ of binding energies per monomer of prism and ring isomers of the hexamer, difference $\Delta E_{\text{Ih-VIII}}$ of lattice energies of ice Ih and VIII, equilibrium O-O distance $R_{\text{OO}}$ in dimer, and equilibrium volumes per monomer $V_{\text{eq}}^{\text{Ih}}$, $V_{\text{eq}}^{\text{VIII}}$ of ice Ih and VIII. The total score in the final column is the average of the individual scores, unmarked values have been computed in this study. 
}\label{tab:gillan-scoring}
\begin{ruledtabular}
\begin{tabular}{l rrrrrrrrrr r}
  & $f_{ss}^{\text{mono}}$ & $E_{b}^{\text{dim}}$ & $E_{b}^{\text{ring}}$ & $E_{\text{Ih}}$ & $\Delta E_{b}^{\text{pr-ring}}$ & $\Delta E_{\text{Ih-VIII}}$ & $R_{\text{OO}}$  & $V_{\text{eq}}^{\text{Ih}}$ & $V_{\text{eq}}^{\text{VIII}}$ & Total\\
Reference & \footnotemark[1]3812\,cm$^{\text{-1}}$ & \footnotemark[1]-216\,meV & \footnotemark[1]-319\,meV &
\footnotemark[2]-615\,meV & \footnotemark[1]13\,meV & \footnotemark[2]21\,meV & \footnotemark[1]2.91\,\AA & \footnotemark[3]30.9\,\AA$^3$ & \footnotemark[3]19.1\,\AA$^3$ & 100\\
Tolerance  &  20\,cm$^{\text{-1}}$ & 10\,meV & 10\,meV &
10\,meV & 10\,meV & 10\,meV & 0.01\,\AA & 1\% & 1\% & \\[0.1cm]
\cline{2-11}
LDA     & 60 &   0 &  0 &  0 & 100 &  0 &   0 & \footnotemark[4]--  &  \footnotemark[4]-- & 23\\
HF      &  0 &  30 &  0 &  0 &  80 & 90 &   0 & \footnotemark[4]--  &  \footnotemark[4]-- & 29\\
PBE     & 50 & 100 & 90 & 80 &   0 &  0 &  90 & \footnotemark[5]100 &  \footnotemark[5]20 & 59\\
PBE-D4  & 50 &  90 & 60 & 0  & 100 & 10 &  80 & \footnotemark[6]50  & \footnotemark[6]100 & 60\\
BLYP-D4 & 30 & 100 & 90 & 60 & 100 & 40 & 100 & \footnotemark[6]90  & \footnotemark[6]80  &	77\\
TPSS-D4 & 50 & 100 & 80 & 50 & 90  & 10 &  80 & \footnotemark[6]60  & \footnotemark[6]70  & 66\\
optB88-vdW
        & \footnotemark[5]60 & \footnotemark[5]100 & \footnotemark[5]90 & \footnotemark[5]20 & \footnotemark[5]100 & \footnotemark[5]100& \footnotemark[5]50  & \footnotemark[5]80  & \footnotemark[5]100 & 78\\
PBE0-D4 & 80 & 100 & 80 & 80 & 90  & 70 & 100 & \footnotemark[7]70  & \footnotemark[7]70  & 82\\
B97-3c  & 70 & 90  & 70 & 30 & 100 & 30 & 80  & 50  & 70  & 66\\
%
\end{tabular}
\footnotetext[1]{References taken as gathered in Ref.~\onlinecite{dftforwater_perspective}.}
\footnotetext[2]{For consistency within this article, references taken from Table~\ref{tab:refs}.}
\footnotetext[3]{Values taken from Ref.~\onlinecite{ice10} to consistently exclude zero-point and thermal effects.}
\footnotetext[4]{Values not computed as expected to be unreliable.}
\footnotetext[5]{Values taken from Ref.~\onlinecite{dftforwater_perspective}.}
\footnotetext[6]{Values taken from Ref.~\onlinecite{ice10} using the D3 dispersion correction.}
\footnotetext[7]{Values taken from Ref.~\onlinecite{dftforwater_perspective} using the TS dispersion correction.}
\end{ruledtabular}
\end{table*}%

In a previous effort to judge \emph{"How good is DFT for water?"}\cite{dftforwater_perspective} a scoring scheme has been devised to judge the performance of approximated methods for the properties of the water monomer, the dimer, the hexamer, and ice structures.
Physical quantities scored are monomer symmetric
stretch frequency $f_{ss}^{\text{mono}}$, dimer binding energy $E_{b}^{\text{dim}}$, ring-hexamer binding energy per monomer $E_{b}^{\text{ring}}$, ice Ih lattice energy $E_{\text{Ih}}$, difference 
$\Delta E_{b}^{\text{pr-ring}}$ of binding energies per monomer of prism and ring isomers of the hexamer, difference $\Delta E_{\text{Ih-VIII}}$ of lattice energies of ice Ih and VIII, equilibrium O-O distance $R_{\text{OO}}$ in dimer, and equilibrium volumes per monomer $V_{\text{eq}}^{\text{Ih}}$, $V_{\text{eq}}^{\text{VIII}}$ of ice Ih and VIII.\cite{gillan}
For more details on the scoring system see Ref.~\onlinecite{dftforwater_perspective}.
For some of the DFAs examined in our benchmark study, we report their performance for this scoring system in Table~\ref{tab:gillan-scoring}. As expected LDA and HF are not reliable to describe water though some individual scores like the relative stability of the prism and ring hexamer are fortuitously good. The same holds for uncorrected as well as dispersion corrected PBE as already recognized in the original study.\cite{dftforwater_perspective}
On the other hand, BLYP-D4, TPSS-D4, optB88-vdW are performing reasonably well. Especially energetic and geometric properties are well reproduced, though the lattice energy of ice Ih seems to be problematic. Symmetric stretch frequencies of the water monomer are as usual underestimated by all (meta-)GGA functionals. Here, it is worth pointing out that out of the non-hybrid functionals, the low-cost method B97-3c yields the best frequencies and overall performs competitively to the dispersion corrected DFAs.
In agreement with our current benchmark, PBE0-D4 is the best performing method yielding an overall score of 82, which is indeed higher than any other DFT method tested on this set so far.

\section{Conclusions}
\label{sec:conclusion}
We have gathered and computed well converged reference interaction energies of water with carbon nanostructures and of water within two- and three-dimensional ice polymorphs compiled in the new WaC18 benchmark set. 
Combined, this gives a challenging set of large  and  complex  systems,  ideally  suited  to benchmark  approximated  methods.
Importantly, those systems are larger than standard noncovalent interaction benchmark sets based on molecular dimers of small to medium sized molecules like S22\cite{s22} and S66.\cite{s66} The 0D, 1D, 2D, and 3D periodicity covered here, also includes new aspects compared to benchmark sets of large supramolecular complexes like L7\cite{l7} and S12l.\cite{s12l}
In contrast to benchmarks using back-corrected experimental references, our high-level theoretical interaction energies make the comparison much more straight-forward 
without the complication of thermal and zero-point energy contributions.
We span a broad range of interaction energies from non-binding (water@benzene, 0-leg motif) to moderately strong binding (ice Ih with lattice energy of by $-615$\,meV).
Fig.~\ref{fig:energy_corr} shows an overview of the different binding strengths by correlating the reference energies with a few considered DFAs.
The correlation highlights that we roughly follow the Jacob's ladder classification of DFAs with HF and LDA being unreliable, and PBE, PBE-D4, and PBE0-D4 step by step increasing the accuracy.
Of all methods considered, BLYP-D4, TPSS-D4, rev-vdW-DF2, and PBE0-D4 are the most accurate within their respective functional class. Replacing D4 with the older D3 or the MBD vdW correction leads to minor deterioration and can be used when D4 is not available.
Our present benchmark focuses on specific noncovelent interactions only, from previous studies it is known that TPSS-D3 and PBE0-D3 yield very good equilibrium geometries\cite{pbeh3c} and organo-metallic reaction energies.\cite{mor41}
In the large GMTKN55 benchmark\cite{gmtkn55} BLYP-D3 and TPSS-D3 are among the best performing GGAs and meta-GGAs, respectively, consistent with our findings.

We see our benchmark results as a guideline for future simulation studies of water in the condensed phase (liquid or solid) and water--carbon nanostructure interfaces. 
Additionally, the provided WaC18 dataset can help as a challenging cross check other DFAs, classical force fields, machine learning potentials, tight-binding Hamiltonians, and to test other many body electronic structure methods like Random Phase Approximation (RPA) or M{\o}ller-Plesset perturbation theory (MP).


\begin{acknowledgments}
We thank Stefan Grimme and Eike Caldeweyher for providing access to the \emph{dftd4} code.
J.G.B acknowledges support by the Alexander von Humboldt foundation.
A.Z. and D.A.’s work is sponsored by the Air Force Office of Scientific Research, 
Air Force Material Command, US Air Force, under Grant FA9550-19-1-7007. 
A.Z. and A.M. were supported by the European Research Council (ERC) under the European Union’s Seventh Framework Program (FP/2007-2013)/ERC Grant Agreement 616121 (HeteroIce project).
This research used resources of the Oak Ridge Leadership Computing Facility, which is a DOE Office of Science User Facility supported under Contract DE-AC05-00OR22725.
We are also grateful, for computational resources, to ARCHER UK
National Supercomputing Service, United Kingdom Car–Parrinello (UKCP)
consortium (EP/F036884/1), the London Centre for Nanotechnology, 
University College London (UCL) Research Computing, 
and the UK Materials and Molecular Modelling Hub, which is partially funded by EPSRC (EP/P020194/1).
\\

\noindent
J.G.B and A.Z. contributed equally to this work.
\end{acknowledgments}


\bibliography{references}

\begin{thebibliography}{170}%
\makeatletter
\providecommand \@ifxundefined [1]{%
 \@ifx{#1\undefined}
}%
\providecommand \@ifnum [1]{%
 \ifnum #1\expandafter \@firstoftwo
 \else \expandafter \@secondoftwo
 \fi
}%
\providecommand \@ifx [1]{%
 \ifx #1\expandafter \@firstoftwo
 \else \expandafter \@secondoftwo
 \fi
}%
\providecommand \natexlab [1]{#1}%
\providecommand \enquote  [1]{``#1''}%
\providecommand \bibnamefont  [1]{#1}%
\providecommand \bibfnamefont [1]{#1}%
\providecommand \citenamefont [1]{#1}%
\providecommand \href@noop [0]{\@secondoftwo}%
\providecommand \href [0]{\begingroup \@sanitize@url \@href}%
\providecommand \@href[1]{\@@startlink{#1}\@@href}%
\providecommand \@@href[1]{\endgroup#1\@@endlink}%
\providecommand \@sanitize@url [0]{\catcode `\\12\catcode `\$12\catcode
  `\&12\catcode `\#12\catcode `\^12\catcode `\_12\catcode `\%12\relax}%
\providecommand \@@startlink[1]{}%
\providecommand \@@endlink[0]{}%
\providecommand \url  [0]{\begingroup\@sanitize@url \@url }%
\providecommand \@url [1]{\endgroup\@href {#1}{\urlprefix }}%
\providecommand \urlprefix  [0]{URL }%
\providecommand \Eprint [0]{\href }%
\providecommand \doibase [0]{http://dx.doi.org/}%
\providecommand \selectlanguage [0]{\@gobble}%
\providecommand \bibinfo  [0]{\@secondoftwo}%
\providecommand \bibfield  [0]{\@secondoftwo}%
\providecommand \translation [1]{[#1]}%
\providecommand \BibitemOpen [0]{}%
\providecommand \bibitemStop [0]{}%
\providecommand \bibitemNoStop [0]{.\EOS\space}%
\providecommand \EOS [0]{\spacefactor3000\relax}%
\providecommand \BibitemShut  [1]{\csname bibitem#1\endcsname}%
\let\auto@bib@innerbib\@empty
\bibitem [{\citenamefont {Fumagalli}\ \emph {et~al.}(2018)\citenamefont
  {Fumagalli}, \citenamefont {Esfandiar}, \citenamefont {Fabregas},
  \citenamefont {Hu}, \citenamefont {Ares}, \citenamefont {Janardanan},
  \citenamefont {Yang}, \citenamefont {Radha}, \citenamefont {Taniguchi},
  \citenamefont {Watanabe}, \citenamefont {Gomila}, \citenamefont {Novoselov},\
  and\ \citenamefont {Geim}}]{Fumagalli2018_dielectric}%
  \BibitemOpen
  \bibfield  {author} {\bibinfo {author} {\bibfnamefont {L.}~\bibnamefont
  {Fumagalli}}, \bibinfo {author} {\bibfnamefont {A.}~\bibnamefont
  {Esfandiar}}, \bibinfo {author} {\bibfnamefont {R.}~\bibnamefont {Fabregas}},
  \bibinfo {author} {\bibfnamefont {S.}~\bibnamefont {Hu}}, \bibinfo {author}
  {\bibfnamefont {P.}~\bibnamefont {Ares}}, \bibinfo {author} {\bibfnamefont
  {A.}~\bibnamefont {Janardanan}}, \bibinfo {author} {\bibfnamefont
  {Q.}~\bibnamefont {Yang}}, \bibinfo {author} {\bibfnamefont {B.}~\bibnamefont
  {Radha}}, \bibinfo {author} {\bibfnamefont {T.}~\bibnamefont {Taniguchi}},
  \bibinfo {author} {\bibfnamefont {K.}~\bibnamefont {Watanabe}}, \bibinfo
  {author} {\bibfnamefont {G.}~\bibnamefont {Gomila}}, \bibinfo {author}
  {\bibfnamefont {K.~S.}\ \bibnamefont {Novoselov}}, \ and\ \bibinfo {author}
  {\bibfnamefont {A.~K.}\ \bibnamefont {Geim}},\ }\href@noop {} {\bibfield
  {journal} {\bibinfo  {journal} {Science (New York, NY)}\ }\textbf {\bibinfo
  {volume} {360}},\ \bibinfo {pages} {1339} (\bibinfo {year}
  {2018})}\BibitemShut {NoStop}%
\bibitem [{\citenamefont {Abraham}\ \emph {et~al.}(2017)\citenamefont
  {Abraham}, \citenamefont {Vasu}, \citenamefont {Williams}, \citenamefont
  {Gopinadhan}, \citenamefont {Su}, \citenamefont {Cherian}, \citenamefont
  {Dix}, \citenamefont {Prestat}, \citenamefont {Haigh}, \citenamefont
  {Grigorieva}, \citenamefont {Carbone}, \citenamefont {Geim},\ and\
  \citenamefont {Nair}}]{Abraham2017}%
  \BibitemOpen
  \bibfield  {author} {\bibinfo {author} {\bibfnamefont {J.}~\bibnamefont
  {Abraham}}, \bibinfo {author} {\bibfnamefont {K.~S.}\ \bibnamefont {Vasu}},
  \bibinfo {author} {\bibfnamefont {C.~D.}\ \bibnamefont {Williams}}, \bibinfo
  {author} {\bibfnamefont {K.}~\bibnamefont {Gopinadhan}}, \bibinfo {author}
  {\bibfnamefont {Y.}~\bibnamefont {Su}}, \bibinfo {author} {\bibfnamefont
  {C.~T.}\ \bibnamefont {Cherian}}, \bibinfo {author} {\bibfnamefont
  {J.}~\bibnamefont {Dix}}, \bibinfo {author} {\bibfnamefont {E.}~\bibnamefont
  {Prestat}}, \bibinfo {author} {\bibfnamefont {S.~J.}\ \bibnamefont {Haigh}},
  \bibinfo {author} {\bibfnamefont {I.~V.}\ \bibnamefont {Grigorieva}},
  \bibinfo {author} {\bibfnamefont {P.}~\bibnamefont {Carbone}}, \bibinfo
  {author} {\bibfnamefont {A.~K.}\ \bibnamefont {Geim}}, \ and\ \bibinfo
  {author} {\bibfnamefont {R.~R.}\ \bibnamefont {Nair}},\ }\href {\doibase
  10.1038/nnano.2017.21} {\bibfield  {journal} {\bibinfo  {journal} {Nature
  Nanotechnology}\ }\textbf {\bibinfo {volume} {12}},\ \bibinfo {pages} {546}
  (\bibinfo {year} {2017})}\BibitemShut {NoStop}%
\bibitem [{\citenamefont {Joshi}\ \emph {et~al.}(2014)\citenamefont {Joshi},
  \citenamefont {Carbone}, \citenamefont {Wang}, \citenamefont {Kravets},
  \citenamefont {Su}, \citenamefont {Grigorieva}, \citenamefont {Wu},
  \citenamefont {Geim},\ and\ \citenamefont {Nair}}]{Joshi2014_MolSieving}%
  \BibitemOpen
  \bibfield  {author} {\bibinfo {author} {\bibfnamefont {R.~K.}\ \bibnamefont
  {Joshi}}, \bibinfo {author} {\bibfnamefont {P.}~\bibnamefont {Carbone}},
  \bibinfo {author} {\bibfnamefont {F.~C.}\ \bibnamefont {Wang}}, \bibinfo
  {author} {\bibfnamefont {V.~G.}\ \bibnamefont {Kravets}}, \bibinfo {author}
  {\bibfnamefont {Y.}~\bibnamefont {Su}}, \bibinfo {author} {\bibfnamefont
  {I.~V.}\ \bibnamefont {Grigorieva}}, \bibinfo {author} {\bibfnamefont
  {H.~A.}\ \bibnamefont {Wu}}, \bibinfo {author} {\bibfnamefont {A.~K.}\
  \bibnamefont {Geim}}, \ and\ \bibinfo {author} {\bibfnamefont {R.~R.}\
  \bibnamefont {Nair}},\ }\href {\doibase 10.1126/science.1245711} {\bibfield
  {journal} {\bibinfo  {journal} {Science}\ }\textbf {\bibinfo {volume}
  {343}},\ \bibinfo {pages} {752} (\bibinfo {year} {2014})}\BibitemShut
  {NoStop}%
\bibitem [{\citenamefont {Nair}\ \emph {et~al.}(2012)\citenamefont {Nair},
  \citenamefont {Wu}, \citenamefont {Jayaram}, \citenamefont {Grigorieva},\
  and\ \citenamefont {Geim}}]{Nair2012}%
  \BibitemOpen
  \bibfield  {author} {\bibinfo {author} {\bibfnamefont {R.~R.}\ \bibnamefont
  {Nair}}, \bibinfo {author} {\bibfnamefont {H.~A.}\ \bibnamefont {Wu}},
  \bibinfo {author} {\bibfnamefont {P.~N.}\ \bibnamefont {Jayaram}}, \bibinfo
  {author} {\bibfnamefont {I.~V.}\ \bibnamefont {Grigorieva}}, \ and\ \bibinfo
  {author} {\bibfnamefont {A.~K.}\ \bibnamefont {Geim}},\ }\href@noop {}
  {\bibfield  {journal} {\bibinfo  {journal} {Science}\ }\textbf {\bibinfo
  {volume} {335}},\ \bibinfo {pages} {442} (\bibinfo {year}
  {2012})}\BibitemShut {NoStop}%
\bibitem [{\citenamefont {Algara-Siller}\ \emph {et~al.}(2015)\citenamefont
  {Algara-Siller}, \citenamefont {Lehtinen}, \citenamefont {Wang},
  \citenamefont {Nair}, \citenamefont {Kaiser}, \citenamefont {Wu},
  \citenamefont {Geim},\ and\ \citenamefont
  {Grigorieva}}]{AlgaraSiller:2015_2Dice}%
  \BibitemOpen
  \bibfield  {author} {\bibinfo {author} {\bibfnamefont {G.}~\bibnamefont
  {Algara-Siller}}, \bibinfo {author} {\bibfnamefont {O.}~\bibnamefont
  {Lehtinen}}, \bibinfo {author} {\bibfnamefont {F.~C.}\ \bibnamefont {Wang}},
  \bibinfo {author} {\bibfnamefont {R.~R.}\ \bibnamefont {Nair}}, \bibinfo
  {author} {\bibfnamefont {U.}~\bibnamefont {Kaiser}}, \bibinfo {author}
  {\bibfnamefont {H.~A.}\ \bibnamefont {Wu}}, \bibinfo {author} {\bibfnamefont
  {A.~K.}\ \bibnamefont {Geim}}, \ and\ \bibinfo {author} {\bibfnamefont
  {I.~V.}\ \bibnamefont {Grigorieva}},\ }\href@noop {} {\bibfield  {journal}
  {\bibinfo  {journal} {Nature}\ }\textbf {\bibinfo {volume} {519}},\ \bibinfo
  {pages} {443} (\bibinfo {year} {2015})}\BibitemShut {NoStop}%
\bibitem [{\citenamefont {Secchi}\ \emph {et~al.}(2016)\citenamefont {Secchi},
  \citenamefont {Marbach}, \citenamefont {Nigu\`{e}s}, \citenamefont {Stein},
  \citenamefont {Siria},\ and\ \citenamefont {Bocquet}}]{water_cnt_secchi}%
  \BibitemOpen
  \bibfield  {author} {\bibinfo {author} {\bibfnamefont {E.}~\bibnamefont
  {Secchi}}, \bibinfo {author} {\bibfnamefont {S.}~\bibnamefont {Marbach}},
  \bibinfo {author} {\bibfnamefont {A.}~\bibnamefont {Nigu\`{e}s}}, \bibinfo
  {author} {\bibfnamefont {D.}~\bibnamefont {Stein}}, \bibinfo {author}
  {\bibfnamefont {A.}~\bibnamefont {Siria}}, \ and\ \bibinfo {author}
  {\bibfnamefont {L.}~\bibnamefont {Bocquet}},\ }\href {\doibase
  10.1038/nature19315} {\bibfield  {journal} {\bibinfo  {journal} {Nature}\
  }\textbf {\bibinfo {volume} {537}},\ \bibinfo {pages} {210} (\bibinfo {year}
  {2016})}\BibitemShut {NoStop}%
\bibitem [{\citenamefont {Radha}\ \emph {et~al.}(2016)\citenamefont {Radha},
  \citenamefont {Esfandiar}, \citenamefont {Wang}, \citenamefont {Rooney},
  \citenamefont {Gopinadhan}, \citenamefont {Keerthi}, \citenamefont
  {Mishchenko}, \citenamefont {Janardanan}, \citenamefont {Blake},
  \citenamefont {Fumagalli}, \citenamefont {Lozada-Hidalgo}, \citenamefont
  {Garaj}, \citenamefont {Haigh}, \citenamefont {Grigorieva}, \citenamefont
  {Wu},\ and\ \citenamefont {Geim}}]{md_w@capillaries}%
  \BibitemOpen
  \bibfield  {author} {\bibinfo {author} {\bibfnamefont {B.}~\bibnamefont
  {Radha}}, \bibinfo {author} {\bibfnamefont {A.}~\bibnamefont {Esfandiar}},
  \bibinfo {author} {\bibfnamefont {F.~C.}\ \bibnamefont {Wang}}, \bibinfo
  {author} {\bibfnamefont {A.~P.}\ \bibnamefont {Rooney}}, \bibinfo {author}
  {\bibfnamefont {K.}~\bibnamefont {Gopinadhan}}, \bibinfo {author}
  {\bibfnamefont {A.}~\bibnamefont {Keerthi}}, \bibinfo {author} {\bibfnamefont
  {A.}~\bibnamefont {Mishchenko}}, \bibinfo {author} {\bibfnamefont
  {A.}~\bibnamefont {Janardanan}}, \bibinfo {author} {\bibfnamefont
  {P.}~\bibnamefont {Blake}}, \bibinfo {author} {\bibfnamefont
  {L.}~\bibnamefont {Fumagalli}}, \bibinfo {author} {\bibfnamefont
  {M.}~\bibnamefont {Lozada-Hidalgo}}, \bibinfo {author} {\bibfnamefont
  {S.}~\bibnamefont {Garaj}}, \bibinfo {author} {\bibfnamefont {S.~J.}\
  \bibnamefont {Haigh}}, \bibinfo {author} {\bibfnamefont {I.~V.}\ \bibnamefont
  {Grigorieva}}, \bibinfo {author} {\bibfnamefont {H.~A.}\ \bibnamefont {Wu}},
  \ and\ \bibinfo {author} {\bibfnamefont {A.~K.}\ \bibnamefont {Geim}},\
  }\href {\doibase 10.1038/nature19363} {\bibfield  {journal} {\bibinfo
  {journal} {Nature}\ }\textbf {\bibinfo {volume} {538}},\ \bibinfo {pages}
  {222} (\bibinfo {year} {2016})}\BibitemShut {NoStop}%
\bibitem [{\citenamefont {Hummer}, \citenamefont {Rasaiah},\ and\ \citenamefont
  {Noworyta}(2001)}]{Hummer:2001gd}%
  \BibitemOpen
  \bibfield  {author} {\bibinfo {author} {\bibfnamefont {G.}~\bibnamefont
  {Hummer}}, \bibinfo {author} {\bibfnamefont {J.~C.}\ \bibnamefont {Rasaiah}},
  \ and\ \bibinfo {author} {\bibfnamefont {J.~P.}\ \bibnamefont {Noworyta}},\
  }\href@noop {} {\bibfield  {journal} {\bibinfo  {journal} {Nature}\ }\textbf
  {\bibinfo {volume} {414}},\ \bibinfo {pages} {188} (\bibinfo {year}
  {2001})}\BibitemShut {NoStop}%
\bibitem [{\citenamefont {Strogatz}\ \emph {et~al.}(2005)\citenamefont
  {Strogatz}, \citenamefont {Abrams}, \citenamefont {McRobie}, \citenamefont
  {Eckhardt},\ and\ \citenamefont {Ott}}]{Strogatz:2005kj}%
  \BibitemOpen
  \bibfield  {author} {\bibinfo {author} {\bibfnamefont {S.~H.}\ \bibnamefont
  {Strogatz}}, \bibinfo {author} {\bibfnamefont {D.~M.}\ \bibnamefont
  {Abrams}}, \bibinfo {author} {\bibfnamefont {A.}~\bibnamefont {McRobie}},
  \bibinfo {author} {\bibfnamefont {B.}~\bibnamefont {Eckhardt}}, \ and\
  \bibinfo {author} {\bibfnamefont {E.}~\bibnamefont {Ott}},\ }\href@noop {}
  {\bibfield  {journal} {\bibinfo  {journal} {Nature}\ }\textbf {\bibinfo
  {volume} {438}},\ \bibinfo {pages} {43} (\bibinfo {year} {2005})}\BibitemShut
  {NoStop}%
\bibitem [{\citenamefont {Holt}(2006)}]{Holt:2006kr}%
  \BibitemOpen
  \bibfield  {author} {\bibinfo {author} {\bibfnamefont {J.~K.}\ \bibnamefont
  {Holt}},\ }\href@noop {} {\bibfield  {journal} {\bibinfo  {journal} {Science
  (New York, NY)}\ }\textbf {\bibinfo {volume} {312}},\ \bibinfo {pages} {1034}
  (\bibinfo {year} {2006})}\BibitemShut {NoStop}%
\bibitem [{\citenamefont {Tocci}, \citenamefont {Joly},\ and\ \citenamefont
  {Michaelides}(2014)}]{Tocci2014_friction}%
  \BibitemOpen
  \bibfield  {author} {\bibinfo {author} {\bibfnamefont {G.}~\bibnamefont
  {Tocci}}, \bibinfo {author} {\bibfnamefont {L.}~\bibnamefont {Joly}}, \ and\
  \bibinfo {author} {\bibfnamefont {A.}~\bibnamefont {Michaelides}},\ }\href
  {\doibase 10.1021/nl502837d} {\bibfield  {journal} {\bibinfo  {journal} {Nano
  Lett.}\ }\textbf {\bibinfo {volume} {14}},\ \bibinfo {pages} {6872} (\bibinfo
  {year} {2014})}\BibitemShut {NoStop}%
\bibitem [{\citenamefont {Michaelides}(2016)}]{Michaelides:2016_Nature_news}%
  \BibitemOpen
  \bibfield  {author} {\bibinfo {author} {\bibfnamefont {A.}~\bibnamefont
  {Michaelides}},\ }\href@noop {} {\bibfield  {journal} {\bibinfo  {journal}
  {Nature}\ }\textbf {\bibinfo {volume} {537}},\ \bibinfo {pages} {171}
  (\bibinfo {year} {2016})}\BibitemShut {NoStop}%
\bibitem [{\citenamefont {Esfandiar}\ \emph {et~al.}(2017)\citenamefont
  {Esfandiar}, \citenamefont {Radha}, \citenamefont {Wang}, \citenamefont
  {Yang}, \citenamefont {Hu}, \citenamefont {Garaj}, \citenamefont {Nair},
  \citenamefont {Geim},\ and\ \citenamefont
  {Gopinadhan}}]{Esfandiar:2017_transport}%
  \BibitemOpen
  \bibfield  {author} {\bibinfo {author} {\bibfnamefont {A.}~\bibnamefont
  {Esfandiar}}, \bibinfo {author} {\bibfnamefont {B.}~\bibnamefont {Radha}},
  \bibinfo {author} {\bibfnamefont {F.~C.}\ \bibnamefont {Wang}}, \bibinfo
  {author} {\bibfnamefont {Q.}~\bibnamefont {Yang}}, \bibinfo {author}
  {\bibfnamefont {S.}~\bibnamefont {Hu}}, \bibinfo {author} {\bibfnamefont
  {S.}~\bibnamefont {Garaj}}, \bibinfo {author} {\bibfnamefont {R.~R.}\
  \bibnamefont {Nair}}, \bibinfo {author} {\bibfnamefont {A.~K.}\ \bibnamefont
  {Geim}}, \ and\ \bibinfo {author} {\bibfnamefont {K.}~\bibnamefont
  {Gopinadhan}},\ }\href@noop {} {\bibfield  {journal} {\bibinfo  {journal}
  {Science (New York, NY)}\ }\textbf {\bibinfo {volume} {358}},\ \bibinfo
  {pages} {511} (\bibinfo {year} {2017})}\BibitemShut {NoStop}%
\bibitem [{\citenamefont {Siria}, \citenamefont {Bocquet},\ and\ \citenamefont
  {Bocquet}(2017)}]{Bocquet-NatRevChem-2017}%
  \BibitemOpen
  \bibfield  {author} {\bibinfo {author} {\bibfnamefont {A.}~\bibnamefont
  {Siria}}, \bibinfo {author} {\bibfnamefont {M.-L.}\ \bibnamefont {Bocquet}},
  \ and\ \bibinfo {author} {\bibfnamefont {L.}~\bibnamefont {Bocquet}},\
  }\href@noop {} {\bibfield  {journal} {\bibinfo  {journal} {Nature Reviews
  Chemistry}\ }\textbf {\bibinfo {volume} {1}},\ \bibinfo {pages} {0091}
  (\bibinfo {year} {2017})}\BibitemShut {NoStop}%
\bibitem [{\citenamefont {Yoshida}\ \emph {et~al.}(2018)\citenamefont
  {Yoshida}, \citenamefont {Kaiser}, \citenamefont {Rotenberg},\ and\
  \citenamefont {Bocquet}}]{bocquet-naturecomm-2018}%
  \BibitemOpen
  \bibfield  {author} {\bibinfo {author} {\bibfnamefont {H.}~\bibnamefont
  {Yoshida}}, \bibinfo {author} {\bibfnamefont {V.}~\bibnamefont {Kaiser}},
  \bibinfo {author} {\bibfnamefont {B.}~\bibnamefont {Rotenberg}}, \ and\
  \bibinfo {author} {\bibfnamefont {L.}~\bibnamefont {Bocquet}},\ }\href
  {\doibase {10.1038/s41467-018-03829-1}} {\bibfield  {journal} {\bibinfo
  {journal} {{Nature Communications}}\ }\textbf {\bibinfo {volume} {{9}}},\
  \bibinfo {pages} {{1496}} (\bibinfo {year} {{2018}})}\BibitemShut {NoStop}%
\bibitem [{\citenamefont {Burke}(2012)}]{Kieron-JCP}%
  \BibitemOpen
  \bibfield  {author} {\bibinfo {author} {\bibfnamefont {K.}~\bibnamefont
  {Burke}},\ }\href@noop {} {\bibfield  {journal} {\bibinfo  {journal} {J.
  Chem. Phys.}\ }\textbf {\bibinfo {volume} {136}},\ \bibinfo {pages} {150901}
  (\bibinfo {year} {2012})}\BibitemShut {NoStop}%
\bibitem [{\citenamefont {Becke}(2014)}]{Becke-JCP}%
  \BibitemOpen
  \bibfield  {author} {\bibinfo {author} {\bibfnamefont {A.~D.}\ \bibnamefont
  {Becke}},\ }\href {\doibase 10.1063/1.4869598} {\bibfield  {journal}
  {\bibinfo  {journal} {J. Chem. Phys.}\ }\textbf {\bibinfo {volume} {140}},\
  \bibinfo {pages} {18A301} (\bibinfo {year} {2014})}\BibitemShut {NoStop}%
\bibitem [{\citenamefont {Yu}, \citenamefont {Li},\ and\ \citenamefont
  {Truhlar}(2016)}]{Truhlar-JCP}%
  \BibitemOpen
  \bibfield  {author} {\bibinfo {author} {\bibfnamefont {H.~S.}\ \bibnamefont
  {Yu}}, \bibinfo {author} {\bibfnamefont {S.~L.}\ \bibnamefont {Li}}, \ and\
  \bibinfo {author} {\bibfnamefont {D.~G.}\ \bibnamefont {Truhlar}},\
  }\href@noop {} {\bibfield  {journal} {\bibinfo  {journal} {J. Chem. Phys.}\
  }\textbf {\bibinfo {volume} {145}},\ \bibinfo {pages} {130901} (\bibinfo
  {year} {2016})}\BibitemShut {NoStop}%
\bibitem [{\citenamefont {Maurer}\ \emph {et~al.}(2019)\citenamefont {Maurer},
  \citenamefont {Freysoldt}, \citenamefont {Reilly}, \citenamefont
  {Brandenburg}, \citenamefont {Hofmann}, \citenamefont {Bj\"{o}orkman},
  \citenamefont {Leb\`{e}gue},\ and\ \citenamefont
  {Tkatchenko}}]{dft-materials-rev}%
  \BibitemOpen
  \bibfield  {author} {\bibinfo {author} {\bibfnamefont {R.~J.}\ \bibnamefont
  {Maurer}}, \bibinfo {author} {\bibfnamefont {C.}~\bibnamefont {Freysoldt}},
  \bibinfo {author} {\bibfnamefont {A.~M.}\ \bibnamefont {Reilly}}, \bibinfo
  {author} {\bibfnamefont {J.~G.}\ \bibnamefont {Brandenburg}}, \bibinfo
  {author} {\bibfnamefont {O.~T.}\ \bibnamefont {Hofmann}}, \bibinfo {author}
  {\bibfnamefont {T.}~\bibnamefont {Bj\"{o}orkman}}, \bibinfo {author}
  {\bibfnamefont {S.}~\bibnamefont {Leb\`{e}gue}}, \ and\ \bibinfo {author}
  {\bibfnamefont {A.}~\bibnamefont {Tkatchenko}},\ }\href {\doibase
  10.1146/annurev-matsci-070218-010143} {\bibfield  {journal} {\bibinfo
  {journal} {{Annu. Rev. Mater. Res.}}\ }\textbf {\bibinfo {volume} {49}},\
  \bibinfo {pages} {3.1} (\bibinfo {year} {2019})}\BibitemShut {NoStop}%
\bibitem [{\citenamefont {Brandenburg}\ \emph {et~al.}(2016)\citenamefont
  {Brandenburg}, \citenamefont {Bates}, \citenamefont {Sun},\ and\
  \citenamefont {Perdew}}]{scand3}%
  \BibitemOpen
  \bibfield  {author} {\bibinfo {author} {\bibfnamefont {J.~G.}\ \bibnamefont
  {Brandenburg}}, \bibinfo {author} {\bibfnamefont {J.~E.}\ \bibnamefont
  {Bates}}, \bibinfo {author} {\bibfnamefont {J.}~\bibnamefont {Sun}}, \ and\
  \bibinfo {author} {\bibfnamefont {J.~P.}\ \bibnamefont {Perdew}},\ }\href
  {\doibase 10.1103/PhysRevB.94.115144} {\bibfield  {journal} {\bibinfo
  {journal} {{Phys. Rev. B}}\ }\textbf {\bibinfo {volume} {94}},\ \bibinfo
  {pages} {115144} (\bibinfo {year} {2016})}\BibitemShut {NoStop}%
\bibitem [{\citenamefont {Hermann}\ and\ \citenamefont
  {Tkatchenko}(2018)}]{dfa_disp_balance}%
  \BibitemOpen
  \bibfield  {author} {\bibinfo {author} {\bibfnamefont {J.}~\bibnamefont
  {Hermann}}\ and\ \bibinfo {author} {\bibfnamefont {A.}~\bibnamefont
  {Tkatchenko}},\ }\href {\doibase 10.1021/acs.jctc.7b01172} {\bibfield
  {journal} {\bibinfo  {journal} {{J. Chem. Theory Comput.}}\ }\textbf
  {\bibinfo {volume} {14}},\ \bibinfo {pages} {1361} (\bibinfo {year}
  {2018})}\BibitemShut {NoStop}%
\bibitem [{\citenamefont {Al-Hamdani}\ \emph {et~al.}(2017)\citenamefont
  {Al-Hamdani}, \citenamefont {Rossi}, \citenamefont {Alf\`{e}}, \citenamefont
  {Tsatsoulis}, \citenamefont {Ramberger}, \citenamefont {Brandenburg},
  \citenamefont {Zen}, \citenamefont {Kresse}, \citenamefont {Gr\"{u}neis},
  \citenamefont {Tkatchenko},\ and\ \citenamefont
  {Michaelides}}]{water_at_hbn}%
  \BibitemOpen
  \bibfield  {author} {\bibinfo {author} {\bibfnamefont {Y.~S.}\ \bibnamefont
  {Al-Hamdani}}, \bibinfo {author} {\bibfnamefont {M.}~\bibnamefont {Rossi}},
  \bibinfo {author} {\bibfnamefont {D.}~\bibnamefont {Alf\`{e}}}, \bibinfo
  {author} {\bibfnamefont {T.}~\bibnamefont {Tsatsoulis}}, \bibinfo {author}
  {\bibfnamefont {B.}~\bibnamefont {Ramberger}}, \bibinfo {author}
  {\bibfnamefont {J.~G.}\ \bibnamefont {Brandenburg}}, \bibinfo {author}
  {\bibfnamefont {A.}~\bibnamefont {Zen}}, \bibinfo {author} {\bibfnamefont
  {G.}~\bibnamefont {Kresse}}, \bibinfo {author} {\bibfnamefont
  {A.}~\bibnamefont {Gr\"{u}neis}}, \bibinfo {author} {\bibfnamefont
  {A.}~\bibnamefont {Tkatchenko}}, \ and\ \bibinfo {author} {\bibfnamefont
  {A.}~\bibnamefont {Michaelides}},\ }\href {\doibase 10.1063/1.4985878}
  {\bibfield  {journal} {\bibinfo  {journal} {J. Chem. Phys.}\ }\textbf
  {\bibinfo {volume} {147}},\ \bibinfo {pages} {044710} (\bibinfo {year}
  {2017})}\BibitemShut {NoStop}%
\bibitem [{\citenamefont {Goerigk}\ \emph {et~al.}(2017)\citenamefont
  {Goerigk}, \citenamefont {Hansen}, \citenamefont {Bauer}, \citenamefont
  {Ehrlich}, \citenamefont {Najibi},\ and\ \citenamefont {Grimme}}]{gmtkn55}%
  \BibitemOpen
  \bibfield  {author} {\bibinfo {author} {\bibfnamefont {L.}~\bibnamefont
  {Goerigk}}, \bibinfo {author} {\bibfnamefont {A.}~\bibnamefont {Hansen}},
  \bibinfo {author} {\bibfnamefont {C.~A.}\ \bibnamefont {Bauer}}, \bibinfo
  {author} {\bibfnamefont {S.}~\bibnamefont {Ehrlich}}, \bibinfo {author}
  {\bibfnamefont {A.}~\bibnamefont {Najibi}}, \ and\ \bibinfo {author}
  {\bibfnamefont {S.}~\bibnamefont {Grimme}},\ }\href@noop {} {\bibfield
  {journal} {\bibinfo  {journal} {Phys. Chem. Chem. Phys.}\ }\textbf {\bibinfo
  {volume} {19}},\ \bibinfo {pages} {32184} (\bibinfo {year}
  {2017})}\BibitemShut {NoStop}%
\bibitem [{\citenamefont {Mardirossian}\ and\ \citenamefont
  {Head-Gordon}(2016{\natexlab{a}})}]{wb97mv}%
  \BibitemOpen
  \bibfield  {author} {\bibinfo {author} {\bibfnamefont {N.}~\bibnamefont
  {Mardirossian}}\ and\ \bibinfo {author} {\bibfnamefont {M.}~\bibnamefont
  {Head-Gordon}},\ }\href {\doibase 10.1063/1.4952647} {\bibfield  {journal}
  {\bibinfo  {journal} {J. Chem. Phys.}\ }\textbf {\bibinfo {volume} {144}},\
  \bibinfo {pages} {214110} (\bibinfo {year} {2016}{\natexlab{a}})}\BibitemShut
  {NoStop}%
\bibitem [{\citenamefont {Mardirossian}\ \emph {et~al.}(2017)\citenamefont
  {Mardirossian}, \citenamefont {Ruiz~Pestana}, \citenamefont {Womack},
  \citenamefont {Skylaris}, \citenamefont {Head-Gordon},\ and\ \citenamefont
  {Head-Gordon}}]{b97mv}%
  \BibitemOpen
  \bibfield  {author} {\bibinfo {author} {\bibfnamefont {N.}~\bibnamefont
  {Mardirossian}}, \bibinfo {author} {\bibfnamefont {L.}~\bibnamefont
  {Ruiz~Pestana}}, \bibinfo {author} {\bibfnamefont {J.~C.}\ \bibnamefont
  {Womack}}, \bibinfo {author} {\bibfnamefont {C.-K.}\ \bibnamefont
  {Skylaris}}, \bibinfo {author} {\bibfnamefont {T.}~\bibnamefont
  {Head-Gordon}}, \ and\ \bibinfo {author} {\bibfnamefont {M.}~\bibnamefont
  {Head-Gordon}},\ }\href {\doibase 10.1021/acs.jpclett.6b02527} {\bibfield
  {journal} {\bibinfo  {journal} {J. Phys. Chem. Lett.}\ }\textbf {\bibinfo
  {volume} {8}},\ \bibinfo {pages} {35} (\bibinfo {year} {2017})}\BibitemShut
  {NoStop}%
\bibitem [{\citenamefont {Mardirossian}\ and\ \citenamefont
  {Head-Gordon}(2017)}]{Mardirossian2017}%
  \BibitemOpen
  \bibfield  {author} {\bibinfo {author} {\bibfnamefont {N.}~\bibnamefont
  {Mardirossian}}\ and\ \bibinfo {author} {\bibfnamefont {M.}~\bibnamefont
  {Head-Gordon}},\ }\href {\doibase 10.1080/00268976.2017.1333644} {\bibfield
  {journal} {\bibinfo  {journal} {Molecular Physics}\ }\textbf {\bibinfo
  {volume} {115}},\ \bibinfo {pages} {2315} (\bibinfo {year}
  {2017})}\BibitemShut {NoStop}%
\bibitem [{\citenamefont {Wang}\ \emph {et~al.}(2017)\citenamefont {Wang},
  \citenamefont {Jin}, \citenamefont {Yu}, \citenamefont {Truhlar},\ and\
  \citenamefont {He}}]{revm06l}%
  \BibitemOpen
  \bibfield  {author} {\bibinfo {author} {\bibfnamefont {Y.}~\bibnamefont
  {Wang}}, \bibinfo {author} {\bibfnamefont {X.}~\bibnamefont {Jin}}, \bibinfo
  {author} {\bibfnamefont {H.~S.}\ \bibnamefont {Yu}}, \bibinfo {author}
  {\bibfnamefont {D.~G.}\ \bibnamefont {Truhlar}}, \ and\ \bibinfo {author}
  {\bibfnamefont {X.}~\bibnamefont {He}},\ }\href {\doibase
  10.1073/pnas.1705670114} {\bibfield  {journal} {\bibinfo  {journal} {Proc.
  Natl. Acad. Sci. U.S.A.}\ }\textbf {\bibinfo {volume} {114}},\ \bibinfo
  {pages} {8487} (\bibinfo {year} {2017})}\BibitemShut {NoStop}%
\bibitem [{\citenamefont {Grimme}\ \emph {et~al.}(2015)\citenamefont {Grimme},
  \citenamefont {Brandenburg}, \citenamefont {Bannwarth},\ and\ \citenamefont
  {Hansen}}]{pbeh3c}%
  \BibitemOpen
  \bibfield  {author} {\bibinfo {author} {\bibfnamefont {S.}~\bibnamefont
  {Grimme}}, \bibinfo {author} {\bibfnamefont {G.}~\bibnamefont {Brandenburg}},
  \bibinfo {author} {\bibfnamefont {C.}~\bibnamefont {Bannwarth}}, \ and\
  \bibinfo {author} {\bibfnamefont {A.}~\bibnamefont {Hansen}},\ }\href@noop {}
  {\bibfield  {journal} {\bibinfo  {journal} {J. Chem. Phys.}\ }\textbf
  {\bibinfo {volume} {143}},\ \bibinfo {pages} {054107} (\bibinfo {year}
  {2015})}\BibitemShut {NoStop}%
\bibitem [{\citenamefont {Witte}\ \emph {et~al.}(2015)\citenamefont {Witte},
  \citenamefont {Goldey}, \citenamefont {Neaton},\ and\ \citenamefont
  {Head-Gordon}}]{headgordan_geom}%
  \BibitemOpen
  \bibfield  {author} {\bibinfo {author} {\bibfnamefont {J.}~\bibnamefont
  {Witte}}, \bibinfo {author} {\bibfnamefont {M.}~\bibnamefont {Goldey}},
  \bibinfo {author} {\bibfnamefont {J.~B.}\ \bibnamefont {Neaton}}, \ and\
  \bibinfo {author} {\bibfnamefont {M.}~\bibnamefont {Head-Gordon}},\
  }\href@noop {} {\bibfield  {journal} {\bibinfo  {journal} {J. Chem. Theory
  Comput.}\ }\textbf {\bibinfo {volume} {11}},\ \bibinfo {pages} {1481}
  (\bibinfo {year} {2015})}\BibitemShut {NoStop}%
\bibitem [{\citenamefont {Piccardo}\ \emph {et~al.}(2015)\citenamefont
  {Piccardo}, \citenamefont {Penocchio}, \citenamefont {Puzzarini},
  \citenamefont {Biczysko},\ and\ \citenamefont {Barone}}]{ccse21}%
  \BibitemOpen
  \bibfield  {author} {\bibinfo {author} {\bibfnamefont {M.}~\bibnamefont
  {Piccardo}}, \bibinfo {author} {\bibfnamefont {E.}~\bibnamefont {Penocchio}},
  \bibinfo {author} {\bibfnamefont {C.}~\bibnamefont {Puzzarini}}, \bibinfo
  {author} {\bibfnamefont {M.}~\bibnamefont {Biczysko}}, \ and\ \bibinfo
  {author} {\bibfnamefont {V.}~\bibnamefont {Barone}},\ }\href@noop {}
  {\bibfield  {journal} {\bibinfo  {journal} {J. Phys. Chem. A}\ }\textbf
  {\bibinfo {volume} {119}},\ \bibinfo {pages} {2058} (\bibinfo {year}
  {2015})}\BibitemShut {NoStop}%
\bibitem [{\citenamefont {Grimme}\ and\ \citenamefont
  {Steinmetz}(2013)}]{rot25}%
  \BibitemOpen
  \bibfield  {author} {\bibinfo {author} {\bibfnamefont {S.}~\bibnamefont
  {Grimme}}\ and\ \bibinfo {author} {\bibfnamefont {M.}~\bibnamefont
  {Steinmetz}},\ }\href@noop {} {\bibfield  {journal} {\bibinfo  {journal}
  {Phys. Chem. Chem. Phys.}\ }\textbf {\bibinfo {volume} {15}},\ \bibinfo
  {pages} {16031} (\bibinfo {year} {2013})}\BibitemShut {NoStop}%
\bibitem [{\citenamefont {Staroverov}\ \emph {et~al.}(2004)\citenamefont
  {Staroverov}, \citenamefont {Scuseria}, \citenamefont {Tao},\ and\
  \citenamefont {Perdew}}]{dftbench-bulk}%
  \BibitemOpen
  \bibfield  {author} {\bibinfo {author} {\bibfnamefont {V.~N.}\ \bibnamefont
  {Staroverov}}, \bibinfo {author} {\bibfnamefont {G.~E.}\ \bibnamefont
  {Scuseria}}, \bibinfo {author} {\bibfnamefont {J.}~\bibnamefont {Tao}}, \
  and\ \bibinfo {author} {\bibfnamefont {J.~P.}\ \bibnamefont {Perdew}},\
  }\href {\doibase 10.1103/PhysRevB.69.075102} {\bibfield  {journal} {\bibinfo
  {journal} {Phys. Rev. B}\ }\textbf {\bibinfo {volume} {69}},\ \bibinfo
  {pages} {075102} (\bibinfo {year} {2004})}\BibitemShut {NoStop}%
\bibitem [{\citenamefont {Pernot}\ \emph {et~al.}(2015)\citenamefont {Pernot},
  \citenamefont {Civalleri}, \citenamefont {Presti},\ and\ \citenamefont
  {Savin}}]{ss20}%
  \BibitemOpen
  \bibfield  {author} {\bibinfo {author} {\bibfnamefont {P.}~\bibnamefont
  {Pernot}}, \bibinfo {author} {\bibfnamefont {B.}~\bibnamefont {Civalleri}},
  \bibinfo {author} {\bibfnamefont {D.}~\bibnamefont {Presti}}, \ and\ \bibinfo
  {author} {\bibfnamefont {A.}~\bibnamefont {Savin}},\ }\href@noop {}
  {\bibfield  {journal} {\bibinfo  {journal} {J. Phys. Chem. A}\ }\textbf
  {\bibinfo {volume} {119}},\ \bibinfo {pages} {5288} (\bibinfo {year}
  {2015})}\BibitemShut {NoStop}%
\bibitem [{\citenamefont {Peintinger}, \citenamefont {Oliveira},\ and\
  \citenamefont {Bredow}(2013)}]{pob}%
  \BibitemOpen
  \bibfield  {author} {\bibinfo {author} {\bibfnamefont {M.~F.}\ \bibnamefont
  {Peintinger}}, \bibinfo {author} {\bibfnamefont {D.~V.}\ \bibnamefont
  {Oliveira}}, \ and\ \bibinfo {author} {\bibfnamefont {T.}~\bibnamefont
  {Bredow}},\ }\href@noop {} {\bibfield  {journal} {\bibinfo  {journal} {J.
  Comp. Chem.}\ }\textbf {\bibinfo {volume} {34}},\ \bibinfo {pages} {451}
  (\bibinfo {year} {2013})}\BibitemShut {NoStop}%
\bibitem [{\citenamefont {Zhang}\ \emph {et~al.}(2018)\citenamefont {Zhang},
  \citenamefont {Reilly}, \citenamefont {Tkatchenko},\ and\ \citenamefont
  {Scheffler}}]{dftbench_bs64}%
  \BibitemOpen
  \bibfield  {author} {\bibinfo {author} {\bibfnamefont {G.-X.}\ \bibnamefont
  {Zhang}}, \bibinfo {author} {\bibfnamefont {A.~M.}\ \bibnamefont {Reilly}},
  \bibinfo {author} {\bibfnamefont {A.}~\bibnamefont {Tkatchenko}}, \ and\
  \bibinfo {author} {\bibfnamefont {M.}~\bibnamefont {Scheffler}},\ }\href
  {\doibase 10.1088/1367-2630/aac7f0} {\bibfield  {journal} {\bibinfo
  {journal} {New J. Phys.}\ }\textbf {\bibinfo {volume} {20}},\ \bibinfo
  {pages} {063020} (\bibinfo {year} {2018})}\BibitemShut {NoStop}%
\bibitem [{\citenamefont {{Otero-de-la-Roza}}\ and\ \citenamefont
  {Johnson}(2012)}]{c21}%
  \BibitemOpen
  \bibfield  {author} {\bibinfo {author} {\bibfnamefont {A.}~\bibnamefont
  {{Otero-de-la-Roza}}}\ and\ \bibinfo {author} {\bibfnamefont {E.~R.}\
  \bibnamefont {Johnson}},\ }\href@noop {} {\bibfield  {journal} {\bibinfo
  {journal} {J. Chem. Phys.}\ }\textbf {\bibinfo {volume} {137}},\ \bibinfo
  {pages} {054103} (\bibinfo {year} {2012})}\BibitemShut {NoStop}%
\bibitem [{\citenamefont {Reilly}\ and\ \citenamefont
  {Tkatchenko}(2013)}]{x23}%
  \BibitemOpen
  \bibfield  {author} {\bibinfo {author} {\bibfnamefont {A.~M.}\ \bibnamefont
  {Reilly}}\ and\ \bibinfo {author} {\bibfnamefont {A.}~\bibnamefont
  {Tkatchenko}},\ }\href@noop {} {\bibfield  {journal} {\bibinfo  {journal} {J.
  Chem. Phys.}\ }\textbf {\bibinfo {volume} {139}},\ \bibinfo {pages} {024705}
  (\bibinfo {year} {2013})}\BibitemShut {NoStop}%
\bibitem [{\citenamefont {Brandenburg}, \citenamefont {Maas},\ and\
  \citenamefont {Grimme}(2015)}]{ice10}%
  \BibitemOpen
  \bibfield  {author} {\bibinfo {author} {\bibfnamefont {J.~G.}\ \bibnamefont
  {Brandenburg}}, \bibinfo {author} {\bibfnamefont {T.}~\bibnamefont {Maas}}, \
  and\ \bibinfo {author} {\bibfnamefont {S.}~\bibnamefont {Grimme}},\
  }\href@noop {} {\bibfield  {journal} {\bibinfo  {journal} {J. Chem. Phys.}\
  }\textbf {\bibinfo {volume} {142}},\ \bibinfo {pages} {124104} (\bibinfo
  {year} {2015})}\BibitemShut {NoStop}%
\bibitem [{\citenamefont {Chickos}(2003)}]{Chickos2003_DHerror}%
  \BibitemOpen
  \bibfield  {author} {\bibinfo {author} {\bibfnamefont {J.~S.}\ \bibnamefont
  {Chickos}},\ }\href@noop {} {\bibfield  {journal} {\bibinfo  {journal} {Netsu
  Sokutei}\ } (\bibinfo {year} {2003})}\BibitemShut {NoStop}%
\bibitem [{\citenamefont {Bj\"{o}rneholm}\ \emph {et~al.}(2016)\citenamefont
  {Bj\"{o}rneholm}, \citenamefont {Hansen}, \citenamefont {Hodgson},
  \citenamefont {Liu}, \citenamefont {Limmer}, \citenamefont {Michaelides},
  \citenamefont {Pedevilla}, \citenamefont {Rossmeisl}, \citenamefont {Shen},
  \citenamefont {Tocci}, \citenamefont {Tyrode}, \citenamefont {Walz},
  \citenamefont {Werner},\ and\ \citenamefont {Bluhm}}]{water-interface}%
  \BibitemOpen
  \bibfield  {author} {\bibinfo {author} {\bibfnamefont {O.}~\bibnamefont
  {Bj\"{o}rneholm}}, \bibinfo {author} {\bibfnamefont {M.~H.}\ \bibnamefont
  {Hansen}}, \bibinfo {author} {\bibfnamefont {A.}~\bibnamefont {Hodgson}},
  \bibinfo {author} {\bibfnamefont {L.-M.}\ \bibnamefont {Liu}}, \bibinfo
  {author} {\bibfnamefont {D.~T.}\ \bibnamefont {Limmer}}, \bibinfo {author}
  {\bibfnamefont {A.}~\bibnamefont {Michaelides}}, \bibinfo {author}
  {\bibfnamefont {P.}~\bibnamefont {Pedevilla}}, \bibinfo {author}
  {\bibfnamefont {J.}~\bibnamefont {Rossmeisl}}, \bibinfo {author}
  {\bibfnamefont {H.}~\bibnamefont {Shen}}, \bibinfo {author} {\bibfnamefont
  {G.}~\bibnamefont {Tocci}}, \bibinfo {author} {\bibfnamefont
  {E.}~\bibnamefont {Tyrode}}, \bibinfo {author} {\bibfnamefont {M.-M.}\
  \bibnamefont {Walz}}, \bibinfo {author} {\bibfnamefont {J.}~\bibnamefont
  {Werner}}, \ and\ \bibinfo {author} {\bibfnamefont {H.}~\bibnamefont
  {Bluhm}},\ }\href {\doibase 10.1021/acs.chemrev.6b00045} {\bibfield
  {journal} {\bibinfo  {journal} {Chem. Rev.}\ }\textbf {\bibinfo {volume}
  {116}},\ \bibinfo {pages} {7698} (\bibinfo {year} {2016})}\BibitemShut
  {NoStop}%
\bibitem [{\citenamefont {Masur}\ \emph {et~al.}(2016)\citenamefont {Masur},
  \citenamefont {Sch\"{u}tz}, \citenamefont {Maschio},\ and\ \citenamefont
  {Usvyat}}]{cryscor_embedding}%
  \BibitemOpen
  \bibfield  {author} {\bibinfo {author} {\bibfnamefont {O.}~\bibnamefont
  {Masur}}, \bibinfo {author} {\bibfnamefont {M.}~\bibnamefont {Sch\"{u}tz}},
  \bibinfo {author} {\bibfnamefont {L.}~\bibnamefont {Maschio}}, \ and\
  \bibinfo {author} {\bibfnamefont {D.}~\bibnamefont {Usvyat}},\ }\href@noop {}
  {\bibfield  {journal} {\bibinfo  {journal} {J. Chem. Theory Comput.}\
  }\textbf {\bibinfo {volume} {12}},\ \bibinfo {pages} {5145} (\bibinfo {year}
  {2016})}\BibitemShut {NoStop}%
\bibitem [{\citenamefont {Bygrave}, \citenamefont {Allan},\ and\ \citenamefont
  {Manby}(2012)}]{manby_embedding}%
  \BibitemOpen
  \bibfield  {author} {\bibinfo {author} {\bibfnamefont {P.~J.}\ \bibnamefont
  {Bygrave}}, \bibinfo {author} {\bibfnamefont {N.~L.}\ \bibnamefont {Allan}},
  \ and\ \bibinfo {author} {\bibfnamefont {F.~R.}\ \bibnamefont {Manby}},\
  }\href@noop {} {\bibfield  {journal} {\bibinfo  {journal} {J. Chem. Phys.}\
  }\textbf {\bibinfo {volume} {137}},\ \bibinfo {pages} {164102} (\bibinfo
  {year} {2012})}\BibitemShut {NoStop}%
\bibitem [{\citenamefont {Beran}\ \emph {et~al.}(2014)\citenamefont {Beran},
  \citenamefont {Wen}, \citenamefont {Nand}, \citenamefont {Huang},\ and\
  \citenamefont {Heit}}]{csp_beran}%
  \BibitemOpen
  \bibfield  {author} {\bibinfo {author} {\bibfnamefont {G.~J.~O.}\
  \bibnamefont {Beran}}, \bibinfo {author} {\bibfnamefont {S.}~\bibnamefont
  {Wen}}, \bibinfo {author} {\bibfnamefont {K.}~\bibnamefont {Nand}}, \bibinfo
  {author} {\bibfnamefont {Y.}~\bibnamefont {Huang}}, \ and\ \bibinfo {author}
  {\bibfnamefont {Y.}~\bibnamefont {Heit}},\ }\href@noop {} {\bibfield
  {journal} {\bibinfo  {journal} {Top. Curr. Chem.}\ }\textbf {\bibinfo
  {volume} {345}},\ \bibinfo {pages} {59} (\bibinfo {year} {2014})}\BibitemShut
  {NoStop}%
\bibitem [{\citenamefont {Riplinger}\ \emph {et~al.}(2013)\citenamefont
  {Riplinger}, \citenamefont {Sandhoefer}, \citenamefont {Hansen},\ and\
  \citenamefont {Neese}}]{dlpnoccsdt}%
  \BibitemOpen
  \bibfield  {author} {\bibinfo {author} {\bibfnamefont {C.}~\bibnamefont
  {Riplinger}}, \bibinfo {author} {\bibfnamefont {B.}~\bibnamefont
  {Sandhoefer}}, \bibinfo {author} {\bibfnamefont {A.}~\bibnamefont {Hansen}},
  \ and\ \bibinfo {author} {\bibfnamefont {F.}~\bibnamefont {Neese}},\
  }\href@noop {} {\bibfield  {journal} {\bibinfo  {journal} {J. Chem. Phys.}\
  }\textbf {\bibinfo {volume} {139}},\ \bibinfo {pages} {134101} (\bibinfo
  {year} {2013})}\BibitemShut {NoStop}%
\bibitem [{\citenamefont {Riplinger}\ \emph {et~al.}(2016)\citenamefont
  {Riplinger}, \citenamefont {Pinski}, \citenamefont {Becker}, \citenamefont
  {Valeev},\ and\ \citenamefont {Neese}}]{dlpno_compact}%
  \BibitemOpen
  \bibfield  {author} {\bibinfo {author} {\bibfnamefont {C.}~\bibnamefont
  {Riplinger}}, \bibinfo {author} {\bibfnamefont {P.}~\bibnamefont {Pinski}},
  \bibinfo {author} {\bibfnamefont {U.}~\bibnamefont {Becker}}, \bibinfo
  {author} {\bibfnamefont {E.~F.}\ \bibnamefont {Valeev}}, \ and\ \bibinfo
  {author} {\bibfnamefont {F.}~\bibnamefont {Neese}},\ }\href {\doibase
  10.1063/1.4939030} {\bibfield  {journal} {\bibinfo  {journal} {J. Chem.
  Phys.}\ }\textbf {\bibinfo {volume} {144}},\ \bibinfo {pages} {024109}
  (\bibinfo {year} {2016})}\BibitemShut {NoStop}%
\bibitem [{\citenamefont {Maurer}\ \emph {et~al.}(2013)\citenamefont {Maurer},
  \citenamefont {Lambrecht}, \citenamefont {Kussmann},\ and\ \citenamefont
  {Ochsenfeld}}]{ao_lmp2}%
  \BibitemOpen
  \bibfield  {author} {\bibinfo {author} {\bibfnamefont {S.~A.}\ \bibnamefont
  {Maurer}}, \bibinfo {author} {\bibfnamefont {D.~S.}\ \bibnamefont
  {Lambrecht}}, \bibinfo {author} {\bibfnamefont {J.}~\bibnamefont {Kussmann}},
  \ and\ \bibinfo {author} {\bibfnamefont {C.}~\bibnamefont {Ochsenfeld}},\
  }\href@noop {} {\bibfield  {journal} {\bibinfo  {journal} {J. Chem. Phys.}\
  }\textbf {\bibinfo {volume} {138}},\ \bibinfo {pages} {014101} (\bibinfo
  {year} {2013})}\BibitemShut {NoStop}%
\bibitem [{\citenamefont {Sch{\"u}tz}\ and\ \citenamefont
  {Werner}(2001)}]{werner_lcc}%
  \BibitemOpen
  \bibfield  {author} {\bibinfo {author} {\bibfnamefont {M.}~\bibnamefont
  {Sch{\"u}tz}}\ and\ \bibinfo {author} {\bibfnamefont {H.~J.}\ \bibnamefont
  {Werner}},\ }\href@noop {} {\bibfield  {journal} {\bibinfo  {journal} {J.
  Chem. Phys.}\ }\textbf {\bibinfo {volume} {114}},\ \bibinfo {pages} {661}
  (\bibinfo {year} {2001})}\BibitemShut {NoStop}%
\bibitem [{\citenamefont {Nagy}, \citenamefont {Samu},\ and\ \citenamefont
  {Kállay}(2018)}]{lno_ccsdt}%
  \BibitemOpen
  \bibfield  {author} {\bibinfo {author} {\bibfnamefont {P.~R.}\ \bibnamefont
  {Nagy}}, \bibinfo {author} {\bibfnamefont {G.}~\bibnamefont {Samu}}, \ and\
  \bibinfo {author} {\bibfnamefont {M.}~\bibnamefont {Kállay}},\ }\href
  {\doibase 10.1021/acs.jctc.8b00442} {\bibfield  {journal} {\bibinfo
  {journal} {J. Chem. Theory Comput.}\ }\textbf {\bibinfo {volume} {14}},\
  \bibinfo {pages} {4193} (\bibinfo {year} {2018})}\BibitemShut {NoStop}%
\bibitem [{\citenamefont {Paulus}(2006)}]{i-ccsdt}%
  \BibitemOpen
  \bibfield  {author} {\bibinfo {author} {\bibfnamefont {B.}~\bibnamefont
  {Paulus}},\ }\href {\doibase 10.1016/j.physrep.2006.01.003} {\bibfield
  {journal} {\bibinfo  {journal} {Physics Reports}\ }\textbf {\bibinfo {volume}
  {428}},\ \bibinfo {pages} {1 } (\bibinfo {year} {2006})}\BibitemShut
  {NoStop}%
\bibitem [{\citenamefont {Beran}(2016)}]{beran_chemrev}%
  \BibitemOpen
  \bibfield  {author} {\bibinfo {author} {\bibfnamefont {G.~J.~O.}\
  \bibnamefont {Beran}},\ }\href@noop {} {\bibfield  {journal} {\bibinfo
  {journal} {Chem. Rev.}\ }\textbf {\bibinfo {volume} {116}},\ \bibinfo {pages}
  {5567} (\bibinfo {year} {2016})}\BibitemShut {NoStop}%
\bibitem [{\citenamefont {Yang}\ \emph {et~al.}(2014)\citenamefont {Yang},
  \citenamefont {Hu}, \citenamefont {Usvyat}, \citenamefont {Matthews},
  \citenamefont {Sch\"{u}tz},\ and\ \citenamefont {Chan}}]{bzsolid_ccsdt}%
  \BibitemOpen
  \bibfield  {author} {\bibinfo {author} {\bibfnamefont {J.}~\bibnamefont
  {Yang}}, \bibinfo {author} {\bibfnamefont {W.}~\bibnamefont {Hu}}, \bibinfo
  {author} {\bibfnamefont {D.}~\bibnamefont {Usvyat}}, \bibinfo {author}
  {\bibfnamefont {D.}~\bibnamefont {Matthews}}, \bibinfo {author}
  {\bibfnamefont {M.}~\bibnamefont {Sch\"{u}tz}}, \ and\ \bibinfo {author}
  {\bibfnamefont {G.~K.~L.}\ \bibnamefont {Chan}},\ }\href@noop {} {\bibfield
  {journal} {\bibinfo  {journal} {Science}\ }\textbf {\bibinfo {volume}
  {345}},\ \bibinfo {pages} {640} (\bibinfo {year} {2014})}\BibitemShut
  {NoStop}%
\bibitem [{\citenamefont {Gruber}\ \emph {et~al.}(2018)\citenamefont {Gruber},
  \citenamefont {Liao}, \citenamefont {Tsatsoulis}, \citenamefont {Hummel},\
  and\ \citenamefont {Gr\"uneis}}]{ccsdt_surfaces}%
  \BibitemOpen
  \bibfield  {author} {\bibinfo {author} {\bibfnamefont {T.}~\bibnamefont
  {Gruber}}, \bibinfo {author} {\bibfnamefont {K.}~\bibnamefont {Liao}},
  \bibinfo {author} {\bibfnamefont {T.}~\bibnamefont {Tsatsoulis}}, \bibinfo
  {author} {\bibfnamefont {F.}~\bibnamefont {Hummel}}, \ and\ \bibinfo {author}
  {\bibfnamefont {A.}~\bibnamefont {Gr\"uneis}},\ }\href {\doibase
  10.1103/PhysRevX.8.021043} {\bibfield  {journal} {\bibinfo  {journal} {Phys.
  Rev. X}\ }\textbf {\bibinfo {volume} {8}},\ \bibinfo {pages} {021043}
  (\bibinfo {year} {2018})}\BibitemShut {NoStop}%
\bibitem [{\citenamefont {Zen}\ \emph {et~al.}(2016)\citenamefont {Zen},
  \citenamefont {Sorella}, \citenamefont {Gillan}, \citenamefont
  {Michaelides},\ and\ \citenamefont {Alf\`e}}]{dmc_sizeconsistent}%
  \BibitemOpen
  \bibfield  {author} {\bibinfo {author} {\bibfnamefont {A.}~\bibnamefont
  {Zen}}, \bibinfo {author} {\bibfnamefont {S.}~\bibnamefont {Sorella}},
  \bibinfo {author} {\bibfnamefont {M.~J.}\ \bibnamefont {Gillan}}, \bibinfo
  {author} {\bibfnamefont {A.}~\bibnamefont {Michaelides}}, \ and\ \bibinfo
  {author} {\bibfnamefont {D.}~\bibnamefont {Alf\`e}},\ }\href {\doibase
  10.1103/PhysRevB.93.241118} {\bibfield  {journal} {\bibinfo  {journal} {Phys.
  Rev. B}\ }\textbf {\bibinfo {volume} {93}},\ \bibinfo {pages} {241118}
  (\bibinfo {year} {2016})}\BibitemShut {NoStop}%
\bibitem [{\citenamefont {Zen}\ \emph {et~al.}(2018)\citenamefont {Zen},
  \citenamefont {Brandenburg}, \citenamefont {Klime\v{s}}, \citenamefont
  {Tkatchenko}, \citenamefont {Alf\`e},\ and\ \citenamefont
  {Michaelides}}]{molcryst_dmc}%
  \BibitemOpen
  \bibfield  {author} {\bibinfo {author} {\bibfnamefont {A.}~\bibnamefont
  {Zen}}, \bibinfo {author} {\bibfnamefont {J.~G.}\ \bibnamefont
  {Brandenburg}}, \bibinfo {author} {\bibfnamefont {J.}~\bibnamefont
  {Klime\v{s}}}, \bibinfo {author} {\bibfnamefont {A.}~\bibnamefont
  {Tkatchenko}}, \bibinfo {author} {\bibfnamefont {D.}~\bibnamefont {Alf\`e}},
  \ and\ \bibinfo {author} {\bibfnamefont {A.}~\bibnamefont {Michaelides}},\
  }\href {\doibase 10.1073/pnas.1715434115} {\bibfield  {journal} {\bibinfo
  {journal} {{Proc. Natl. Acad. Sci. U.S.A.}}\ }\textbf {\bibinfo {volume}
  {115}},\ \bibinfo {pages} {1724} (\bibinfo {year} {2018})}\BibitemShut
  {NoStop}%
\bibitem [{\citenamefont {Michaelides}\ \emph {et~al.}(2015)\citenamefont
  {Michaelides}, \citenamefont {Martinez}, \citenamefont {Alavi}, \citenamefont
  {Kresse},\ and\ \citenamefont {Manby}}]{advancesQC_editorial}%
  \BibitemOpen
  \bibfield  {author} {\bibinfo {author} {\bibfnamefont {A.}~\bibnamefont
  {Michaelides}}, \bibinfo {author} {\bibfnamefont {T.~J.}\ \bibnamefont
  {Martinez}}, \bibinfo {author} {\bibfnamefont {A.}~\bibnamefont {Alavi}},
  \bibinfo {author} {\bibfnamefont {G.}~\bibnamefont {Kresse}}, \ and\ \bibinfo
  {author} {\bibfnamefont {F.~R.}\ \bibnamefont {Manby}},\ }\href {\doibase
  10.1063/1.4930182} {\bibfield  {journal} {\bibinfo  {journal} {J. Chem.
  Phys.}\ }\textbf {\bibinfo {volume} {143}},\ \bibinfo {pages} {102601}
  (\bibinfo {year} {2015})}\BibitemShut {NoStop}%
\bibitem [{\citenamefont {Al-Hamdani}\ and\ \citenamefont
  {Tkatchenko}(2019{\natexlab{a}})}]{largeNCI_perspective}%
  \BibitemOpen
  \bibfield  {author} {\bibinfo {author} {\bibfnamefont {Y.~S.}\ \bibnamefont
  {Al-Hamdani}}\ and\ \bibinfo {author} {\bibfnamefont {A.}~\bibnamefont
  {Tkatchenko}},\ }\href {\doibase 10.1063/1.5075487} {\bibfield  {journal}
  {\bibinfo  {journal} {J. Chem. Phys.}\ }\textbf {\bibinfo {volume} {150}},\
  \bibinfo {pages} {010901} (\bibinfo {year} {2019}{\natexlab{a}})}\BibitemShut
  {NoStop}%
\bibitem [{\citenamefont {Brandenburg}\ \emph {et~al.}(2019)\citenamefont
  {Brandenburg}, \citenamefont {Zen}, \citenamefont {Fitzner}, \citenamefont
  {Ramberger}, \citenamefont {Kresse}, \citenamefont {Tsatsoulis},
  \citenamefont {Gr\"{u}neis}, \citenamefont {Michaelides},\ and\ \citenamefont
  {Alf\`{e}}}]{water-graphene-dmc}%
  \BibitemOpen
  \bibfield  {author} {\bibinfo {author} {\bibfnamefont {J.~G.}\ \bibnamefont
  {Brandenburg}}, \bibinfo {author} {\bibfnamefont {A.}~\bibnamefont {Zen}},
  \bibinfo {author} {\bibfnamefont {M.}~\bibnamefont {Fitzner}}, \bibinfo
  {author} {\bibfnamefont {B.}~\bibnamefont {Ramberger}}, \bibinfo {author}
  {\bibfnamefont {G.}~\bibnamefont {Kresse}}, \bibinfo {author} {\bibfnamefont
  {T.}~\bibnamefont {Tsatsoulis}}, \bibinfo {author} {\bibfnamefont
  {A.}~\bibnamefont {Gr\"{u}neis}}, \bibinfo {author} {\bibfnamefont
  {A.}~\bibnamefont {Michaelides}}, \ and\ \bibinfo {author} {\bibfnamefont
  {D.}~\bibnamefont {Alf\`{e}}},\ }\href {\doibase 10.1021/acs.jpclett.8b03679}
  {\bibfield  {journal} {\bibinfo  {journal} {{J. Phys. Chem. Lett.}}\ }\textbf
  {\bibinfo {volume} {10}},\ \bibinfo {pages} {358} (\bibinfo {year}
  {2019})}\BibitemShut {NoStop}%
\bibitem [{\citenamefont {Al-Hamdani}, \citenamefont {Alf\`{e}},\ and\
  \citenamefont {Michaelides}(2017{\natexlab{a}})}]{yasmine_water_cnt}%
  \BibitemOpen
  \bibfield  {author} {\bibinfo {author} {\bibfnamefont {Y.~S.}\ \bibnamefont
  {Al-Hamdani}}, \bibinfo {author} {\bibfnamefont {D.}~\bibnamefont
  {Alf\`{e}}}, \ and\ \bibinfo {author} {\bibfnamefont {A.}~\bibnamefont
  {Michaelides}},\ }\href {\doibase 10.1063/1.4977180} {\bibfield  {journal}
  {\bibinfo  {journal} {J. Chem. Phys.}\ }\textbf {\bibinfo {volume} {146}},\
  \bibinfo {pages} {094701} (\bibinfo {year} {2017}{\natexlab{a}})}\BibitemShut
  {NoStop}%
\bibitem [{\citenamefont {Chen}\ \emph {et~al.}(2016)\citenamefont {Chen},
  \citenamefont {Zen}, \citenamefont {Brandenburg}, \citenamefont {Alf{\`e}},\
  and\ \citenamefont {Michaelides}}]{2dice-dmc}%
  \BibitemOpen
  \bibfield  {author} {\bibinfo {author} {\bibfnamefont {J.}~\bibnamefont
  {Chen}}, \bibinfo {author} {\bibfnamefont {A.}~\bibnamefont {Zen}}, \bibinfo
  {author} {\bibfnamefont {J.~G.}\ \bibnamefont {Brandenburg}}, \bibinfo
  {author} {\bibfnamefont {D.}~\bibnamefont {Alf{\`e}}}, \ and\ \bibinfo
  {author} {\bibfnamefont {A.}~\bibnamefont {Michaelides}},\ }\href {\doibase
  10.1103/PhysRevB.94.220102} {\bibfield  {journal} {\bibinfo  {journal}
  {{Phys. Rev. B}}\ }\textbf {\bibinfo {volume} {94}},\ \bibinfo {pages}
  {220102} (\bibinfo {year} {2016})}\BibitemShut {NoStop}%
\bibitem [{\citenamefont {Ma}\ \emph {et~al.}(2011)\citenamefont {Ma},
  \citenamefont {Michaelides}, \citenamefont {Alf\`e}, \citenamefont {Schimka},
  \citenamefont {Kresse},\ and\ \citenamefont {Wang}}]{water_graphene_dmc_old}%
  \BibitemOpen
  \bibfield  {author} {\bibinfo {author} {\bibfnamefont {J.}~\bibnamefont
  {Ma}}, \bibinfo {author} {\bibfnamefont {A.}~\bibnamefont {Michaelides}},
  \bibinfo {author} {\bibfnamefont {D.}~\bibnamefont {Alf\`e}}, \bibinfo
  {author} {\bibfnamefont {L.}~\bibnamefont {Schimka}}, \bibinfo {author}
  {\bibfnamefont {G.}~\bibnamefont {Kresse}}, \ and\ \bibinfo {author}
  {\bibfnamefont {E.}~\bibnamefont {Wang}},\ }\href {\doibase
  10.1103/PhysRevB.84.033402} {\bibfield  {journal} {\bibinfo  {journal} {Phys.
  Rev. B}\ }\textbf {\bibinfo {volume} {84}},\ \bibinfo {pages} {033402}
  (\bibinfo {year} {2011})}\BibitemShut {NoStop}%
\bibitem [{\citenamefont {Ajala}\ \emph {et~al.}(2019)\citenamefont {Ajala},
  \citenamefont {Voora}, \citenamefont {Mardirossian}, \citenamefont {Furche},\
  and\ \citenamefont {Paesani}}]{water_graphene_paesani}%
  \BibitemOpen
  \bibfield  {author} {\bibinfo {author} {\bibfnamefont {A.~O.}\ \bibnamefont
  {Ajala}}, \bibinfo {author} {\bibfnamefont {V.}~\bibnamefont {Voora}},
  \bibinfo {author} {\bibfnamefont {N.}~\bibnamefont {Mardirossian}}, \bibinfo
  {author} {\bibfnamefont {F.}~\bibnamefont {Furche}}, \ and\ \bibinfo {author}
  {\bibfnamefont {F.}~\bibnamefont {Paesani}},\ }\href {\doibase
  10.1021/acs.jctc.9b00110} {\bibfield  {journal} {\bibinfo  {journal} {J.
  Chem. Theory Comput.}\ }\textbf {\bibinfo {volume} {15}},\ \bibinfo {pages}
  {2359} (\bibinfo {year} {2019})}\BibitemShut {NoStop}%
\bibitem [{\citenamefont {Santra}\ \emph {et~al.}(2013)\citenamefont {Santra},
  \citenamefont {Klimeš}, \citenamefont {Tkatchenko}, \citenamefont {Alfè},
  \citenamefont {Slater}, \citenamefont {Michaelides}, \citenamefont {Car},\
  and\ \citenamefont {Scheffler}}]{ice_dmc_dft}%
  \BibitemOpen
  \bibfield  {author} {\bibinfo {author} {\bibfnamefont {B.}~\bibnamefont
  {Santra}}, \bibinfo {author} {\bibfnamefont {J.}~\bibnamefont {Klimeš}},
  \bibinfo {author} {\bibfnamefont {A.}~\bibnamefont {Tkatchenko}}, \bibinfo
  {author} {\bibfnamefont {D.}~\bibnamefont {Alfè}}, \bibinfo {author}
  {\bibfnamefont {B.}~\bibnamefont {Slater}}, \bibinfo {author} {\bibfnamefont
  {A.}~\bibnamefont {Michaelides}}, \bibinfo {author} {\bibfnamefont
  {R.}~\bibnamefont {Car}}, \ and\ \bibinfo {author} {\bibfnamefont
  {M.}~\bibnamefont {Scheffler}},\ }\href {\doibase 10.1063/1.4824481}
  {\bibfield  {journal} {\bibinfo  {journal} {J. Chem. Phys.}\ }\textbf
  {\bibinfo {volume} {139}},\ \bibinfo {pages} {154702} (\bibinfo {year}
  {2013})}\BibitemShut {NoStop}%
\bibitem [{\citenamefont {He}\ \emph {et~al.}(2012)\citenamefont {He},
  \citenamefont {Sode}, \citenamefont {Xantheas},\ and\ \citenamefont
  {Hirata}}]{hirata-ice}%
  \BibitemOpen
  \bibfield  {author} {\bibinfo {author} {\bibfnamefont {X.}~\bibnamefont
  {He}}, \bibinfo {author} {\bibfnamefont {O.}~\bibnamefont {Sode}}, \bibinfo
  {author} {\bibfnamefont {S.~S.}\ \bibnamefont {Xantheas}}, \ and\ \bibinfo
  {author} {\bibfnamefont {S.}~\bibnamefont {Hirata}},\ }\href {\doibase
  10.1063/1.4767898} {\bibfield  {journal} {\bibinfo  {journal} {J. Chem.
  Phys.}\ }\textbf {\bibinfo {volume} {137}},\ \bibinfo {pages} {204505}
  (\bibinfo {year} {2012})}\BibitemShut {NoStop}%
\bibitem [{\citenamefont {Hirata}\ \emph {et~al.}(2017)\citenamefont {Hirata},
  \citenamefont {Gilliard}, \citenamefont {He}, \citenamefont {Keçeli},
  \citenamefont {Li}, \citenamefont {A.~Salim}, \citenamefont {Sode},\ and\
  \citenamefont {Yagi}}]{hirata-fragmentation-ice}%
  \BibitemOpen
  \bibfield  {author} {\bibinfo {author} {\bibfnamefont {S.}~\bibnamefont
  {Hirata}}, \bibinfo {author} {\bibfnamefont {K.}~\bibnamefont {Gilliard}},
  \bibinfo {author} {\bibfnamefont {X.}~\bibnamefont {He}}, \bibinfo {author}
  {\bibfnamefont {M.}~\bibnamefont {Keçeli}}, \bibinfo {author} {\bibfnamefont
  {J.}~\bibnamefont {Li}}, \bibinfo {author} {\bibfnamefont {M.}~\bibnamefont
  {A.~Salim}}, \bibinfo {author} {\bibfnamefont {O.}~\bibnamefont {Sode}}, \
  and\ \bibinfo {author} {\bibfnamefont {K.}~\bibnamefont {Yagi}},\ }\enquote
  {\bibinfo {title} {Ab initio ice, dry ice, and liquid water},}\ in\ \href
  {\doibase 10.1002/9781119129271.ch9} {\emph {\bibinfo {booktitle}
  {Fragmentation}}}\ (\bibinfo  {publisher} {John Wiley \& Sons, Ltd},\
  \bibinfo {year} {2017})\ Chap.~\bibinfo {chapter} {9}, pp.\ \bibinfo {pages}
  {245--296}\BibitemShut {NoStop}%
\bibitem [{\citenamefont {Hamada}\ and\ \citenamefont
  {Otani}(2010)}]{graphene-metal-dft}%
  \BibitemOpen
  \bibfield  {author} {\bibinfo {author} {\bibfnamefont {I.}~\bibnamefont
  {Hamada}}\ and\ \bibinfo {author} {\bibfnamefont {M.}~\bibnamefont {Otani}},\
  }\href {\doibase 10.1103/PhysRevB.82.153412} {\bibfield  {journal} {\bibinfo
  {journal} {Phys. Rev. B}\ }\textbf {\bibinfo {volume} {82}},\ \bibinfo
  {pages} {153412} (\bibinfo {year} {2010})}\BibitemShut {NoStop}%
\bibitem [{\citenamefont {Voloshina}\ \emph {et~al.}(2011)\citenamefont
  {Voloshina}, \citenamefont {Usvyat}, \citenamefont {Schutz}, \citenamefont
  {Dedkov},\ and\ \citenamefont {Paulus}}]{water@graphene_paulus}%
  \BibitemOpen
  \bibfield  {author} {\bibinfo {author} {\bibfnamefont {E.}~\bibnamefont
  {Voloshina}}, \bibinfo {author} {\bibfnamefont {D.}~\bibnamefont {Usvyat}},
  \bibinfo {author} {\bibfnamefont {M.}~\bibnamefont {Schutz}}, \bibinfo
  {author} {\bibfnamefont {Y.}~\bibnamefont {Dedkov}}, \ and\ \bibinfo {author}
  {\bibfnamefont {B.}~\bibnamefont {Paulus}},\ }\href {\doibase
  10.1039/C1CP20609E} {\bibfield  {journal} {\bibinfo  {journal} {Phys. Chem.
  Chem. Phys.}\ }\textbf {\bibinfo {volume} {13}},\ \bibinfo {pages} {12041}
  (\bibinfo {year} {2011})}\BibitemShut {NoStop}%
\bibitem [{\citenamefont {Lei}\ \emph {et~al.}(2016)\citenamefont {Lei},
  \citenamefont {Paulus}, \citenamefont {Li},\ and\ \citenamefont
  {Schmidt}}]{water-cnt_ccsdt}%
  \BibitemOpen
  \bibfield  {author} {\bibinfo {author} {\bibfnamefont {S.}~\bibnamefont
  {Lei}}, \bibinfo {author} {\bibfnamefont {B.}~\bibnamefont {Paulus}},
  \bibinfo {author} {\bibfnamefont {S.}~\bibnamefont {Li}}, \ and\ \bibinfo
  {author} {\bibfnamefont {B.}~\bibnamefont {Schmidt}},\ }\href {\doibase
  10.1002/jcc.24342} {\bibfield  {journal} {\bibinfo  {journal} {J. Comput.
  Chem.}\ }\textbf {\bibinfo {volume} {37}},\ \bibinfo {pages} {1313} (\bibinfo
  {year} {2016})}\BibitemShut {NoStop}%
\bibitem [{\citenamefont {Hamada}(2012)}]{water@graphene_dft1}%
  \BibitemOpen
  \bibfield  {author} {\bibinfo {author} {\bibfnamefont {I.}~\bibnamefont
  {Hamada}},\ }\href {\doibase 10.1103/PhysRevB.86.195436} {\bibfield
  {journal} {\bibinfo  {journal} {Phys. Rev. B}\ }\textbf {\bibinfo {volume}
  {86}},\ \bibinfo {pages} {195436} (\bibinfo {year} {2012})}\BibitemShut
  {NoStop}%
\bibitem [{\citenamefont {Silvestrelli}\ and\ \citenamefont
  {Ambrosetti}(2014)}]{water@graphene_dft2}%
  \BibitemOpen
  \bibfield  {author} {\bibinfo {author} {\bibfnamefont {P.~L.}\ \bibnamefont
  {Silvestrelli}}\ and\ \bibinfo {author} {\bibfnamefont {A.}~\bibnamefont
  {Ambrosetti}},\ }\href {\doibase 10.1063/1.4869330} {\bibfield  {journal}
  {\bibinfo  {journal} {J. Chem. Phys.}\ }\textbf {\bibinfo {volume} {140}},\
  \bibinfo {pages} {124107} (\bibinfo {year} {2014})}\BibitemShut {NoStop}%
\bibitem [{\citenamefont {Ambrosetti}\ and\ \citenamefont
  {Silvestrelli}(2011)}]{water@graphene_dft2b}%
  \BibitemOpen
  \bibfield  {author} {\bibinfo {author} {\bibfnamefont {A.}~\bibnamefont
  {Ambrosetti}}\ and\ \bibinfo {author} {\bibfnamefont {P.~L.}\ \bibnamefont
  {Silvestrelli}},\ }\href {\doibase 10.1021/jp110669p} {\bibfield  {journal}
  {\bibinfo  {journal} {J. Phys. Chem. C}\ }\textbf {\bibinfo {volume} {115}},\
  \bibinfo {pages} {3695} (\bibinfo {year} {2011})}\BibitemShut {NoStop}%
\bibitem [{\citenamefont {Lorenz}\ \emph {et~al.}(2014)\citenamefont {Lorenz},
  \citenamefont {Civalleri}, \citenamefont {Maschio}, \citenamefont {Sgroi},\
  and\ \citenamefont {Pullini}}]{water@graphene_dft3}%
  \BibitemOpen
  \bibfield  {author} {\bibinfo {author} {\bibfnamefont {M.}~\bibnamefont
  {Lorenz}}, \bibinfo {author} {\bibfnamefont {B.}~\bibnamefont {Civalleri}},
  \bibinfo {author} {\bibfnamefont {L.}~\bibnamefont {Maschio}}, \bibinfo
  {author} {\bibfnamefont {M.}~\bibnamefont {Sgroi}}, \ and\ \bibinfo {author}
  {\bibfnamefont {D.}~\bibnamefont {Pullini}},\ }\href {\doibase
  10.1002/jcc.23686} {\bibfield  {journal} {\bibinfo  {journal} {J. Comput.
  Chem.}\ }\textbf {\bibinfo {volume} {35}},\ \bibinfo {pages} {1789} (\bibinfo
  {year} {2014})}\BibitemShut {NoStop}%
\bibitem [{\citenamefont {Leenaerts}, \citenamefont {Partoens},\ and\
  \citenamefont {Peeters}(2009)}]{water-graphene-pbe}%
  \BibitemOpen
  \bibfield  {author} {\bibinfo {author} {\bibfnamefont {O.}~\bibnamefont
  {Leenaerts}}, \bibinfo {author} {\bibfnamefont {B.}~\bibnamefont {Partoens}},
  \ and\ \bibinfo {author} {\bibfnamefont {F.~M.}\ \bibnamefont {Peeters}},\
  }\href {\doibase 10.1103/PhysRevB.79.235440} {\bibfield  {journal} {\bibinfo
  {journal} {Phys. Rev. B}\ }\textbf {\bibinfo {volume} {79}},\ \bibinfo
  {pages} {235440} (\bibinfo {year} {2009})}\BibitemShut {NoStop}%
\bibitem [{\citenamefont {He{\ss}elmann}(2013)}]{water-graphene-hesselmann}%
  \BibitemOpen
  \bibfield  {author} {\bibinfo {author} {\bibfnamefont {A.}~\bibnamefont
  {He{\ss}elmann}},\ }\href {\doibase 10.1021/ct300735g} {\bibfield  {journal}
  {\bibinfo  {journal} {J. Chem. Theory Comput.}\ }\textbf {\bibinfo {volume}
  {9}},\ \bibinfo {pages} {273} (\bibinfo {year} {2013})}\BibitemShut {NoStop}%
\bibitem [{\citenamefont {Rube\v{s}}\ \emph {et~al.}(2009)\citenamefont
  {Rube\v{s}}, \citenamefont {Nachtigall}, \citenamefont {Vondr\'{a}\v{s}ek},\
  and\ \citenamefont {Bludsk\'{y}}}]{water@graphene_dft/cc}%
  \BibitemOpen
  \bibfield  {author} {\bibinfo {author} {\bibfnamefont {M.}~\bibnamefont
  {Rube\v{s}}}, \bibinfo {author} {\bibfnamefont {P.}~\bibnamefont
  {Nachtigall}}, \bibinfo {author} {\bibfnamefont {J.}~\bibnamefont
  {Vondr\'{a}\v{s}ek}}, \ and\ \bibinfo {author} {\bibfnamefont
  {O.}~\bibnamefont {Bludsk\'{y}}},\ }\href {\doibase 10.1021/jp901410m}
  {\bibfield  {journal} {\bibinfo  {journal} {J. Phys. Chem. C}\ }\textbf
  {\bibinfo {volume} {113}},\ \bibinfo {pages} {8412} (\bibinfo {year}
  {2009})}\BibitemShut {NoStop}%
\bibitem [{\citenamefont {Jenness}, \citenamefont {Karalti},\ and\
  \citenamefont {Jordan}(2010)}]{water@graphene_sapt2}%
  \BibitemOpen
  \bibfield  {author} {\bibinfo {author} {\bibfnamefont {G.~R.}\ \bibnamefont
  {Jenness}}, \bibinfo {author} {\bibfnamefont {O.}~\bibnamefont {Karalti}}, \
  and\ \bibinfo {author} {\bibfnamefont {K.~D.}\ \bibnamefont {Jordan}},\
  }\href {\doibase 10.1039/C000988A} {\bibfield  {journal} {\bibinfo  {journal}
  {Phys. Chem. Chem. Phys.}\ }\textbf {\bibinfo {volume} {12}},\ \bibinfo
  {pages} {6375} (\bibinfo {year} {2010})}\BibitemShut {NoStop}%
\bibitem [{\citenamefont {Jordan}\ and\ \citenamefont
  {Heßelmann}(2019)}]{water_graphene_jordancomment}%
  \BibitemOpen
  \bibfield  {author} {\bibinfo {author} {\bibfnamefont {K.~D.}\ \bibnamefont
  {Jordan}}\ and\ \bibinfo {author} {\bibfnamefont {A.}~\bibnamefont
  {Heßelmann}},\ }\href {\doibase 10.1021/acs.jpcc.9b02326} {\bibfield
  {journal} {\bibinfo  {journal} {J. Phys. Chem. C}\ }\textbf {\bibinfo
  {volume} {123}},\ \bibinfo {pages} {10163} (\bibinfo {year}
  {2019})}\BibitemShut {NoStop}%
\bibitem [{\citenamefont {Gillan}, \citenamefont {Alfè},\ and\ \citenamefont
  {Michaelides}(2016)}]{dftforwater_perspective}%
  \BibitemOpen
  \bibfield  {author} {\bibinfo {author} {\bibfnamefont {M.~J.}\ \bibnamefont
  {Gillan}}, \bibinfo {author} {\bibfnamefont {D.}~\bibnamefont {Alfè}}, \
  and\ \bibinfo {author} {\bibfnamefont {A.}~\bibnamefont {Michaelides}},\
  }\href {\doibase 10.1063/1.4944633} {\bibfield  {journal} {\bibinfo
  {journal} {J. Chem. Phys.}\ }\textbf {\bibinfo {volume} {144}},\ \bibinfo
  {pages} {130901} (\bibinfo {year} {2016})}\BibitemShut {NoStop}%
\bibitem [{sup()}]{suppinfo}%
  \BibitemOpen
  \href@noop {} {}\bibinfo {note} {See supplemental material for geometries and
  reference energies of all considered structures of the WaC18 benchmark set,
  including water@graphene, water@CNT, water@AH, 2D-ice, and 3D-ice. Geometries
  are provided in {\sc Vasp} POSCAR and XYZ format.}\BibitemShut {Stop}%
\bibitem [{geo()}]{geoms}%
  \BibitemOpen
  \href@noop {} {}\bibinfo {note} {Note that due to different geometry
  optimization strategies, the water structures are slightly different and it
  is important to use the geometries as provided in the supporting
  information.\cite{suppinfo}}\BibitemShut {NoStop}%
\bibitem [{\citenamefont {Al-Hamdani}, \citenamefont {Alf\`{e}},\ and\
  \citenamefont {Michaelides}(2017{\natexlab{b}})}]{water-cnt-dmc}%
  \BibitemOpen
  \bibfield  {author} {\bibinfo {author} {\bibfnamefont {Y.~S.}\ \bibnamefont
  {Al-Hamdani}}, \bibinfo {author} {\bibfnamefont {D.}~\bibnamefont
  {Alf\`{e}}}, \ and\ \bibinfo {author} {\bibfnamefont {A.}~\bibnamefont
  {Michaelides}},\ }\href {\doibase 10.1063/1.4977180} {\bibfield  {journal}
  {\bibinfo  {journal} {J. Chem. Phys.}\ }\textbf {\bibinfo {volume} {146}},\
  \bibinfo {pages} {094701} (\bibinfo {year} {2017}{\natexlab{b}})}\BibitemShut
  {NoStop}%
\bibitem [{\citenamefont {Dubeck{\'y}}, \citenamefont {Mitas},\ and\
  \citenamefont {Jurecka}(2016)}]{Dubecky_chemrev_2016}%
  \BibitemOpen
  \bibfield  {author} {\bibinfo {author} {\bibfnamefont {M.}~\bibnamefont
  {Dubeck{\'y}}}, \bibinfo {author} {\bibfnamefont {L.}~\bibnamefont {Mitas}},
  \ and\ \bibinfo {author} {\bibfnamefont {P.}~\bibnamefont {Jurecka}},\
  }\href@noop {} {\bibfield  {journal} {\bibinfo  {journal} {Chem. Rev.}\
  }\textbf {\bibinfo {volume} {116}},\ \bibinfo {pages} {5188} (\bibinfo {year}
  {2016})}\BibitemShut {NoStop}%
\bibitem [{\citenamefont {Al-Hamdani}\ and\ \citenamefont
  {Tkatchenko}(2019{\natexlab{b}})}]{AlHamdani_JCP-perspective_2019}%
  \BibitemOpen
  \bibfield  {author} {\bibinfo {author} {\bibfnamefont {Y.~S.}\ \bibnamefont
  {Al-Hamdani}}\ and\ \bibinfo {author} {\bibfnamefont {A.}~\bibnamefont
  {Tkatchenko}},\ }\href@noop {} {\bibfield  {journal} {\bibinfo  {journal} {J.
  Chem. Phys.}\ }\textbf {\bibinfo {volume} {150}},\ \bibinfo {pages} {010901}
  (\bibinfo {year} {2019}{\natexlab{b}})}\BibitemShut {NoStop}%
\bibitem [{\citenamefont {Feller}(1999)}]{feller-wbz}%
  \BibitemOpen
  \bibfield  {author} {\bibinfo {author} {\bibfnamefont {D.}~\bibnamefont
  {Feller}},\ }\href {\doibase 10.1021/jp991932w} {\bibfield  {journal}
  {\bibinfo  {journal} {J. Phys. Chem. A}\ }\textbf {\bibinfo {volume} {103}},\
  \bibinfo {pages} {7558} (\bibinfo {year} {1999})}\BibitemShut {NoStop}%
\bibitem [{\citenamefont {Slipchenko}\ and\ \citenamefont
  {Gordon}(2009)}]{slipchenko-wbz}%
  \BibitemOpen
  \bibfield  {author} {\bibinfo {author} {\bibfnamefont {L.~V.}\ \bibnamefont
  {Slipchenko}}\ and\ \bibinfo {author} {\bibfnamefont {M.~S.}\ \bibnamefont
  {Gordon}},\ }\href {\doibase 10.1021/jp808845b} {\bibfield  {journal}
  {\bibinfo  {journal} {J. Phys. Chem. A}\ }\textbf {\bibinfo {volume} {113}},\
  \bibinfo {pages} {2092} (\bibinfo {year} {2009})}\BibitemShut {NoStop}%
\bibitem [{\citenamefont {Needs}\ \emph {et~al.}(2010)\citenamefont {Needs},
  \citenamefont {Towler}, \citenamefont {Drummond},\ and\ \citenamefont
  {Rios}}]{casino}%
  \BibitemOpen
  \bibfield  {author} {\bibinfo {author} {\bibfnamefont {R.~J.}\ \bibnamefont
  {Needs}}, \bibinfo {author} {\bibfnamefont {M.~D.}\ \bibnamefont {Towler}},
  \bibinfo {author} {\bibfnamefont {N.~D.}\ \bibnamefont {Drummond}}, \ and\
  \bibinfo {author} {\bibfnamefont {P.~L.}\ \bibnamefont {Rios}},\ }\href@noop
  {} {\bibfield  {journal} {\bibinfo  {journal} {J. Phys.: Condens. Matter}\
  }\textbf {\bibinfo {volume} {22}},\ \bibinfo {pages} {023201} (\bibinfo
  {year} {2010})}\BibitemShut {NoStop}%
\bibitem [{\citenamefont {Trail}\ and\ \citenamefont
  {Needs}(2005{\natexlab{a}})}]{trail05_NCHF}%
  \BibitemOpen
  \bibfield  {author} {\bibinfo {author} {\bibfnamefont {J.~R.}\ \bibnamefont
  {Trail}}\ and\ \bibinfo {author} {\bibfnamefont {R.~J.}\ \bibnamefont
  {Needs}},\ }\href@noop {} {\bibfield  {journal} {\bibinfo  {journal} {J.
  Chem. Phys.}\ }\textbf {\bibinfo {volume} {122}},\ \bibinfo {pages} {014112}
  (\bibinfo {year} {2005}{\natexlab{a}})}\BibitemShut {NoStop}%
\bibitem [{\citenamefont {Trail}\ and\ \citenamefont
  {Needs}(2005{\natexlab{b}})}]{trail05_SRHF}%
  \BibitemOpen
  \bibfield  {author} {\bibinfo {author} {\bibfnamefont {J.~R.}\ \bibnamefont
  {Trail}}\ and\ \bibinfo {author} {\bibfnamefont {R.~J.}\ \bibnamefont
  {Needs}},\ }\href@noop {} {\bibfield  {journal} {\bibinfo  {journal} {J.
  Chem. Phys.}\ }\textbf {\bibinfo {volume} {122}},\ \bibinfo {pages} {174109}
  (\bibinfo {year} {2005}{\natexlab{b}})}\BibitemShut {NoStop}%
\bibitem [{\citenamefont {Mitas}, \citenamefont {Shirley},\ and\ \citenamefont
  {Ceperley}(1991)}]{mitas91}%
  \BibitemOpen
  \bibfield  {author} {\bibinfo {author} {\bibfnamefont {L.}~\bibnamefont
  {Mitas}}, \bibinfo {author} {\bibfnamefont {E.~L.}\ \bibnamefont {Shirley}},
  \ and\ \bibinfo {author} {\bibfnamefont {D.~M.}\ \bibnamefont {Ceperley}},\
  }\href@noop {} {\bibfield  {journal} {\bibinfo  {journal} {J. Chem. Phys.}\
  }\textbf {\bibinfo {volume} {95}},\ \bibinfo {pages} {3467} (\bibinfo {year}
  {1991})}\BibitemShut {NoStop}%
\bibitem [{\citenamefont {Giannozzi}\ \emph {et~al.}(2009)\citenamefont
  {Giannozzi} \emph {et~al.}}]{espresso}%
  \BibitemOpen
  \bibfield  {author} {\bibinfo {author} {\bibfnamefont {P.}~\bibnamefont
  {Giannozzi}} \emph {et~al.},\ }\href@noop {} {\bibfield  {journal} {\bibinfo
  {journal} {J. Phys.: Condens. Matter}\ }\textbf {\bibinfo {volume} {21}},\
  \bibinfo {pages} {395502 (19pp)} (\bibinfo {year} {2009})}\BibitemShut
  {NoStop}%
\bibitem [{\citenamefont {Alf\`{e}}\ and\ \citenamefont
  {Gillan}(2004)}]{alfe04}%
  \BibitemOpen
  \bibfield  {author} {\bibinfo {author} {\bibfnamefont {D.}~\bibnamefont
  {Alf\`{e}}}\ and\ \bibinfo {author} {\bibfnamefont {M.~J.}\ \bibnamefont
  {Gillan}},\ }\href@noop {} {\bibfield  {journal} {\bibinfo  {journal} {Phys.
  Rev. B}\ }\textbf {\bibinfo {volume} {70}},\ \bibinfo {pages} {161101}
  (\bibinfo {year} {2004})}\BibitemShut {NoStop}%
\bibitem [{\citenamefont {Fraser}\ \emph {et~al.}(1996)\citenamefont {Fraser},
  \citenamefont {Foulkes}, \citenamefont {Rajagopal}, \citenamefont {Needs},
  \citenamefont {Kenny},\ and\ \citenamefont {Williamson}}]{MPC:Fraser1996}%
  \BibitemOpen
  \bibfield  {author} {\bibinfo {author} {\bibfnamefont {L.~M.}\ \bibnamefont
  {Fraser}}, \bibinfo {author} {\bibfnamefont {W.~M.~C.}\ \bibnamefont
  {Foulkes}}, \bibinfo {author} {\bibfnamefont {G.}~\bibnamefont {Rajagopal}},
  \bibinfo {author} {\bibfnamefont {R.~J.}\ \bibnamefont {Needs}}, \bibinfo
  {author} {\bibfnamefont {S.~D.}\ \bibnamefont {Kenny}}, \ and\ \bibinfo
  {author} {\bibfnamefont {A.~J.}\ \bibnamefont {Williamson}},\ }\href@noop {}
  {\bibfield  {journal} {\bibinfo  {journal} {Phys. Rev. B}\ }\textbf {\bibinfo
  {volume} {53}},\ \bibinfo {pages} {1814} (\bibinfo {year}
  {1996})}\BibitemShut {NoStop}%
\bibitem [{\citenamefont {Williamson}\ \emph {et~al.}(1997)\citenamefont
  {Williamson}, \citenamefont {Rajagopal}, \citenamefont {Needs}, \citenamefont
  {Fraser}, \citenamefont {Foulkes}, \citenamefont {Wang},\ and\ \citenamefont
  {Chou}}]{MPC:Will1997}%
  \BibitemOpen
  \bibfield  {author} {\bibinfo {author} {\bibfnamefont {A.~J.}\ \bibnamefont
  {Williamson}}, \bibinfo {author} {\bibfnamefont {G.}~\bibnamefont
  {Rajagopal}}, \bibinfo {author} {\bibfnamefont {R.~J.}\ \bibnamefont
  {Needs}}, \bibinfo {author} {\bibfnamefont {L.~M.}\ \bibnamefont {Fraser}},
  \bibinfo {author} {\bibfnamefont {W.~M.~C.}\ \bibnamefont {Foulkes}},
  \bibinfo {author} {\bibfnamefont {Y.}~\bibnamefont {Wang}}, \ and\ \bibinfo
  {author} {\bibfnamefont {M.-Y.}\ \bibnamefont {Chou}},\ }\href@noop {}
  {\bibfield  {journal} {\bibinfo  {journal} {Phys. Rev. B}\ }\textbf {\bibinfo
  {volume} {55}},\ \bibinfo {pages} {R4851} (\bibinfo {year}
  {1997})}\BibitemShut {NoStop}%
\bibitem [{\citenamefont {Kent}\ \emph {et~al.}(1999)\citenamefont {Kent},
  \citenamefont {Hood}, \citenamefont {Williamson}, \citenamefont {Needs},
  \citenamefont {Foulkes},\ and\ \citenamefont {Rajagopal}}]{MPC:Kent1999}%
  \BibitemOpen
  \bibfield  {author} {\bibinfo {author} {\bibfnamefont {P.~R.~C.}\
  \bibnamefont {Kent}}, \bibinfo {author} {\bibfnamefont {R.~Q.}\ \bibnamefont
  {Hood}}, \bibinfo {author} {\bibfnamefont {A.~J.}\ \bibnamefont
  {Williamson}}, \bibinfo {author} {\bibfnamefont {R.~J.}\ \bibnamefont
  {Needs}}, \bibinfo {author} {\bibfnamefont {W.~M.~C.}\ \bibnamefont
  {Foulkes}}, \ and\ \bibinfo {author} {\bibfnamefont {G.}~\bibnamefont
  {Rajagopal}},\ }\href@noop {} {\bibfield  {journal} {\bibinfo  {journal}
  {Phys. Rev. B}\ }\textbf {\bibinfo {volume} {59}},\ \bibinfo {pages} {1917}
  (\bibinfo {year} {1999})}\BibitemShut {NoStop}%
\bibitem [{\citenamefont {Kwee}, \citenamefont {Zhang},\ and\ \citenamefont
  {Krakauer}(2008)}]{KZK:prl2008}%
  \BibitemOpen
  \bibfield  {author} {\bibinfo {author} {\bibfnamefont {H.}~\bibnamefont
  {Kwee}}, \bibinfo {author} {\bibfnamefont {S.}~\bibnamefont {Zhang}}, \ and\
  \bibinfo {author} {\bibfnamefont {H.}~\bibnamefont {Krakauer}},\ }\href@noop
  {} {\bibfield  {journal} {\bibinfo  {journal} {Phys. Rev. Lett.}\ }\textbf
  {\bibinfo {volume} {100}},\ \bibinfo {pages} {126404} (\bibinfo {year}
  {2008})}\BibitemShut {NoStop}%
\bibitem [{\citenamefont {Zen}\ \emph {et~al.}(2019)\citenamefont {Zen},
  \citenamefont {Brandenburg}, \citenamefont {Michaelides},\ and\ \citenamefont
  {Alf\`e}}]{zen_DMCwDLA}%
  \BibitemOpen
  \bibfield  {author} {\bibinfo {author} {\bibfnamefont {A.}~\bibnamefont
  {Zen}}, \bibinfo {author} {\bibfnamefont {J.~G.}\ \bibnamefont
  {Brandenburg}}, \bibinfo {author} {\bibfnamefont {A.}~\bibnamefont
  {Michaelides}}, \ and\ \bibinfo {author} {\bibfnamefont {D.}~\bibnamefont
  {Alf\`e}},\ }\href@noop {} {\bibfield  {journal} {\bibinfo  {journal}
  {arXiv:1907.04432 [physics.comp-ph]}\ } (\bibinfo {year} {2019})}\BibitemShut
  {NoStop}%
\bibitem [{\citenamefont {Trail}\ and\ \citenamefont {Needs}(2017)}]{eCEPP}%
  \BibitemOpen
  \bibfield  {author} {\bibinfo {author} {\bibfnamefont {J.~R.}\ \bibnamefont
  {Trail}}\ and\ \bibinfo {author} {\bibfnamefont {R.~J.}\ \bibnamefont
  {Needs}},\ }\href@noop {} {\bibfield  {journal} {\bibinfo  {journal} {J.
  Chem. Phys.}\ }\textbf {\bibinfo {volume} {146}},\ \bibinfo {pages} {204107}
  (\bibinfo {year} {2017})}\BibitemShut {NoStop}%
\bibitem [{\citenamefont {Perdew}, \citenamefont {Burke},\ and\ \citenamefont
  {Ernzerhof}(1996)}]{pbe}%
  \BibitemOpen
  \bibfield  {author} {\bibinfo {author} {\bibfnamefont {J.~P.}\ \bibnamefont
  {Perdew}}, \bibinfo {author} {\bibfnamefont {K.}~\bibnamefont {Burke}}, \
  and\ \bibinfo {author} {\bibfnamefont {M.}~\bibnamefont {Ernzerhof}},\
  }\href@noop {} {\bibfield  {journal} {\bibinfo  {journal} {Phys. Rev. Lett.}\
  }\textbf {\bibinfo {volume} {77}},\ \bibinfo {pages} {3865} (\bibinfo {year}
  {1996})},\ \bibinfo {note} {erratum {\it Phys. Rev. Lett.} {\bf 78}, 1396
  (1997)}\BibitemShut {NoStop}%
\bibitem [{\citenamefont {Neese}(2012)}]{orca}%
  \BibitemOpen
  \bibfield  {author} {\bibinfo {author} {\bibfnamefont {F.}~\bibnamefont
  {Neese}},\ }\href@noop {} {\bibfield  {journal} {\bibinfo  {journal} {WIREs
  Comput. Mol. Sci.}\ }\textbf {\bibinfo {volume} {2}},\ \bibinfo {pages} {73}
  (\bibinfo {year} {2012})}\BibitemShut {NoStop}%
\bibitem [{\citenamefont {Dunning}(1989)}]{dunning}%
  \BibitemOpen
  \bibfield  {author} {\bibinfo {author} {\bibfnamefont {T.~H.}\ \bibnamefont
  {Dunning}, \bibfnamefont {Jr.}},\ }\href@noop {} {\bibfield  {journal}
  {\bibinfo  {journal} {J. Chem. Phys.}\ }\textbf {\bibinfo {volume} {90}},\
  \bibinfo {pages} {1007} (\bibinfo {year} {1989})}\BibitemShut {NoStop}%
\bibitem [{\citenamefont {Kendall}, \citenamefont {Dunning},\ and\
  \citenamefont {Harrison}(1992)}]{augdunning}%
  \BibitemOpen
  \bibfield  {author} {\bibinfo {author} {\bibfnamefont {R.~A.}\ \bibnamefont
  {Kendall}}, \bibinfo {author} {\bibfnamefont {T.~H.}\ \bibnamefont {Dunning},
  \bibfnamefont {Jr.}}, \ and\ \bibinfo {author} {\bibfnamefont {R.~J.}\
  \bibnamefont {Harrison}},\ }\href@noop {} {\bibfield  {journal} {\bibinfo
  {journal} {J. Chem. Phys.}\ }\textbf {\bibinfo {volume} {96}},\ \bibinfo
  {pages} {6796} (\bibinfo {year} {1992})}\BibitemShut {NoStop}%
\bibitem [{\citenamefont {Weigend}, \citenamefont {Furche},\ and\ \citenamefont
  {Ahlrichs}(2003)}]{qzvp}%
  \BibitemOpen
  \bibfield  {author} {\bibinfo {author} {\bibfnamefont {F.}~\bibnamefont
  {Weigend}}, \bibinfo {author} {\bibfnamefont {F.}~\bibnamefont {Furche}}, \
  and\ \bibinfo {author} {\bibfnamefont {R.}~\bibnamefont {Ahlrichs}},\
  }\href@noop {} {\bibfield  {journal} {\bibinfo  {journal} {J. Chem. Phys.}\
  }\textbf {\bibinfo {volume} {119}},\ \bibinfo {pages} {12753} (\bibinfo
  {year} {2003})}\BibitemShut {NoStop}%
\bibitem [{\citenamefont {Erba}\ \emph {et~al.}(2017)\citenamefont {Erba},
  \citenamefont {Baima}, \citenamefont {Bush}, \citenamefont {Orlando},\ and\
  \citenamefont {Dovesi}}]{crystal17}%
  \BibitemOpen
  \bibfield  {author} {\bibinfo {author} {\bibfnamefont {A.}~\bibnamefont
  {Erba}}, \bibinfo {author} {\bibfnamefont {J.}~\bibnamefont {Baima}},
  \bibinfo {author} {\bibfnamefont {I.}~\bibnamefont {Bush}}, \bibinfo {author}
  {\bibfnamefont {R.}~\bibnamefont {Orlando}}, \ and\ \bibinfo {author}
  {\bibfnamefont {R.}~\bibnamefont {Dovesi}},\ }\href {\doibase
  10.1021/acs.jctc.7b00687} {\bibfield  {journal} {\bibinfo  {journal} {J.
  Chem. Theory Comput.}\ }\textbf {\bibinfo {volume} {13}},\ \bibinfo {pages}
  {5019} (\bibinfo {year} {2017})}\BibitemShut {NoStop}%
\bibitem [{\citenamefont {Dovesi}\ \emph {et~al.}(2018)\citenamefont {Dovesi},
  \citenamefont {Erba}, \citenamefont {Orlando}, \citenamefont
  {Zicovich-Wilson}, \citenamefont {Civalleri}, \citenamefont {Maschio},
  \citenamefont {R\'{e}rat}, \citenamefont {Casassa}, \citenamefont {Baima},
  \citenamefont {Salustro},\ and\ \citenamefont {Kirtman}}]{crystal17_wire}%
  \BibitemOpen
  \bibfield  {author} {\bibinfo {author} {\bibfnamefont {R.}~\bibnamefont
  {Dovesi}}, \bibinfo {author} {\bibfnamefont {A.}~\bibnamefont {Erba}},
  \bibinfo {author} {\bibfnamefont {R.}~\bibnamefont {Orlando}}, \bibinfo
  {author} {\bibfnamefont {C.~M.}\ \bibnamefont {Zicovich-Wilson}}, \bibinfo
  {author} {\bibfnamefont {B.}~\bibnamefont {Civalleri}}, \bibinfo {author}
  {\bibfnamefont {L.}~\bibnamefont {Maschio}}, \bibinfo {author} {\bibfnamefont
  {M.}~\bibnamefont {R\'{e}rat}}, \bibinfo {author} {\bibfnamefont
  {S.}~\bibnamefont {Casassa}}, \bibinfo {author} {\bibfnamefont
  {J.}~\bibnamefont {Baima}}, \bibinfo {author} {\bibfnamefont
  {S.}~\bibnamefont {Salustro}}, \ and\ \bibinfo {author} {\bibfnamefont
  {B.}~\bibnamefont {Kirtman}},\ }\href@noop {} {\bibfield  {journal} {\bibinfo
   {journal} {WIREs Comput. Mol. Sci.}\ }\textbf {\bibinfo {volume} {8}},\
  \bibinfo {pages} {e1360} (\bibinfo {year} {2018})}\BibitemShut {NoStop}%
\bibitem [{\citenamefont {Kresse}\ and\ \citenamefont
  {Furthm\"uller}(1996{\natexlab{a}})}]{vasp2}%
  \BibitemOpen
  \bibfield  {author} {\bibinfo {author} {\bibfnamefont {G.}~\bibnamefont
  {Kresse}}\ and\ \bibinfo {author} {\bibfnamefont {J.}~\bibnamefont
  {Furthm\"uller}},\ }\href@noop {} {\bibfield  {journal} {\bibinfo  {journal}
  {J. Comp. Mat. Sci.}\ }\textbf {\bibinfo {volume} {6}},\ \bibinfo {pages}
  {15} (\bibinfo {year} {1996}{\natexlab{a}})}\BibitemShut {NoStop}%
\bibitem [{\citenamefont {Kresse}\ and\ \citenamefont
  {Furthm\"uller}(1996{\natexlab{b}})}]{vasp3}%
  \BibitemOpen
  \bibfield  {author} {\bibinfo {author} {\bibfnamefont {G.}~\bibnamefont
  {Kresse}}\ and\ \bibinfo {author} {\bibfnamefont {J.}~\bibnamefont
  {Furthm\"uller}},\ }\href@noop {} {\bibfield  {journal} {\bibinfo  {journal}
  {Phys. Rev. B}\ }\textbf {\bibinfo {volume} {54}},\ \bibinfo {pages} {11169}
  (\bibinfo {year} {1996}{\natexlab{b}})}\BibitemShut {NoStop}%
\bibitem [{\citenamefont {Bl\"ochl}(1994)}]{paw1}%
  \BibitemOpen
  \bibfield  {author} {\bibinfo {author} {\bibfnamefont {P.~E.}\ \bibnamefont
  {Bl\"ochl}},\ }\href {\doibase 10.1103/PhysRevB.50.17953} {\bibfield
  {journal} {\bibinfo  {journal} {Phys. Rev. B}\ }\textbf {\bibinfo {volume}
  {50}},\ \bibinfo {pages} {17953} (\bibinfo {year} {1994})}\BibitemShut
  {NoStop}%
\bibitem [{\citenamefont {Kresse}\ and\ \citenamefont {Joubert}(1999)}]{paw2}%
  \BibitemOpen
  \bibfield  {author} {\bibinfo {author} {\bibfnamefont {G.}~\bibnamefont
  {Kresse}}\ and\ \bibinfo {author} {\bibfnamefont {J.}~\bibnamefont
  {Joubert}},\ }\href {\doibase 10.1103/PhysRevB.59.1758} {\bibfield  {journal}
  {\bibinfo  {journal} {Phys. Rev. B}\ }\textbf {\bibinfo {volume} {59}},\
  \bibinfo {pages} {1758} (\bibinfo {year} {1999})}\BibitemShut {NoStop}%
\bibitem [{paw({\natexlab{a}})}]{pawconvergence}%
  \BibitemOpen
  \href@noop {} {}\bibinfo {note} {In
  addition to the reported PBE calculations, we tested the PAW convergence for
  TPSS and rev-vdW-DF2 and the meta-GGA convergence is a bit slower. Still,
  comparison with PW cutoff of 1500 eV showed that the employed production run
  settings are converged within 2\,meV.}\BibitemShut {Stop}%
\bibitem [{paw({\natexlab{b}})}]{pawimpact}%
  \BibitemOpen
  \href@noop {} {}\bibinfo {note} {As
  common practice, PBE based PAW potentials have been used for all GGA,
  meta-GGA, vdW-DF and hybrid DF calculations. Several studies addressed
  possible discrepancies w.r.t. all electron
  calculations.\cite{scan_paw_numerics,pbe0_pp_numerics,vdwdf_pp_numerics} To
  test its impact for weakly bonded systems, we computed the PBE lattice energy
  for ice Ih with LDA and PBE based PAWs yielding a very small difference of
  1\,meV.}\BibitemShut {Stop}%
\bibitem [{\citenamefont {Neese}, \citenamefont {Hansen},\ and\ \citenamefont
  {Liakos}(2009)}]{extrapolation}%
  \BibitemOpen
  \bibfield  {author} {\bibinfo {author} {\bibfnamefont {F.}~\bibnamefont
  {Neese}}, \bibinfo {author} {\bibfnamefont {A.}~\bibnamefont {Hansen}}, \
  and\ \bibinfo {author} {\bibfnamefont {D.~G.}\ \bibnamefont {Liakos}},\
  }\href@noop {} {\bibfield  {journal} {\bibinfo  {journal} {J. Chem. Phys.}\
  }\textbf {\bibinfo {volume} {131}},\ \bibinfo {pages} {064103} (\bibinfo
  {year} {2009})}\BibitemShut {NoStop}%
\bibitem [{\citenamefont {Grimme}(2006)}]{dftd2}%
  \BibitemOpen
  \bibfield  {author} {\bibinfo {author} {\bibfnamefont {S.}~\bibnamefont
  {Grimme}},\ }\href@noop {} {\bibfield  {journal} {\bibinfo  {journal} {{J.
  Comput. Chem.}}\ }\textbf {\bibinfo {volume} {27}},\ \bibinfo {pages} {1787}
  (\bibinfo {year} {2006})}\BibitemShut {NoStop}%
\bibitem [{\citenamefont {Grimme}\ \emph {et~al.}(2010)\citenamefont {Grimme},
  \citenamefont {Antony}, \citenamefont {Ehrlich},\ and\ \citenamefont
  {Krieg}}]{dftd3}%
  \BibitemOpen
  \bibfield  {author} {\bibinfo {author} {\bibfnamefont {S.}~\bibnamefont
  {Grimme}}, \bibinfo {author} {\bibfnamefont {J.}~\bibnamefont {Antony}},
  \bibinfo {author} {\bibfnamefont {S.}~\bibnamefont {Ehrlich}}, \ and\
  \bibinfo {author} {\bibfnamefont {H.}~\bibnamefont {Krieg}},\ }\href
  {\doibase 10.1063/1.3382344} {\bibfield  {journal} {\bibinfo  {journal} {J.
  Chem. Phys.}\ }\textbf {\bibinfo {volume} {132}},\ \bibinfo {pages} {154104}
  (\bibinfo {year} {2010})}\BibitemShut {NoStop}%
\bibitem [{\citenamefont {Grimme}, \citenamefont {Ehrlich},\ and\ \citenamefont
  {Goerigk}(2011)}]{dftd3bj}%
  \BibitemOpen
  \bibfield  {author} {\bibinfo {author} {\bibfnamefont {S.}~\bibnamefont
  {Grimme}}, \bibinfo {author} {\bibfnamefont {S.}~\bibnamefont {Ehrlich}}, \
  and\ \bibinfo {author} {\bibfnamefont {L.}~\bibnamefont {Goerigk}},\ }\href
  {\doibase 10.1002/jcc.21759} {\bibfield  {journal} {\bibinfo  {journal} {J.
  Comput. Chem.}\ }\textbf {\bibinfo {volume} {32}},\ \bibinfo {pages} {1456}
  (\bibinfo {year} {2011})}\BibitemShut {NoStop}%
\bibitem [{\citenamefont {Caldeweyher}, \citenamefont {Bannwarth},\ and\
  \citenamefont {Grimme}(2017)}]{dftd3.5}%
  \BibitemOpen
  \bibfield  {author} {\bibinfo {author} {\bibfnamefont {E.}~\bibnamefont
  {Caldeweyher}}, \bibinfo {author} {\bibfnamefont {C.}~\bibnamefont
  {Bannwarth}}, \ and\ \bibinfo {author} {\bibfnamefont {S.}~\bibnamefont
  {Grimme}},\ }\href {\doibase 10.1063/1.4993215} {\bibfield  {journal}
  {\bibinfo  {journal} {J. Chem. Phys.}\ }\textbf {\bibinfo {volume} {147}},\
  \bibinfo {pages} {034112} (\bibinfo {year} {2017})}\BibitemShut {NoStop}%
\bibitem [{\citenamefont {Caldeweyher}\ \emph {et~al.}(2019)\citenamefont
  {Caldeweyher}, \citenamefont {Ehlert}, \citenamefont {Hansen}, \citenamefont
  {Neugebauer}, \citenamefont {Spicher}, \citenamefont {Bannwarth},\ and\
  \citenamefont {Grimme}}]{dftd4}%
  \BibitemOpen
  \bibfield  {author} {\bibinfo {author} {\bibfnamefont {E.}~\bibnamefont
  {Caldeweyher}}, \bibinfo {author} {\bibfnamefont {S.}~\bibnamefont {Ehlert}},
  \bibinfo {author} {\bibfnamefont {A.}~\bibnamefont {Hansen}}, \bibinfo
  {author} {\bibfnamefont {H.}~\bibnamefont {Neugebauer}}, \bibinfo {author}
  {\bibfnamefont {S.}~\bibnamefont {Spicher}}, \bibinfo {author} {\bibfnamefont
  {C.}~\bibnamefont {Bannwarth}}, \ and\ \bibinfo {author} {\bibfnamefont
  {S.}~\bibnamefont {Grimme}},\ }\href {\doibase 10.26434/chemrxiv.7430216.v2}
  {\  (\bibinfo {year} {2019}),\ 10.26434/chemrxiv.7430216.v2},\ \bibinfo
  {note} {preprint: chemrxiv.7430216.v2}\BibitemShut {NoStop}%
\bibitem [{\citenamefont {Tkatchenko}\ and\ \citenamefont
  {Scheffler}(2009)}]{ts}%
  \BibitemOpen
  \bibfield  {author} {\bibinfo {author} {\bibfnamefont {A.}~\bibnamefont
  {Tkatchenko}}\ and\ \bibinfo {author} {\bibfnamefont {M.}~\bibnamefont
  {Scheffler}},\ }\href {\doibase 10.1103/PhysRevLett.102.073005} {\bibfield
  {journal} {\bibinfo  {journal} {Phys. Rev. Lett.}\ }\textbf {\bibinfo
  {volume} {102}},\ \bibinfo {pages} {073005} (\bibinfo {year}
  {2009})}\BibitemShut {NoStop}%
\bibitem [{\citenamefont {Tkatchenko}\ \emph {et~al.}(2012)\citenamefont
  {Tkatchenko}, \citenamefont {DiStasio.}, \citenamefont {Car},\ and\
  \citenamefont {Scheffler}}]{ts-mbd}%
  \BibitemOpen
  \bibfield  {author} {\bibinfo {author} {\bibfnamefont {A.}~\bibnamefont
  {Tkatchenko}}, \bibinfo {author} {\bibfnamefont {R.~A.}\ \bibnamefont
  {DiStasio.}}, \bibinfo {author} {\bibfnamefont {R.}~\bibnamefont {Car}}, \
  and\ \bibinfo {author} {\bibfnamefont {M.}~\bibnamefont {Scheffler}},\ }\href
  {\doibase 10.1103/PhysRevLett.108.236402} {\bibfield  {journal} {\bibinfo
  {journal} {Phys. Rev. Lett.}\ }\textbf {\bibinfo {volume} {108}},\ \bibinfo
  {pages} {236402} (\bibinfo {year} {2012})}\BibitemShut {NoStop}%
\bibitem [{\citenamefont {Vydrov}\ and\ \citenamefont {{Van
  Voorhis}}(2010)}]{vv10}%
  \BibitemOpen
  \bibfield  {author} {\bibinfo {author} {\bibfnamefont {O.~A.}\ \bibnamefont
  {Vydrov}}\ and\ \bibinfo {author} {\bibfnamefont {T.}~\bibnamefont {{Van
  Voorhis}}},\ }\href@noop {} {\bibfield  {journal} {\bibinfo  {journal} {J.
  Chem. Phys.}\ }\textbf {\bibinfo {volume} {133}},\ \bibinfo {pages} {244103}
  (\bibinfo {year} {2010})}\BibitemShut {NoStop}%
\bibitem [{\citenamefont {Steinmann}\ and\ \citenamefont
  {Corminboeuf}(2010)}]{ddsc}%
  \BibitemOpen
  \bibfield  {author} {\bibinfo {author} {\bibfnamefont {S.~N.}\ \bibnamefont
  {Steinmann}}\ and\ \bibinfo {author} {\bibfnamefont {C.~A.}\ \bibnamefont
  {Corminboeuf}},\ }\href@noop {} {\bibfield  {journal} {\bibinfo  {journal}
  {{J. Chem. Theory Comput.}}\ }\textbf {\bibinfo {volume} {6}},\ \bibinfo
  {pages} {1990} (\bibinfo {year} {2010})}\BibitemShut {NoStop}%
\bibitem [{\citenamefont {Dion}\ \emph
  {et~al.}(2004{\natexlab{a}})\citenamefont {Dion}, \citenamefont {Rydberg},
  \citenamefont {Schr{\"o}der}, \citenamefont {Langreth},\ and\ \citenamefont
  {Lundqvist}}]{vdwdf}%
  \BibitemOpen
  \bibfield  {author} {\bibinfo {author} {\bibfnamefont {M.}~\bibnamefont
  {Dion}}, \bibinfo {author} {\bibfnamefont {H.}~\bibnamefont {Rydberg}},
  \bibinfo {author} {\bibfnamefont {E.}~\bibnamefont {Schr{\"o}der}}, \bibinfo
  {author} {\bibfnamefont {D.~C.}\ \bibnamefont {Langreth}}, \ and\ \bibinfo
  {author} {\bibfnamefont {B.~I.}\ \bibnamefont {Lundqvist}},\ }\href {\doibase
  10.1103/PhysRevLett.92.246401} {\bibfield  {journal} {\bibinfo  {journal}
  {Phys. Rev. Lett.}\ }\textbf {\bibinfo {volume} {92}},\ \bibinfo {pages}
  {246401} (\bibinfo {year} {2004}{\natexlab{a}})}\BibitemShut {NoStop}%
\bibitem [{\citenamefont {Klime\v{s}}\ and\ \citenamefont
  {Michaelides}(2012)}]{vdw_perspective}%
  \BibitemOpen
  \bibfield  {author} {\bibinfo {author} {\bibfnamefont {J.}~\bibnamefont
  {Klime\v{s}}}\ and\ \bibinfo {author} {\bibfnamefont {A.}~\bibnamefont
  {Michaelides}},\ }\href {\doibase http://dx.doi.org/10.1063/1.4754130}
  {\bibfield  {journal} {\bibinfo  {journal} {J. Chem. Phys.}\ }\textbf
  {\bibinfo {volume} {137}},\ \bibinfo {pages} {120901} (\bibinfo {year}
  {2012})}\BibitemShut {NoStop}%
\bibitem [{\citenamefont {Grimme}\ \emph {et~al.}(2016)\citenamefont {Grimme},
  \citenamefont {Hansen}, \citenamefont {Brandenburg},\ and\ \citenamefont
  {Bannwarth}}]{disp_chemrev}%
  \BibitemOpen
  \bibfield  {author} {\bibinfo {author} {\bibfnamefont {S.}~\bibnamefont
  {Grimme}}, \bibinfo {author} {\bibfnamefont {A.}~\bibnamefont {Hansen}},
  \bibinfo {author} {\bibfnamefont {J.~G.}\ \bibnamefont {Brandenburg}}, \ and\
  \bibinfo {author} {\bibfnamefont {C.}~\bibnamefont {Bannwarth}},\ }\href
  {\doibase 10.1021/acs.chemrev.5b00533} {\bibfield  {journal} {\bibinfo
  {journal} {{Chem. Rev.}}\ }\textbf {\bibinfo {volume} {116}},\ \bibinfo
  {pages} {5105} (\bibinfo {year} {2016})}\BibitemShut {NoStop}%
\bibitem [{\citenamefont {Hermann}, \citenamefont {DiStasio},\ and\
  \citenamefont {Tkatchenko}(2017)}]{chemrev_tkatchenko}%
  \BibitemOpen
  \bibfield  {author} {\bibinfo {author} {\bibfnamefont {J.}~\bibnamefont
  {Hermann}}, \bibinfo {author} {\bibfnamefont {R.~A.}\ \bibnamefont
  {DiStasio}}, \ and\ \bibinfo {author} {\bibfnamefont {A.}~\bibnamefont
  {Tkatchenko}},\ }\href {\doibase 10.1021/acs.chemrev.6b00446} {\bibfield
  {journal} {\bibinfo  {journal} {Chem. Rev.}\ }\textbf {\bibinfo {volume}
  {117}},\ \bibinfo {pages} {4714} (\bibinfo {year} {2017})}\BibitemShut
  {NoStop}%
\bibitem [{\citenamefont {Berland}\ \emph {et~al.}(2015)\citenamefont
  {Berland}, \citenamefont {Cooper}, \citenamefont {Lee}, \citenamefont
  {Schr\"{o}der}, \citenamefont {Thonhauser}, \citenamefont {Hyldgaard},\ and\
  \citenamefont {Lundqvist}}]{vdwdf_review}%
  \BibitemOpen
  \bibfield  {author} {\bibinfo {author} {\bibfnamefont {K.}~\bibnamefont
  {Berland}}, \bibinfo {author} {\bibfnamefont {V.~R.}\ \bibnamefont {Cooper}},
  \bibinfo {author} {\bibfnamefont {K.}~\bibnamefont {Lee}}, \bibinfo {author}
  {\bibfnamefont {E.}~\bibnamefont {Schr\"{o}der}}, \bibinfo {author}
  {\bibfnamefont {T.}~\bibnamefont {Thonhauser}}, \bibinfo {author}
  {\bibfnamefont {P.}~\bibnamefont {Hyldgaard}}, \ and\ \bibinfo {author}
  {\bibfnamefont {B.~I.}\ \bibnamefont {Lundqvist}},\ }\href@noop {} {\bibfield
   {journal} {\bibinfo  {journal} {Rep. Prog. Phys.}\ }\textbf {\bibinfo
  {volume} {78}},\ \bibinfo {pages} {066501} (\bibinfo {year}
  {2015})}\BibitemShut {NoStop}%
\bibitem [{\citenamefont {Hammer}, \citenamefont {Hansen},\ and\ \citenamefont
  {Norskov}(1999)}]{rpbe}%
  \BibitemOpen
  \bibfield  {author} {\bibinfo {author} {\bibfnamefont {B.}~\bibnamefont
  {Hammer}}, \bibinfo {author} {\bibfnamefont {L.~B.}\ \bibnamefont {Hansen}},
  \ and\ \bibinfo {author} {\bibfnamefont {J.~K.}\ \bibnamefont {Norskov}},\
  }\href@noop {} {\bibfield  {journal} {\bibinfo  {journal} {Phys. Rev. B}\
  }\textbf {\bibinfo {volume} {59}},\ \bibinfo {pages} {7413} (\bibinfo {year}
  {1999})}\BibitemShut {NoStop}%
\bibitem [{\citenamefont {Zhang}\ and\ \citenamefont {Yang}(1998)}]{revpbe}%
  \BibitemOpen
  \bibfield  {author} {\bibinfo {author} {\bibfnamefont {Y.}~\bibnamefont
  {Zhang}}\ and\ \bibinfo {author} {\bibfnamefont {W.}~\bibnamefont {Yang}},\
  }\href@noop {} {\bibfield  {journal} {\bibinfo  {journal} {Phys. Rev. Lett.}\
  }\textbf {\bibinfo {volume} {80}},\ \bibinfo {pages} {890} (\bibinfo {year}
  {1998})}\BibitemShut {NoStop}%
\bibitem [{\citenamefont {Becke}(1988)}]{b88}%
  \BibitemOpen
  \bibfield  {author} {\bibinfo {author} {\bibfnamefont {A.~D.}\ \bibnamefont
  {Becke}},\ }\href@noop {} {\bibfield  {journal} {\bibinfo  {journal} {Phys.
  Rev. A}\ }\textbf {\bibinfo {volume} {38}},\ \bibinfo {pages} {3098}
  (\bibinfo {year} {1988})}\BibitemShut {NoStop}%
\bibitem [{\citenamefont {Lee}, \citenamefont {Yang},\ and\ \citenamefont
  {Parr}(1988)}]{lyp}%
  \BibitemOpen
  \bibfield  {author} {\bibinfo {author} {\bibfnamefont {C.}~\bibnamefont
  {Lee}}, \bibinfo {author} {\bibfnamefont {W.}~\bibnamefont {Yang}}, \ and\
  \bibinfo {author} {\bibfnamefont {R.~G.}\ \bibnamefont {Parr}},\ }\href@noop
  {} {\bibfield  {journal} {\bibinfo  {journal} {Phys. Rev. B}\ }\textbf
  {\bibinfo {volume} {37}},\ \bibinfo {pages} {785} (\bibinfo {year}
  {1988})}\BibitemShut {NoStop}%
\bibitem [{\citenamefont {Zhao}\ and\ \citenamefont {Truhlar}(2006)}]{m06l}%
  \BibitemOpen
  \bibfield  {author} {\bibinfo {author} {\bibfnamefont {Y.}~\bibnamefont
  {Zhao}}\ and\ \bibinfo {author} {\bibfnamefont {D.~G.}\ \bibnamefont
  {Truhlar}},\ }\href@noop {} {\bibfield  {journal} {\bibinfo  {journal} {J.
  Chem. Phys.}\ }\textbf {\bibinfo {volume} {125}},\ \bibinfo {pages} {194101}
  (\bibinfo {year} {2006})}\BibitemShut {NoStop}%
\bibitem [{\citenamefont {Sun}, \citenamefont {Ruzsinszky},\ and\ \citenamefont
  {Perdew}(2015)}]{scan}%
  \BibitemOpen
  \bibfield  {author} {\bibinfo {author} {\bibfnamefont {J.}~\bibnamefont
  {Sun}}, \bibinfo {author} {\bibfnamefont {A.}~\bibnamefont {Ruzsinszky}}, \
  and\ \bibinfo {author} {\bibfnamefont {J.~P.}\ \bibnamefont {Perdew}},\
  }\href@noop {} {\bibfield  {journal} {\bibinfo  {journal} {Phys. Rev. Lett.}\
  }\textbf {\bibinfo {volume} {115}},\ \bibinfo {pages} {036402} (\bibinfo
  {year} {2015})}\BibitemShut {NoStop}%
\bibitem [{\citenamefont {Tao}\ \emph {et~al.}(2003)\citenamefont {Tao},
  \citenamefont {Perdew}, \citenamefont {Staroverov},\ and\ \citenamefont
  {Scuseria}}]{tpss}%
  \BibitemOpen
  \bibfield  {author} {\bibinfo {author} {\bibfnamefont {J.}~\bibnamefont
  {Tao}}, \bibinfo {author} {\bibfnamefont {J.~P.}\ \bibnamefont {Perdew}},
  \bibinfo {author} {\bibfnamefont {V.~N.}\ \bibnamefont {Staroverov}}, \ and\
  \bibinfo {author} {\bibfnamefont {G.~E.}\ \bibnamefont {Scuseria}},\
  }\href@noop {} {\bibfield  {journal} {\bibinfo  {journal} {Phys. Rev. Lett.}\
  }\textbf {\bibinfo {volume} {91}},\ \bibinfo {pages} {146401} (\bibinfo
  {year} {2003})}\BibitemShut {NoStop}%
\bibitem [{\citenamefont {Klime\v{s}}, \citenamefont {Bowler},\ and\
  \citenamefont {Michaelides}(2010)}]{optpbevdw}%
  \BibitemOpen
  \bibfield  {author} {\bibinfo {author} {\bibfnamefont {J.}~\bibnamefont
  {Klime\v{s}}}, \bibinfo {author} {\bibfnamefont {D.~R.}\ \bibnamefont
  {Bowler}}, \ and\ \bibinfo {author} {\bibfnamefont {A.}~\bibnamefont
  {Michaelides}},\ }\href {\doibase 10.1088/0953-8984/22/2/022201} {\bibfield
  {journal} {\bibinfo  {journal} {J. Phys.: Condens. Matter}\ }\textbf
  {\bibinfo {volume} {22}},\ \bibinfo {pages} {022201} (\bibinfo {year}
  {2010})}\BibitemShut {NoStop}%
\bibitem [{\citenamefont {Lee}\ \emph {et~al.}(2010)\citenamefont {Lee},
  \citenamefont {Murray}, \citenamefont {Kong}, \citenamefont {Lundqvist},\
  and\ \citenamefont {Langreth}}]{vdwdf2}%
  \BibitemOpen
  \bibfield  {author} {\bibinfo {author} {\bibfnamefont {K.}~\bibnamefont
  {Lee}}, \bibinfo {author} {\bibfnamefont {E.~D.}\ \bibnamefont {Murray}},
  \bibinfo {author} {\bibfnamefont {L.}~\bibnamefont {Kong}}, \bibinfo {author}
  {\bibfnamefont {B.~I.}\ \bibnamefont {Lundqvist}}, \ and\ \bibinfo {author}
  {\bibfnamefont {D.~C.}\ \bibnamefont {Langreth}},\ }\href {\doibase
  10.1103/PhysRevB.82.081101} {\bibfield  {journal} {\bibinfo  {journal} {Phys.
  Rev. B}\ }\textbf {\bibinfo {volume} {82}},\ \bibinfo {pages} {081101}
  (\bibinfo {year} {2010})}\BibitemShut {NoStop}%
\bibitem [{\citenamefont {Hamada}(2014)}]{revvdwdf2}%
  \BibitemOpen
  \bibfield  {author} {\bibinfo {author} {\bibfnamefont {I.}~\bibnamefont
  {Hamada}},\ }\href@noop {} {\bibfield  {journal} {\bibinfo  {journal} {{Phys.
  Rev. B}}\ }\textbf {\bibinfo {volume} {89}},\ \bibinfo {pages} {121103(R)}
  (\bibinfo {year} {2014})}\BibitemShut {NoStop}%
\bibitem [{\citenamefont {Adamo}\ and\ \citenamefont {Barone}(1999)}]{pbe0}%
  \BibitemOpen
  \bibfield  {author} {\bibinfo {author} {\bibfnamefont {C.}~\bibnamefont
  {Adamo}}\ and\ \bibinfo {author} {\bibfnamefont {V.}~\bibnamefont {Barone}},\
  }\href@noop {} {\bibfield  {journal} {\bibinfo  {journal} {J. Chem. Phys.}\
  }\textbf {\bibinfo {volume} {110}},\ \bibinfo {pages} {6158} (\bibinfo {year}
  {1999})}\BibitemShut {NoStop}%
\bibitem [{\citenamefont {Becke}(1993)}]{b3lypa}%
  \BibitemOpen
  \bibfield  {author} {\bibinfo {author} {\bibfnamefont {A.~D.}\ \bibnamefont
  {Becke}},\ }\href@noop {} {\bibfield  {journal} {\bibinfo  {journal} {J.
  Chem. Phys.}\ }\textbf {\bibinfo {volume} {98}},\ \bibinfo {pages} {5648}
  (\bibinfo {year} {1993})}\BibitemShut {NoStop}%
\bibitem [{\citenamefont {Stephens}\ \emph {et~al.}(1994)\citenamefont
  {Stephens}, \citenamefont {Devlin}, \citenamefont {Chabalowski},\ and\
  \citenamefont {Frisch}}]{b3lypb}%
  \BibitemOpen
  \bibfield  {author} {\bibinfo {author} {\bibfnamefont {P.~J.}\ \bibnamefont
  {Stephens}}, \bibinfo {author} {\bibfnamefont {F.~J.}\ \bibnamefont
  {Devlin}}, \bibinfo {author} {\bibfnamefont {C.~F.}\ \bibnamefont
  {Chabalowski}}, \ and\ \bibinfo {author} {\bibfnamefont {M.~J.}\ \bibnamefont
  {Frisch}},\ }\href@noop {} {\bibfield  {journal} {\bibinfo  {journal} {J.
  Phys. Chem.}\ }\textbf {\bibinfo {volume} {98}},\ \bibinfo {pages} {11623}
  (\bibinfo {year} {1994})}\BibitemShut {NoStop}%
\bibitem [{\citenamefont {Sure}\ and\ \citenamefont {Grimme}(2013)}]{hf3c}%
  \BibitemOpen
  \bibfield  {author} {\bibinfo {author} {\bibfnamefont {R.}~\bibnamefont
  {Sure}}\ and\ \bibinfo {author} {\bibfnamefont {S.}~\bibnamefont {Grimme}},\
  }\href@noop {} {\bibfield  {journal} {\bibinfo  {journal} {J. Comput. Chem.}\
  }\textbf {\bibinfo {volume} {34}},\ \bibinfo {pages} {1672} (\bibinfo {year}
  {2013})}\BibitemShut {NoStop}%
\bibitem [{\citenamefont {Cutini}\ \emph {et~al.}(2016)\citenamefont {Cutini},
  \citenamefont {Civalleri}, \citenamefont {Corno}, \citenamefont {Orlando},
  \citenamefont {Brandenburg}, \citenamefont {Maschio},\ and\ \citenamefont
  {Ugliengoa}}]{shf3c}%
  \BibitemOpen
  \bibfield  {author} {\bibinfo {author} {\bibfnamefont {M.}~\bibnamefont
  {Cutini}}, \bibinfo {author} {\bibfnamefont {B.}~\bibnamefont {Civalleri}},
  \bibinfo {author} {\bibfnamefont {M.}~\bibnamefont {Corno}}, \bibinfo
  {author} {\bibfnamefont {R.}~\bibnamefont {Orlando}}, \bibinfo {author}
  {\bibfnamefont {J.~G.}\ \bibnamefont {Brandenburg}}, \bibinfo {author}
  {\bibfnamefont {L.}~\bibnamefont {Maschio}}, \ and\ \bibinfo {author}
  {\bibfnamefont {P.}~\bibnamefont {Ugliengoa}},\ }\href {\doibase
  10.1021/acs.jctc.6b00304} {\bibfield  {journal} {\bibinfo  {journal} {{J.
  Chem. Theory Comput.}}\ }\textbf {\bibinfo {volume} {12}},\ \bibinfo {pages}
  {3340} (\bibinfo {year} {2016})}\BibitemShut {NoStop}%
\bibitem [{\citenamefont {Brandenburg}, \citenamefont {Caldeweyher},\ and\
  \citenamefont {Grimme}(2016)}]{hse3c}%
  \BibitemOpen
  \bibfield  {author} {\bibinfo {author} {\bibfnamefont {J.~G.}\ \bibnamefont
  {Brandenburg}}, \bibinfo {author} {\bibfnamefont {E.}~\bibnamefont
  {Caldeweyher}}, \ and\ \bibinfo {author} {\bibfnamefont {S.}~\bibnamefont
  {Grimme}},\ }\href {\doibase 10.1039/C6CP01697A} {\bibfield  {journal}
  {\bibinfo  {journal} {{Phys. Chem. Chem. Phys.}}\ }\textbf {\bibinfo {volume}
  {18}},\ \bibinfo {pages} {15519} (\bibinfo {year} {2016})}\BibitemShut
  {NoStop}%
\bibitem [{\citenamefont {Brandenburg}\ \emph {et~al.}(2018)\citenamefont
  {Brandenburg}, \citenamefont {Bannwarth}, \citenamefont {Hansen},\ and\
  \citenamefont {Grimme}}]{b973c}%
  \BibitemOpen
  \bibfield  {author} {\bibinfo {author} {\bibfnamefont {J.~G.}\ \bibnamefont
  {Brandenburg}}, \bibinfo {author} {\bibfnamefont {C.}~\bibnamefont
  {Bannwarth}}, \bibinfo {author} {\bibfnamefont {A.}~\bibnamefont {Hansen}}, \
  and\ \bibinfo {author} {\bibfnamefont {S.}~\bibnamefont {Grimme}},\ }\href
  {\doibase 10.1063/1.5012601} {\bibfield  {journal} {\bibinfo  {journal} {{J.
  Chem. Phys.}}\ }\textbf {\bibinfo {volume} {148}},\ \bibinfo {pages} {064104}
  (\bibinfo {year} {2018})}\BibitemShut {NoStop}%
\bibitem [{\citenamefont {London}(1937)}]{london1937}%
  \BibitemOpen
  \bibfield  {author} {\bibinfo {author} {\bibfnamefont {F.}~\bibnamefont
  {London}},\ }\href {\doibase 10.1039/TF937330008B} {\bibfield  {journal}
  {\bibinfo  {journal} {Trans. Faraday Soc.}\ }\textbf {\bibinfo {volume}
  {33}},\ \bibinfo {pages} {8} (\bibinfo {year} {1937})}\BibitemShut {NoStop}%
\bibitem [{\citenamefont {Stone}(1997)}]{stone}%
  \BibitemOpen
  \bibfield  {author} {\bibinfo {author} {\bibfnamefont {A.~J.}\ \bibnamefont
  {Stone}},\ }\href@noop {} {\emph {\bibinfo {title} {{The Theory of
  Intermolecular Forces}}}}\ (\bibinfo  {publisher} {Oxford University Press},\
  \bibinfo {address} {Oxford},\ \bibinfo {year} {1997})\BibitemShut {NoStop}%
\bibitem [{\citenamefont {Jure\v{c}ka}\ \emph {et~al.}(2006)\citenamefont
  {Jure\v{c}ka}, \citenamefont {\v{S}poner}, \citenamefont {Cerny},\ and\
  \citenamefont {Hobza}}]{s22}%
  \BibitemOpen
  \bibfield  {author} {\bibinfo {author} {\bibfnamefont {P.}~\bibnamefont
  {Jure\v{c}ka}}, \bibinfo {author} {\bibfnamefont {J.}~\bibnamefont
  {\v{S}poner}}, \bibinfo {author} {\bibfnamefont {J.}~\bibnamefont {Cerny}}, \
  and\ \bibinfo {author} {\bibfnamefont {P.}~\bibnamefont {Hobza}},\
  }\href@noop {} {\bibfield  {journal} {\bibinfo  {journal} {Phys. Chem. Chem.
  Phys.}\ }\textbf {\bibinfo {volume} {8}},\ \bibinfo {pages} {1985} (\bibinfo
  {year} {2006})}\BibitemShut {NoStop}%
\bibitem [{\citenamefont {Kristy{\'a}n}\ and\ \citenamefont
  {Pulay}(1994)}]{kristyan1994}%
  \BibitemOpen
  \bibfield  {author} {\bibinfo {author} {\bibfnamefont {S.}~\bibnamefont
  {Kristy{\'a}n}}\ and\ \bibinfo {author} {\bibfnamefont {P.}~\bibnamefont
  {Pulay}},\ }\href@noop {} {\bibfield  {journal} {\bibinfo  {journal} {Chem.
  Phys. Lett.}\ }\textbf {\bibinfo {volume} {229}},\ \bibinfo {pages} {175}
  (\bibinfo {year} {1994})}\BibitemShut {NoStop}%
\bibitem [{\citenamefont {Axilrod}\ and\ \citenamefont {Teller}(1943)}]{atm1}%
  \BibitemOpen
  \bibfield  {author} {\bibinfo {author} {\bibfnamefont {B.~M.}\ \bibnamefont
  {Axilrod}}\ and\ \bibinfo {author} {\bibfnamefont {E.}~\bibnamefont
  {Teller}},\ }\href@noop {} {\bibfield  {journal} {\bibinfo  {journal} {J.
  Chem. Phys.}\ }\textbf {\bibinfo {volume} {11}},\ \bibinfo {pages} {299}
  (\bibinfo {year} {1943})}\BibitemShut {NoStop}%
\bibitem [{\citenamefont {Muto}(1944)}]{atm2}%
  \BibitemOpen
  \bibfield  {author} {\bibinfo {author} {\bibfnamefont {Y.}~\bibnamefont
  {Muto}},\ }\href@noop {} {\bibfield  {journal} {\bibinfo  {journal} {Proc.
  Phys. Math. Soc. Jpn.}\ }\textbf {\bibinfo {volume} {17}},\ \bibinfo {pages}
  {629} (\bibinfo {year} {1944})}\BibitemShut {NoStop}%
\bibitem [{\citenamefont {Peach}, \citenamefont {Williamson},\ and\
  \citenamefont {Tozer}(2011)}]{peach2011}%
  \BibitemOpen
  \bibfield  {author} {\bibinfo {author} {\bibfnamefont {M.~J.~G.}\
  \bibnamefont {Peach}}, \bibinfo {author} {\bibfnamefont {M.~J.}\ \bibnamefont
  {Williamson}}, \ and\ \bibinfo {author} {\bibfnamefont {D.~J.}\ \bibnamefont
  {Tozer}},\ }\href@noop {} {\bibfield  {journal} {\bibinfo  {journal} {J.
  Chem. Theory Comput.}\ }\textbf {\bibinfo {volume} {7}},\ \bibinfo {pages}
  {3578} (\bibinfo {year} {2011})}\BibitemShut {NoStop}%
\bibitem [{\citenamefont {Johnson}, \citenamefont {Otero-de-la Roza},\ and\
  \citenamefont {Dale}(2013)}]{model-overpolarization}%
  \BibitemOpen
  \bibfield  {author} {\bibinfo {author} {\bibfnamefont {E.~R.}\ \bibnamefont
  {Johnson}}, \bibinfo {author} {\bibfnamefont {A.}~\bibnamefont {Otero-de-la
  Roza}}, \ and\ \bibinfo {author} {\bibfnamefont {S.~G.}\ \bibnamefont
  {Dale}},\ }\href {\doibase 10.1063/1.4829642} {\bibfield  {journal} {\bibinfo
   {journal} {J. Chem. Phys.}\ }\textbf {\bibinfo {volume} {139}},\ \bibinfo
  {pages} {184116} (\bibinfo {year} {2013})}\BibitemShut {NoStop}%
\bibitem [{\citenamefont {Bryantsev}\ \emph {et~al.}(2009)\citenamefont
  {Bryantsev}, \citenamefont {Diallo}, \citenamefont {van Duin},\ and\
  \citenamefont {Goddard}}]{water27}%
  \BibitemOpen
  \bibfield  {author} {\bibinfo {author} {\bibfnamefont {V.~S.}\ \bibnamefont
  {Bryantsev}}, \bibinfo {author} {\bibfnamefont {M.~S.}\ \bibnamefont
  {Diallo}}, \bibinfo {author} {\bibfnamefont {A.~C.~T.}\ \bibnamefont {van
  Duin}}, \ and\ \bibinfo {author} {\bibfnamefont {W.~A.}\ \bibnamefont
  {Goddard}},\ }\href@noop {} {\bibfield  {journal} {\bibinfo  {journal} {J.
  Chem. Theory Comput.}\ }\textbf {\bibinfo {volume} {5}},\ \bibinfo {pages}
  {1016} (\bibinfo {year} {2009})}\BibitemShut {NoStop}%
\bibitem [{\citenamefont {Goerigk}, \citenamefont {Kruse},\ and\ \citenamefont
  {Grimme}(2011)}]{s66-bench}%
  \BibitemOpen
  \bibfield  {author} {\bibinfo {author} {\bibfnamefont {L.}~\bibnamefont
  {Goerigk}}, \bibinfo {author} {\bibfnamefont {H.}~\bibnamefont {Kruse}}, \
  and\ \bibinfo {author} {\bibfnamefont {S.}~\bibnamefont {Grimme}},\
  }\href@noop {} {\bibfield  {journal} {\bibinfo  {journal} {Chem. Phys. Chem}\
  }\textbf {\bibinfo {volume} {12}},\ \bibinfo {pages} {3421} (\bibinfo {year}
  {2011})}\BibitemShut {NoStop}%
\bibitem [{\citenamefont {\v{R}ez\'{a}\v{c}}\ \emph {et~al.}(2015)\citenamefont
  {\v{R}ez\'{a}\v{c}}, \citenamefont {Huang}, \citenamefont {Hobza},\ and\
  \citenamefont {Beran}}]{beran-threebody}%
  \BibitemOpen
  \bibfield  {author} {\bibinfo {author} {\bibfnamefont {J.}~\bibnamefont
  {\v{R}ez\'{a}\v{c}}}, \bibinfo {author} {\bibfnamefont {J.}~\bibnamefont
  {Huang}}, \bibinfo {author} {\bibfnamefont {P.}~\bibnamefont {Hobza}}, \ and\
  \bibinfo {author} {\bibfnamefont {G.~J.~O.}\ \bibnamefont {Beran}},\ }\href
  {\doibase 10.1021/acs.jctc.5b00281} {\bibfield  {journal} {\bibinfo
  {journal} {J. Chem. Theory. Comput.}\ }\textbf {\bibinfo {volume} {11}},\
  \bibinfo {pages} {3065} (\bibinfo {year} {2015})}\BibitemShut {NoStop}%
\bibitem [{\citenamefont {Dion}\ \emph
  {et~al.}(2004{\natexlab{b}})\citenamefont {Dion}, \citenamefont {Rydberg},
  \citenamefont {Schr{\"o}der}, \citenamefont {Langreth},\ and\ \citenamefont
  {Lundqvist}}]{dion2004}%
  \BibitemOpen
  \bibfield  {author} {\bibinfo {author} {\bibfnamefont {M.}~\bibnamefont
  {Dion}}, \bibinfo {author} {\bibfnamefont {H.}~\bibnamefont {Rydberg}},
  \bibinfo {author} {\bibfnamefont {E.}~\bibnamefont {Schr{\"o}der}}, \bibinfo
  {author} {\bibfnamefont {D.~C.}\ \bibnamefont {Langreth}}, \ and\ \bibinfo
  {author} {\bibfnamefont {B.~I.}\ \bibnamefont {Lundqvist}},\ }\href {\doibase
  10.1103/PhysRevLett.92.246401} {\bibfield  {journal} {\bibinfo  {journal}
  {Phys. Rev. Lett.}\ }\textbf {\bibinfo {volume} {92}},\ \bibinfo {pages}
  {246401} (\bibinfo {year} {2004}{\natexlab{b}})}\BibitemShut {NoStop}%
\bibitem [{\citenamefont {Dion}\ \emph {et~al.}(2005)\citenamefont {Dion},
  \citenamefont {Rydberg}, \citenamefont {Schr{\"o}der}, \citenamefont
  {Langreth},\ and\ \citenamefont {Lundqvist}}]{dion2005}%
  \BibitemOpen
  \bibfield  {author} {\bibinfo {author} {\bibfnamefont {M.}~\bibnamefont
  {Dion}}, \bibinfo {author} {\bibfnamefont {H.}~\bibnamefont {Rydberg}},
  \bibinfo {author} {\bibfnamefont {E.}~\bibnamefont {Schr{\"o}der}}, \bibinfo
  {author} {\bibfnamefont {D.~C.}\ \bibnamefont {Langreth}}, \ and\ \bibinfo
  {author} {\bibfnamefont {B.~I.}\ \bibnamefont {Lundqvist}},\ }\href {\doibase
  10.1103/PhysRevLett.95.109902} {\bibfield  {journal} {\bibinfo  {journal}
  {Phys. Rev. Lett.}\ }\textbf {\bibinfo {volume} {95}},\ \bibinfo {pages}
  {109902} (\bibinfo {year} {2005})}\BibitemShut {NoStop}%
\bibitem [{ice()}]{ice-errorcompensation}%
  \BibitemOpen
  \href@noop {} {}\bibinfo {note} {Replacing (a) the hard PAWs in VASP\,5.4 by
  the standard ones or (b) using CP2K with GPW and TZV2P basis set both results
  in a systematic shift towards stronger bound systems, e.g. (a) 28 or (b) 29
  meV for ice Ih, see Table~\ref{tab:convergence}. For RPBE-D4 or revPBE-D4,
  this would systematically remove most of the systematic shift (MDs of 24 and
  35\,meV for 2D/3D-ice) giving artificially excellent results. This might
  partially explain why revPBE-D3 has been one of the most popular GGA
  functionals applied to water and ice in recent
  years.\cite{revpbed3_water,revpbe0d3_water, watermd_mgga}}\BibitemShut
  {NoStop}%
\bibitem [{\citenamefont {Mardirossian}\ and\ \citenamefont
  {Head-Gordon}(2016{\natexlab{b}})}]{minnesota_grid}%
  \BibitemOpen
  \bibfield  {author} {\bibinfo {author} {\bibfnamefont {N.}~\bibnamefont
  {Mardirossian}}\ and\ \bibinfo {author} {\bibfnamefont {M.}~\bibnamefont
  {Head-Gordon}},\ }\href {\doibase 10.1021/acs.jctc.6b00637} {\bibfield
  {journal} {\bibinfo  {journal} {J. Chem. Theory Comput.}\ }\textbf {\bibinfo
  {volume} {12}},\ \bibinfo {pages} {4303} (\bibinfo {year}
  {2016}{\natexlab{b}})}\BibitemShut {NoStop}%
\bibitem [{\citenamefont {Sun}\ \emph {et~al.}(2016)\citenamefont {Sun},
  \citenamefont {Remsing}, \citenamefont {Zhang}, \citenamefont {Sun},
  \citenamefont {Ruzsinszky}, \citenamefont {Peng}, \citenamefont {Yang},
  \citenamefont {Paul}, \citenamefont {Waghmare}, \citenamefont {Wu},
  \citenamefont {Klein},\ and\ \citenamefont {Perdew}}]{scan_natchem}%
  \BibitemOpen
  \bibfield  {author} {\bibinfo {author} {\bibfnamefont {J.}~\bibnamefont
  {Sun}}, \bibinfo {author} {\bibfnamefont {R.}~\bibnamefont {Remsing}},
  \bibinfo {author} {\bibfnamefont {Y.}~\bibnamefont {Zhang}}, \bibinfo
  {author} {\bibfnamefont {Z.}~\bibnamefont {Sun}}, \bibinfo {author}
  {\bibfnamefont {A.}~\bibnamefont {Ruzsinszky}}, \bibinfo {author}
  {\bibfnamefont {H.}~\bibnamefont {Peng}}, \bibinfo {author} {\bibfnamefont
  {Z.}~\bibnamefont {Yang}}, \bibinfo {author} {\bibfnamefont {A.}~\bibnamefont
  {Paul}}, \bibinfo {author} {\bibfnamefont {U.}~\bibnamefont {Waghmare}},
  \bibinfo {author} {\bibfnamefont {X.}~\bibnamefont {Wu}}, \bibinfo {author}
  {\bibfnamefont {M.~L.}\ \bibnamefont {Klein}}, \ and\ \bibinfo {author}
  {\bibfnamefont {J.~P.}\ \bibnamefont {Perdew}},\ }\href {\doibase
  10.1038/nchem.2535} {\bibfield  {journal} {\bibinfo  {journal} {Nat. Chem.}\
  }\textbf {\bibinfo {volume} {8}},\ \bibinfo {pages} {831} (\bibinfo {year}
  {2016})}\BibitemShut {NoStop}%
\bibitem [{\citenamefont {Caldeweyher}\ and\ \citenamefont
  {Brandenburg}(2018)}]{3c_review}%
  \BibitemOpen
  \bibfield  {author} {\bibinfo {author} {\bibfnamefont {E.}~\bibnamefont
  {Caldeweyher}}\ and\ \bibinfo {author} {\bibfnamefont {J.~G.}\ \bibnamefont
  {Brandenburg}},\ }\href {\doibase 10.1088/1361-648X/aabcfb} {\bibfield
  {journal} {\bibinfo  {journal} {{J. Phys.: Condens. Matter}}\ }\textbf
  {\bibinfo {volume} {30}},\ \bibinfo {pages} {213001} (\bibinfo {year}
  {2018})}\BibitemShut {NoStop}%
\bibitem [{\citenamefont {Whalley}(1984)}]{whalley_ice}%
  \BibitemOpen
  \bibfield  {author} {\bibinfo {author} {\bibfnamefont {E.}~\bibnamefont
  {Whalley}},\ }\href {\doibase 10.1063/1.448153} {\bibfield  {journal}
  {\bibinfo  {journal} {J. Chem. Phys.}\ }\textbf {\bibinfo {volume} {81}},\
  \bibinfo {pages} {4087} (\bibinfo {year} {1984})}\BibitemShut {NoStop}%
\bibitem [{gil()}]{gillan}%
  \BibitemOpen
  \href@noop {} {}\bibinfo {note} {For explicit definition of all quantities
  and scoring procedures, please read section IX\,D of
  Ref.~\onlinecite{dftforwater_perspective}. Small numerical differences to the
  original scores might occur due to updated references and slightly different
  numerical settings. Here, all molecular properties are computed with {\sc
  Orca} in a larger def2-QZVPDD basis set, large integration grind, and tight
  convergence thresholds.}\BibitemShut {Stop}%
\bibitem [{\citenamefont {\v{R}ez\'a\v{c}}, \citenamefont {Riley},\ and\
  \citenamefont {Hobza}(2011)}]{s66}%
  \BibitemOpen
  \bibfield  {author} {\bibinfo {author} {\bibfnamefont {J.}~\bibnamefont
  {\v{R}ez\'a\v{c}}}, \bibinfo {author} {\bibfnamefont {K.~E.}\ \bibnamefont
  {Riley}}, \ and\ \bibinfo {author} {\bibfnamefont {P.}~\bibnamefont
  {Hobza}},\ }\href@noop {} {\bibfield  {journal} {\bibinfo  {journal} {J.
  Chem. Theory Comput.}\ }\textbf {\bibinfo {volume} {7}},\ \bibinfo {pages}
  {2427} (\bibinfo {year} {2011})}\BibitemShut {NoStop}%
\bibitem [{\citenamefont {Sedlak}\ \emph {et~al.}(2013)\citenamefont {Sedlak},
  \citenamefont {Janowski}, \citenamefont {Pito\v{n}\'{a}k}, \citenamefont
  {\v{R}ez\'{a}\v{c}}, \citenamefont {Pulay},\ and\ \citenamefont
  {Hobza}}]{l7}%
  \BibitemOpen
  \bibfield  {author} {\bibinfo {author} {\bibfnamefont {R.}~\bibnamefont
  {Sedlak}}, \bibinfo {author} {\bibfnamefont {T.}~\bibnamefont {Janowski}},
  \bibinfo {author} {\bibfnamefont {M.}~\bibnamefont {Pito\v{n}\'{a}k}},
  \bibinfo {author} {\bibfnamefont {J.}~\bibnamefont {\v{R}ez\'{a}\v{c}}},
  \bibinfo {author} {\bibfnamefont {P.}~\bibnamefont {Pulay}}, \ and\ \bibinfo
  {author} {\bibfnamefont {P.}~\bibnamefont {Hobza}},\ }\href@noop {}
  {\bibfield  {journal} {\bibinfo  {journal} {J. Chem. Theory Comput.}\
  }\textbf {\bibinfo {volume} {9}},\ \bibinfo {pages} {3364} (\bibinfo {year}
  {2013})}\BibitemShut {NoStop}%
\bibitem [{\citenamefont {Grimme}(2012)}]{s12l}%
  \BibitemOpen
  \bibfield  {author} {\bibinfo {author} {\bibfnamefont {S.}~\bibnamefont
  {Grimme}},\ }\href@noop {} {\bibfield  {journal} {\bibinfo  {journal} {Chem.
  Eur. J.}\ }\textbf {\bibinfo {volume} {18}},\ \bibinfo {pages} {9955}
  (\bibinfo {year} {2012})}\BibitemShut {NoStop}%
\bibitem [{\citenamefont {Dohm}\ \emph {et~al.}(2018)\citenamefont {Dohm},
  \citenamefont {Hansen}, \citenamefont {Steinmetz}, \citenamefont {Grimme},\
  and\ \citenamefont {Checinski}}]{mor41}%
  \BibitemOpen
  \bibfield  {author} {\bibinfo {author} {\bibfnamefont {S.}~\bibnamefont
  {Dohm}}, \bibinfo {author} {\bibfnamefont {A.}~\bibnamefont {Hansen}},
  \bibinfo {author} {\bibfnamefont {M.}~\bibnamefont {Steinmetz}}, \bibinfo
  {author} {\bibfnamefont {S.}~\bibnamefont {Grimme}}, \ and\ \bibinfo {author}
  {\bibfnamefont {M.~P.}\ \bibnamefont {Checinski}},\ }\href {\doibase
  10.1021/acs.jctc.7b01183} {\bibfield  {journal} {\bibinfo  {journal} {J.
  Chem. Theory Comput.}\ }\textbf {\bibinfo {volume} {14}},\ \bibinfo {pages}
  {2596} (\bibinfo {year} {2018})}\BibitemShut {NoStop}%
\bibitem [{\citenamefont {Yao}\ and\ \citenamefont
  {Kanai}(2017)}]{scan_paw_numerics}%
  \BibitemOpen
  \bibfield  {author} {\bibinfo {author} {\bibfnamefont {Y.}~\bibnamefont
  {Yao}}\ and\ \bibinfo {author} {\bibfnamefont {Y.}~\bibnamefont {Kanai}},\
  }\href {\doibase 10.1063/1.4984939} {\bibfield  {journal} {\bibinfo
  {journal} {J. Chem. Phys.}\ }\textbf {\bibinfo {volume} {146}},\ \bibinfo
  {pages} {224105} (\bibinfo {year} {2017})}\BibitemShut {NoStop}%
\bibitem [{\citenamefont {Yang}, \citenamefont {Tan},\ and\ \citenamefont
  {Rappe}(2018)}]{pbe0_pp_numerics}%
  \BibitemOpen
  \bibfield  {author} {\bibinfo {author} {\bibfnamefont {J.}~\bibnamefont
  {Yang}}, \bibinfo {author} {\bibfnamefont {L.~Z.}\ \bibnamefont {Tan}}, \
  and\ \bibinfo {author} {\bibfnamefont {A.~M.}\ \bibnamefont {Rappe}},\ }\href
  {\doibase 10.1103/PhysRevB.97.085130} {\bibfield  {journal} {\bibinfo
  {journal} {Phys. Rev. B}\ }\textbf {\bibinfo {volume} {97}},\ \bibinfo
  {pages} {085130} (\bibinfo {year} {2018})}\BibitemShut {NoStop}%
\bibitem [{\citenamefont {Hamada}\ and\ \citenamefont
  {Yanagisawa}(2011)}]{vdwdf_pp_numerics}%
  \BibitemOpen
  \bibfield  {author} {\bibinfo {author} {\bibfnamefont {I.}~\bibnamefont
  {Hamada}}\ and\ \bibinfo {author} {\bibfnamefont {S.}~\bibnamefont
  {Yanagisawa}},\ }\href {\doibase 10.1103/PhysRevB.84.153104} {\bibfield
  {journal} {\bibinfo  {journal} {Phys. Rev. B}\ }\textbf {\bibinfo {volume}
  {84}},\ \bibinfo {pages} {153104} (\bibinfo {year} {2011})}\BibitemShut
  {NoStop}%
\bibitem [{\citenamefont {Dodia}\ \emph {et~al.}(2019)\citenamefont {Dodia},
  \citenamefont {Ohto}, \citenamefont {Imoto},\ and\ \citenamefont
  {Nagata}}]{revpbed3_water}%
  \BibitemOpen
  \bibfield  {author} {\bibinfo {author} {\bibfnamefont {M.}~\bibnamefont
  {Dodia}}, \bibinfo {author} {\bibfnamefont {T.}~\bibnamefont {Ohto}},
  \bibinfo {author} {\bibfnamefont {S.}~\bibnamefont {Imoto}}, \ and\ \bibinfo
  {author} {\bibfnamefont {Y.}~\bibnamefont {Nagata}},\ }\href {\doibase
  10.1021/acs.jctc.9b00253} {\bibfield  {journal} {\bibinfo  {journal} {J.
  Chem. Theory Comput.}\ ,\ \bibinfo {pages} {in press}} (\bibinfo {year}
  {2019})}\BibitemShut {NoStop}%
\bibitem [{\citenamefont {Ruiz~Pestana}\ \emph {et~al.}(2018)\citenamefont
  {Ruiz~Pestana}, \citenamefont {Marsalek}, \citenamefont {Markland},\ and\
  \citenamefont {Head-Gordon}}]{revpbe0d3_water}%
  \BibitemOpen
  \bibfield  {author} {\bibinfo {author} {\bibfnamefont {L.}~\bibnamefont
  {Ruiz~Pestana}}, \bibinfo {author} {\bibfnamefont {O.}~\bibnamefont
  {Marsalek}}, \bibinfo {author} {\bibfnamefont {T.~E.}\ \bibnamefont
  {Markland}}, \ and\ \bibinfo {author} {\bibfnamefont {T.}~\bibnamefont
  {Head-Gordon}},\ }\href {\doibase 10.1021/acs.jpclett.8b02400} {\bibfield
  {journal} {\bibinfo  {journal} {J. Phys. Chem. Lett.}\ }\textbf {\bibinfo
  {volume} {9}},\ \bibinfo {pages} {5009} (\bibinfo {year} {2018})}\BibitemShut
  {NoStop}%
\bibitem [{\citenamefont {Ruiz~Pestana}\ \emph {et~al.}(2017)\citenamefont
  {Ruiz~Pestana}, \citenamefont {Mardirossian}, \citenamefont {Head-Gordon},\
  and\ \citenamefont {Head-Gordon}}]{watermd_mgga}%
  \BibitemOpen
  \bibfield  {author} {\bibinfo {author} {\bibfnamefont {L.}~\bibnamefont
  {Ruiz~Pestana}}, \bibinfo {author} {\bibfnamefont {N.}~\bibnamefont
  {Mardirossian}}, \bibinfo {author} {\bibfnamefont {M.}~\bibnamefont
  {Head-Gordon}}, \ and\ \bibinfo {author} {\bibfnamefont {T.}~\bibnamefont
  {Head-Gordon}},\ }\href {\doibase 10.1039/C6SC04711D} {\bibfield  {journal}
  {\bibinfo  {journal} {Chem. Sci.}\ }\textbf {\bibinfo {volume} {8}},\
  \bibinfo {pages} {3554} (\bibinfo {year} {2017})}\BibitemShut {NoStop}%
\end{thebibliography}%



\end{document}